\documentclass[aps,prx,superscriptaddress,nofootinbib,nobibnotes,floatfix,showpacs,reprint,longbibliography]{revtex4-2}
\usepackage{siunitx}
\usepackage[utf8]{inputenc}
\usepackage[T1]{fontenc}
\usepackage{lmodern}
\usepackage{orcidlink}
\usepackage{amsmath,amsfonts,amssymb,amsthm} \allowdisplaybreaks
\usepackage{mathtools}
\usepackage{nicematrix} 
\usepackage{pifont} 
\usepackage{bm} 
\usepackage{xcolor} 
\usepackage{xfrac} 
\usepackage{enumitem} 
\usepackage[normalem]{ulem} 
\usepackage{soul} 
\usepackage{subfigure}
\usepackage{multirow}
\usepackage{dcolumn} 
\usepackage{graphicx} 
\usepackage{hyperref} 
\hypersetup{
    unicode={true},
    colorlinks={true},
    linkcolor={blue},
    citecolor={blue},
    urlcolor={blue}
}
\usepackage[sectionbib]{bibunits}
\makeatletter
\def\maketitle{
\@author@finish
\title@column\titleblock@produce
\suppressfloats[t]}
\makeatother

\newcommand{\nocontentsline}[3]{}
\let\origcontentsline\addcontentsline
\newcommand\stoptoc{\let\addcontentsline\nocontentsline}
\newcommand\resumetoc{\let\addcontentsline\origcontentsline}

\makeatletter
\newcounter{savesection}
\newcounter{apdxsection}
\renewcommand\appendix{\par
  \setcounter{savesection}{\value{section}}%
  \setcounter{section}{\value{apdxsection}}%
  \setcounter{subsection}{0}%
  \gdef\thesubsection{\@Alph\c@section}}
\newcommand\unappendix{\par
  \setcounter{apdxsection}{\value{subsection}}%
  \setcounter{section}{\value{savesection}}%
  \setcounter{subsection}{0}%
  \gdef\thesubsection{\@arabic\c@section}}
\makeatother

\newcommand{\createsa}[1][i]{a_{#1,\sigma}^{\dagger}}
\newcommand{\annihilatesa}[1][i]{a_{#1,\sigma}}
\newcommand{\createsb}[1][j]{b_{#1,\sigma}^{\dagger}}
\newcommand{\annihilatesb}[1][j]{b_{#1,\sigma}}

\newcommand{\fcreatesa}[1][\mathbf{k}]{a_{#1,\sigma}^{\dagger}}
\newcommand{\fannihilatesa}[1][\mathbf{k}]{a_{#1,\sigma}}
\newcommand{\fcreatesb}[1][\mathbf{k}]{b_{#1,\sigma}^{\dagger}}
\newcommand{\fannihilatesb}[1][\mathbf{k}]{b_{#1,\sigma}}

\newcommand{\createta}[1][i]{a_{#1,\tau}^{\dagger}}
\newcommand{\annihilateta}[1][i]{a_{#1,\tau}}
\newcommand{\createtb}[1][j]{b_{#1,\tau}^{\dagger}}
\newcommand{\annihilatetb}[1][j]{b_{#1,\tau}}

\newcommand{\fcreateta}[1][\mathbf{k}]{a_{#1,\tau}^{\dagger}}
\newcommand{\fannihilateta}[1][\mathbf{k}]{a_{#1,\tau}}
\newcommand{\fcreatetb}[1][\mathbf{k}]{b_{#1,\tau}^{\dagger}}
\newcommand{\fannihilatetb}[1][\mathbf{k}]{b_{#1,\tau}}

\newcommand{\fbcreatesa}[1][\mathbf{k}]{{\alpha}_{#1,\sigma}^{\dagger}}
\newcommand{\fbannihilatesa}[1][\mathbf{k}]{{\alpha}_{#1,\sigma}}
\newcommand{\fbcreatesb}[1][\mathbf{k}]{{\beta}_{#1,\sigma}^{\dagger}}
\newcommand{\fbannihilatesb}[1][\mathbf{k}]{{\beta}_{#1,\sigma}}

\newcommand{\fbcreateta}[1][\mathbf{k}]{{\alpha}_{#1,\tau}^{\dagger}}
\newcommand{\fbannihilateta}[1][\mathbf{k}]{{\alpha}_{#1,\tau}}
\newcommand{\fbcreatetb}[1][\mathbf{k}]{{\beta}_{#1,\tau}^{\dagger}}
\newcommand{\fbannihilatetb}[1][\mathbf{k}]{{\beta}_{#1,\tau}}

\newcommand{\nasigma}[2]{\createsa[#1]\annihilatesa[#2]}
\newcommand{\nbsigma}[2]{\createsb[#1]\annihilatesb[#2]}
\newcommand{\natau}[2]{\createta[#1]\annihilateta[#2]}
\newcommand{\nbtau}[2]{\createtb[#1]\annihilatetb[#2]}

\newcommand{\fnasigma}[2]{\fcreatesa[#1]\fannihilatesa[#2]}
\newcommand{\fnbsigma}[2]{\fcreatesb[#1]\fannihilatesb[#2]}
\newcommand{\fnatau}[2]{\fcreateta[#1]\fannihilateta[#2]}
\newcommand{\fnbtau}[2]{\fcreatetb[#1]\fannihilatetb[#2]}

\newcommand{\fbnasigma}[2]{\fbcreatesa[#1]\fbannihilatesa[#2]}
\newcommand{\fbnbsigma}[2]{\fbcreatesb[#1]\fbannihilatesb[#2]}
\newcommand{\fbnatau}[2]{\fbcreateta[#1]\fbannihilateta[#2]}
\newcommand{\fbnbtau}[2]{\fbcreatetb[#1]\fbannihilatetb[#2]}

\begin{document}

\title{A Unified Theory of Collective Magnon and Orbiton Excitations in Altermagnets}

\author{Bishal Das}
\email[Corresponding author: ]{dasbishal98@gmail.com}
\affiliation{Department of Physics, Indian Institute of Technology Bombay, Mumbai 400076, India}

\author{Chanchal K. Barman}
\email{arckb2@gmail.com}
\affiliation{Institute of Physics, Polish Academy of Sciences, Aleja Lotnik\'{o}w 32/46, 02668 Warsaw, Poland}

\author{Aftab Alam}
\email{aftab@iitb.ac.in}
\affiliation{Department of Physics, Indian Institute of Technology Bombay, Mumbai 400076, India}

\begin{abstract}
    Altermagnetism has recently emerged as a distinct collinear magnetic phase exhibiting momentum-dependent spin-splitting despite vanishing net magnetization, as a consequence of inequivalent non-magnetic environments. Lately, it has been proposed that strong electronic correlations may yield spontaneous altermagnetism due to orbital ordering even for equivalent non-magnetic environments. While previous studies have largely focused on the electronic structure, a unified understanding of the collective excitations associated with these two different microscopic mechanisms stabilizing altermagnetism remains absent. Here, we develop an extended Kugel\textquotesingle-Khomski\u{\i} spin-orbital model on a decorated square lattice that simultaneously incorporates inequivalent non-magnetic environments and correlation-driven orbital ordering within a common theoretical framework. Employing a self-consistent mean-field spin-wave orbital-wave formalism, we demonstrate the emergence of mutually unhybridized but interdependent magnon and orbiton excitations exhibiting characteristic chiral-splitting. We show that the splitting originates from two distinct microscopic contributions: a lattice-dependent term arising from inequivalent non-magnetic environments and an orbital-order-induced exchange-anisotropy term that survives even for crystallographically equivalent non-magnetic environments. The proposed framework therefore unifies the collective excitation spectra associated with both crystalline symmetry-mediated and orbitally-ordered altermagnetism. We further investigate the finite-temperature evolution of the coupled spin-orbital system, revealing the breakdown of spin-wave and orbital-wave approximations through spurious first-order transitions, while complementary classical Monte Carlo simulations recover the expected continuous second-order behaviour. Our work establishes a unified microscopic framework for understanding collective spin-wave and orbital-wave excitations in altermagnets and highlights the decisive role of the non-magnetic lattice environment in controlling their thermodynamic properties.
\end{abstract}
\date{\today}
\maketitle
\stoptoc

\section{Introduction} \label{intro}

Emergent collective phenomena constitute one of the defining characteristics of interacting many-body systems. The interplay among microscopic degrees of freedom often drives spontaneous symmetry breaking, giving rise to ordered phases accompanied by characteristic collective excitations. These excitations are conveniently described as quasiparticles associated with the broken-symmetry state and provide a direct window into the underlying many-body interactions. In magnetically ordered crystals, the spin degree of freedom gives rise to collective spin-wave (SW) excitations \cite{Bloch1930, Slater1930, Pomeranchuk1941, Zhitomirsky2013}, known as magnons, which transport spin angular momentum without charge and have consequently become central to modern spintronics and spin-caloritronics \cite{Bauer2012, Boona2014, Uchida2014, Bose2019, Kikkawa2023}.

Besides spin, the electronic orbital constitutes another fundamental, yet comparatively less explored, degree of freedom in strongly correlated materials. In orbitally degenerate systems, orbital ordering frequently develops alongside magnetic ordering \cite{Roth1966, Pokrovskii1972, KugelKhomskii1972, KugelKomskii1973, KugelKomskii1976, KugelKhomskii1982, Brink1998, Brink2001, Khomskii2014} through \textit{superexchange} interactions \cite{Kramers1934, Anderson1950}. Within the superexchange mechanism, virtual electronic transitions (hoppings) between orbitals of neighbouring magnetic ions via intermediate non-magnetic ligand orbitals lowers the total energy of the system. These virtual processes depend sensitively on the overlap and relative orientation of the participating orbitals and lead to the Goodenough-Kanamori-Anderson (GKA) rules governing magnetic exchange \cite{Goodenough1955, Goodenough1958, Kanamori1959, Anderson1959, Khomskii2014}. The combined effects of the Pauli exclusion principle, on-site Coulomb repulsion, and Hund's exchange interaction lift the orbital degeneracy by favouring specific orbital occupations, thereby establishing long-range orbital order. The corresponding collective orbital-wave (OW) excitations are known as orbitons \cite{Cyrot1975, Komarov1975, Khaliullin1997, Brink1998, Khomskii2014}, whose experimental observation by resonant inelastic X-ray scattering has firmly established orbital dynamics as an experimentally accessible collective phenomenon in correlated quantum materials \cite{Inami2003, Martinelli2024}. Since both magnetic and orbital order originate from the same correlated electronic degrees of freedom, they are generally strongly intertwined, although their ordering temperatures may differ substantially \cite{KugelKomskii1976, KugelKhomskii1982, Khomskii2014}.

The recent discovery of altermagnetism \cite{Liu2022, Smejkal2022a, Smejkal2022b} has introduced a fundamentally new paradigm in magnetic materials, characterized by momentum-dependent spin-splitting of electronic bands despite a vanishing net magnetization. Subsequent experimental observations have firmly established altermagnetism as a genuine magnetic phase, stimulating extensive efforts to understand its microscopic origin and emergent physical properties \cite{Lee2024, Krempasky2024, Osumi2024, Reimers2024, Ding2024, Zhou2025}. Altermagnets are distinguished from conventional collinear antiferromagnets by their spin space group symmetries \cite{Litvin1974, Litvin1977}. Microscopically, the electronic spin-splitting originates from inequivalent non-magnetic crystalline environments surrounding the spin-up and spin-down magnetic sublattices. When the \textit{opposite-spin} sublattices are related by rotational or mirror symmetries (symmorphic or nonsymmorphic), but not by inversion or non-trivial lattice translations, the resulting collinear staggered spin configuration realizes an altermagnetic (AM) state; otherwise, a conventional antiferromagnetic (AFM) state is stabilized. Minimal microscopic models capture this physics through anisotropic electronic hoppings either between magnetic and neighbouring non-magnetic sites or between magnetic sites belonging to the \textit{same-spin} sublattice \cite{Brekke2023, Maier2023, Roig2024}, whereas isotropic hopping restores the conventional antiferromagnetic state.

Beyond their electronic structure, altermagnets exhibit equally intriguing collective excitations. It has been shown that inequivalent non-magnetic environments surrounding the \textit{opposite-spin} sublattices generate chiral-splitting of magnon bands due to anisotropic exchange interactions connecting the \textit{same-spin} sublattices \cite{Brekke2023, Maier2023, Smejkal2023, Cui2023, Roig2024, Cichutek2025a, Cichutek2025b, Eto2025, Yan2025, Xie2026}. More recently, an alternative route has been identified in which orbital ordering driven by strong electronic correlations results in spontaneous altermagnetism and induces anisotropic electronic hopping between orbitals on the \textit{opposite-spin} sublattices even in crystallographically equivalent non-magnetic environments \cite{Cuono2023, Leeb2024, Meier2026, Kaushal2026, Ji2026, Jana2026, dOrnellas2026, Daghofer2026}. These developments reveal that both lattice asymmetry and correlation-driven orbital physics can independently stabilize the altermagnetic state.
A theoretical framework capable of describing both crystal-symmetry-driven and orbital-order-driven altermagnetism within the same unified model is still lacking. In particular, it is presently unclear how lattice-induced anisotropy and correlation-driven orbital ordering cooperate to determine the spectra and finite-temperature properties of collective excitations in altermagnets.

In this work, we address this issue by developing a minimal spin-orbital model that incorporates both mechanisms within a common theoretical framework. Our aim is to obtain a simplified picture of collective spin-wave and orbital-wave excitations in altermagnets, guided by the physically relevant mechanisms described in existing literature. We propose an extended Kugel\textquotesingle-Khomski\u{\i} model \cite{KugelKhomskii1972, KugelKomskii1973, KugelKomskii1976, KugelKhomskii1982} on a two-dimensional square lattice decorated with inequivalent non-magnetic sites. The model consists of second-nearest and third-nearest neighbour anisotropic Heisenberg-like spins and Ising-like orbital isospins, with an additional biquadratic interaction between them. To explicitly account for the underlying non-magnetic environment, we further introduce non-fluctuating lattice isospins that directly modulate the exchange couplings of our model.
We treat the spins and orbital isospins of our unified model within the spin-wave orbital-wave (SW-OW) theory by describing the spin-orbital interactions in a self-consistent mean-field approach, effectively decoupling the spin and orbital sectors that produces unhybridized magnons and orbitons but nevertheless heavily influence each other.
We demonstrate that both excitation spectra exhibit characteristic chiral-splitting governed by two distinct microscopic contributions. The first originates from inequivalent non-magnetic lattice environments, while the second survives even in crystallographically equivalent magnetic lattices and arises solely from orbital-order-induced exchange anisotropy. The present work therefore establishes a unified microscopic framework for collective spin-wave and orbital-wave excitations in altermagnets by demonstrating that the two seemingly distinct microscopic routes towards altermagnetism manifest as two complementary mechanisms for magnon and orbiton chiral-splitting within a single spin-orbital theory.

Finally, we investigate the evolution of the excitation spectra and their order parameters with increasing temperature. The self-consistent mean-field SW-OW theory predicts an abrupt suppression of both magnetic and orbital order parameters, signalling spurious first-order phase transitions that reflect the breakdown of SW-OW approximations in the vicinity of the critical temperatures. To establish the true thermodynamic behaviour, we perform classical Monte-Carlo simulations, which recover the expected continuous second-order phase transitions in the thermodynamic limit. Furthermore, we demonstrate that the explicit modification of exchange couplings by the non-magnetic lattice provides an effective route for tuning the magnetic and orbital ordering temperatures, highlighting the important role of the underlying lattice environment in determining the thermodynamic properties of altermagnetic systems.

\section{The Unified Model Framework}

\subsection{Physical Considerations} \label{model_physics}

The essential microscopic ingredient responsible for altermagnetic spin-splitting in a collinear magnetic system is the anisotropy between the two spin sublattices. Such anisotropy may originate from two distinct mechanisms. The first is intrinsic to the crystal structure, where inequivalent non-magnetic environments surrounding the magnetic ions produce the anisotropic electronic hoppings responsible for crystalline symmetry-mediated altermagnetism (CS-AM) \cite{Liu2022, Smejkal2022a, Smejkal2022b}. The second mechanism is correlation-driven, in which spontaneous orbital ordering generates anisotropic electronic hoppings that differentiates otherwise equivalent spin sublattices, giving rise to the recently proposed orbitally-ordered altermagnetism (OO-AM) \cite{Cuono2023, Leeb2024, Meier2026, Kaushal2026, Ji2026, Jana2026}. Since both mechanisms may coexist in realistic materials, disentangling their individual contributions within a fully microscopic description is generally challenging.

\begin{figure*}
    \centering
    \includegraphics[width=\textwidth]{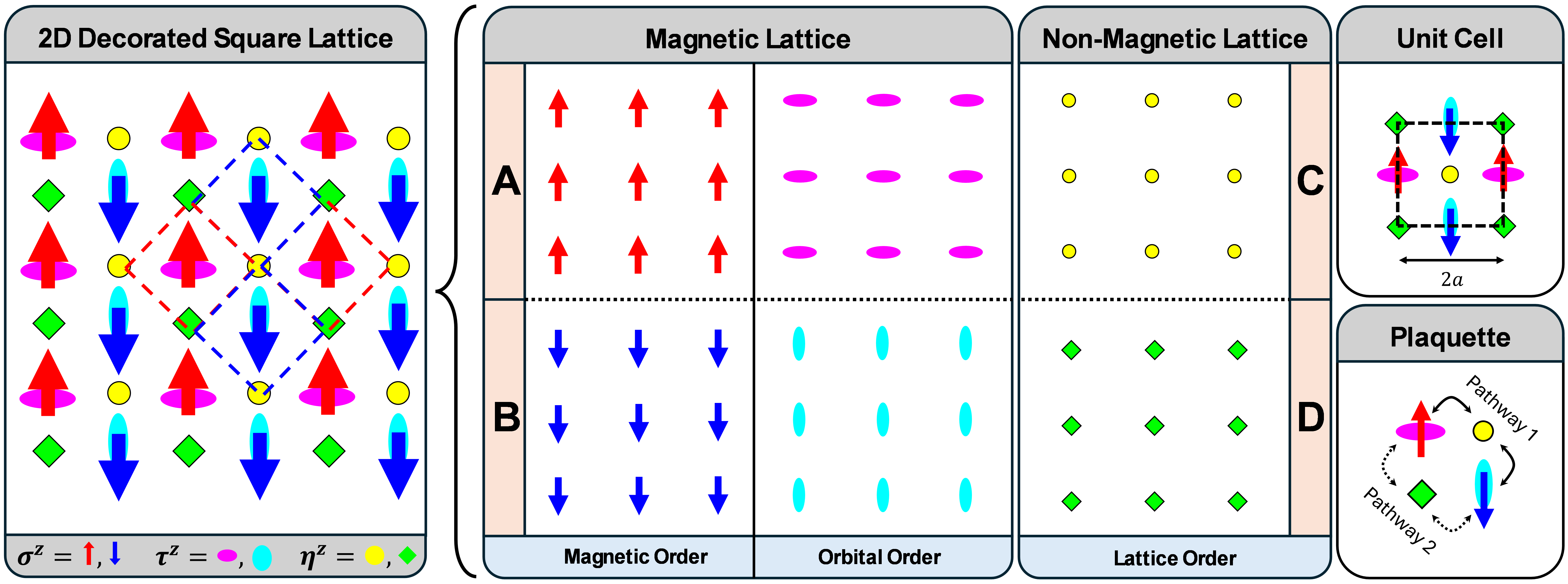}
    \caption{Decorated square lattice model with various motifs including magnetic order, orbital order and lattice order.}
    \label{fig1}
\end{figure*}

An alternative approach is to construct an effective model that captures the essential physical ingredients responsible for stabilizing altermagnetism while remaining sufficiently simple to permit an analytical description of its collective excitations. Existing studies of collective excitations in altermagnets have almost exclusively focused on SW excitations or magnons by employing effective quantum Heisenberg models that inherit the symmetry of the underlying lattice \cite{Brekke2023, Smejkal2023, Cui2023, Cichutek2025a, Cichutek2025b, Eto2025, Yan2025, Xie2026}. While such descriptions are adequate for CS-AMs, they are incomplete for OO-AMs, where orbital degrees of freedom are equally indispensable and can give rise to OW excitations or orbitons.

As discussed in Sec.~\ref{intro}, orbital ordering in strongly correlated systems is predominantly governed by \textit{superexchange} interactions that can also influence spin ordering. According to the GKA rules \cite{Anderson1950, Goodenough1955, Goodenough1958, Kanamori1959, Anderson1959}, ferro-type (aligned) spin order is generally accompanied by antiferro-type (staggered) orbital order, whereas staggered spin order favours aligned orbital order. The coexistence of staggered spin and staggered orbital order predicted for OO-AMs therefore appears to contradict the conventional GKA picture. Although few microscopic explanations have recently been proposed for this apparent violation \cite{Meier2026, Kaushal2026, Ji2026}, they consistently demonstrate that a minimal two-orbital description is insufficient to stabilize this unconventional ordered state of spins and orbitals. Instead, a minimal three-orbital model has been shown to be necessary \cite{Meier2026, Kaushal2026}, in which two orbitals actively participate in orbital ordering while the third orbital mediates the superexchange processes required to stabilize the unconventional order.

This scenario naturally arises in transition-metal compounds, where magnetism predominantly originates from partially filled $d$-orbitals. Under crystal-field splitting, the 5-fold degenerate $d$-manifold separates into the doubly degenerate $e_g$ and triply degenerate $t_{2g}$ subspaces \cite{Khomskii2014}. The latter provides the minimal setting for realizing OO-AM. In particular, a $t_{2g}^{2}$ electronic configuration has been proposed \cite{Cuono2023, Meier2026, Kaushal2026, Jana2026}, in which the $d_{yz}$ and $d_{zx}$ orbitals constitute the \textit{active} orbital sector responsible for orbital ordering, whereas the half-filled $d_{xy}$ orbital facilitates the superexchange pathways that stabilize the ordered state. Motivated by this microscopic picture, we retain only the two \textit{active} orbital degrees of freedom in the present work, as they alone determine the low-energy collective OW excitations. We emphasize, however, that a complete microscopic description of the electronic structure and the stability of the unconventional staggered spin staggered orbital ground state requires the full three-orbital model and lies beyond the scope of the present study \cite{Meier2026, Kaushal2026, Ji2026}.

Another important aspect that has received comparatively little attention in existing theories of collective excitations in altermagnets, particularly for CS-AMs \cite{Brekke2023, Smejkal2023, Cui2023, Cichutek2025a, Cichutek2025b, Eto2025, Yan2025, Xie2026}, is the explicit role of the non-magnetic environment surrounding the magnetic ions.
In conventional effective spin models, the intermediate non-magnetic ligands are incorporated only implicitly through anisotropic exchange interactions between magnetic centres.
Here, we adopt a different perspective by treating the non-magnetic environment itself as an ordered sublattice that directly influences the magnetic interactions. Physically, this sublattice represents the ordered arrangement of the intermediate ligands connecting neighbouring magnetic ions. We therefore introduce an effective lattice order parameter, analogous to the spin and orbital order parameters, whose configuration explicitly modulates the exchange interactions. We restrict our attention to the ordered lattice phase and neglect fluctuations of this lattice degree of freedom. Such fluctuations would eventually lead to lattice disorder, modify the underlying crystalline symmetry, and consequently suppress altermagnetism \cite{Li2026}. The interplay between lattice disorder and collective excitations constitutes an interesting problem in its own right and will be addressed elsewhere.

\subsection{Building an Extended Model involving Spins and Isospins} \label{model_setup}

To capture within a unified framework both crystalline-symmetry-driven and orbital-order-driven mechanisms of altermagnetism, we consider a two-dimensional (2D) decorated square lattice shown in Fig.~\ref{fig1}, which may be viewed as a modified Lieb lattice. The lattice consists of two interpenetrating bipartite sublattices. One sublattice is occupied by magnetic ions and constitutes the magnetic lattice, while the other is occupied by intermediate ligands constituting the non-magnetic lattice that decorate the magnetic lattice. The magnetic lattice hosts two coupled internal degrees of freedom: localized magnetic spins $\hat{\boldsymbol{\sigma}}$ (red and blue arrows) forming a collinear staggered magnetic order, and orbital isospins $\hat{\boldsymbol{\tau}}$ (magenta and cyan ellipses) describing the staggered occupation of two active orbitals. The decorating non-magnetic lattice is composed of two inequivalent ligand species arranged in a staggered pattern, represented by lattice isospins $\hat{\boldsymbol{\eta}}$ (yellow circles and green diamonds). Consequently, the magnetic lattice is divided into the A and B magnetic sublattices, whereas the non-magnetic lattice forms the C and D sublattices, as illustrated in the middle panels of Fig.~\ref{fig1}.

The staggered arrangement of ligand species generates inequivalent local crystalline environments around the magnetic ions, thereby partitioning the magnetic lattice into two fully compensated \textit{spin sublattices}, highlighted by the red (spin-up) and blue (spin-down) dashed boxes in the left panel of Fig.~\ref{fig1}. The decorated lattice belongs to the wallpaper group $p4mm$, whose symmetry operations are $\{\mathbb{I}, C_{2z}, C_{4z}, C_{4z}^{3}, m_x, m_y, m_{x+y}, m_{x-y}\}$, where $\mathbb{I}$ denotes the identity operation, $C_{nz}$ represents $n$-fold rotation about the axis perpendicular to the lattice plane, and $m_c$ denotes mirror reflection about the plane normal to the direction $c$. The corresponding primitive unit cell is shown in the upper-right panel of Fig.~\ref{fig1}. Owing to the decoration, the lattice constant becomes $2a$. The magnetic ions occupy the Wyckoff position $2c:(0.5,0),(0,0.5)$ with site symmetry $mm2$, whereas the non-magnetic ligands occupy the $1a:(0,0)$ and $1b:(0.5,0.5)$ Wyckoff positions possessing the full point-group symmetry $4mm$.

The magnetic and non-magnetic sites are nearest neighbours separated by a distance $a$. Magnetic ions belonging to \textit{opposite-spin} sublattices are second-nearest neighbours at a distance $\sqrt{2}a$, whereas those belonging to the same-spin sublattice are third-nearest neighbours separated by $2a$. Although the decorated lattice is centrosymmetric because $C_{2z}$ is equivalent to inversion ($\mathcal{P}$) in the lattice plane, the two \textit{opposite-spin} sublattices are related only through four-fold rotations ($C_{4z}$, $C_{4z}^{3}$) and diagonal mirror operations ($m_{x+y}$, $m_{x-y}$), since the C and D sublattices are occupied by chemically distinct ligand species. This symmetry relation naturally realizes altermagnetism. In contrast, if the same ligand species occupies both C and D sublattices, the opposite-spin sublattices additionally become related by inversion symmetry, reducing the system to a conventional antiferromagnet. In this limit the wallpaper group changes from $p4mm$ to the higher-symmetry group $p4gm$. The inequivalent ligand species also generate two distinct superexchange pathways (indicated by the solid and dashed arrows in the lower-right panel of Fig.~\ref{fig1}) connecting neighbouring magnetic ions within each four-site plaquette consisting of one site from each of the A, B, C and D sublattices.
Motivated by these considerations, we introduce the following extended anisotropic Heisenberg Hamiltonian describing the coupled spin and orbital degrees of freedom on the decorated square lattice:
\begin{widetext}
    \begin{equation}       
        \begin{gathered}
            \hat{\mathcal{H}} = -\sum_{\langle\langle i,j \rangle\rangle}^{\text{mag}} \mathcal{J}_{1,ij}^{\sigma} \left[ \hat{\sigma}_{i}^{z} \hat{\sigma}_{j}^{z} + \frac{\lambda_{1}^{\sigma}}{2} \left(\hat{\sigma}_{i}^{+} \hat{\sigma}_{j}^{-} + \hat{\sigma}_{i}^{-} \hat{\sigma}_{j}^{+}\right) \right] -\sum_{\langle\langle i,j \rangle\rangle}^{\text{mag}} \mathcal{J}_{1,ij}^{\tau} \left[ \hat{\tau}_{i}^{z}\hat{\tau}_{j}^{z} + \frac{\lambda_{1}^{\tau}}{2} \left(\hat{\tau}_{i}^{+}\hat{\tau}_{j}^{-} + \hat{\tau}_{i}^{-}\hat{\tau}_{j}^{+}\right) \right] \\
            -\sum_{\langle\langle i,j \rangle\rangle}^{\text{mag}} \mathcal{Q}_{1,ij} \left[ \hat{\sigma}_{i}^{z} \hat{\sigma}_{j}^{z} + \frac{\lambda_{1}^{\sigma}}{2} \left(\hat{\sigma}_{i}^{+} \hat{\sigma}_{j}^{-} + \hat{\sigma}_{i}^{-} \hat{\sigma}_{j}^{+}\right) \right]\!\! \left[ \hat{\tau}_{i}^{z}\hat{\tau}_{j}^{z} + \frac{\lambda_{1}^{\tau}}{2} \left(\hat{\tau}_{i}^{+}\hat{\tau}_{j}^{-} + \hat{\tau}_{i}^{-}\hat{\tau}_{j}^{+}\right) \right] \\
            -\sum_{\langle\langle\langle i,j \rangle\rangle\rangle}^{\text{mag}} \mathcal{J}_{2,ij}^{\sigma} \left[ \hat{\sigma}_{i}^{z} \hat{\sigma}_{j}^{z} + \frac{\lambda_{2}^{\sigma}}{2} \left(\hat{\sigma}_{i}^{+} \hat{\sigma}_{j}^{-} + \hat{\sigma}_{i}^{-} \hat{\sigma}_{j}^{+}\right) \right] -\sum_{\langle\langle\langle i,j \rangle\rangle\rangle}^{\text{mag}} \mathcal{J}_{2,ij}^{\tau} \left[ \hat{\tau}_{i}^{z}\hat{\tau}_{j}^{z} + \frac{\lambda_{2}^{\tau}}{2} \left(\hat{\tau}_{i}^{+}\hat{\tau}_{j}^{-} + \hat{\tau}_{i}^{-}\hat{\tau}_{j}^{+}\right) \right] \\
            -\sum_{\langle\langle\langle i,j \rangle\rangle\rangle}^{\text{mag}} \mathcal{Q}_{2,ij} \left[ \hat{\sigma}_{i}^{z} \hat{\sigma}_{j}^{z} + \frac{\lambda_{2}^{\sigma}}{2} \left(\hat{\sigma}_{i}^{+} \hat{\sigma}_{j}^{-} + \hat{\sigma}_{i}^{-} \hat{\sigma}_{j}^{+}\right) \right]\!\! \left[ \hat{\tau}_{i}^{z}\hat{\tau}_{j}^{z} + \frac{\lambda_{2}^{\tau}}{2} \left(\hat{\tau}_{i}^{+}\hat{\tau}_{j}^{-} + \hat{\tau}_{i}^{-}\hat{\tau}_{j}^{+}\right) \right] \\
            - \mathcal{K}_{z}^{\sigma} \sum_{i}^{\text{mag}} (\hat{\sigma}_{i}^{z})^{2} - \mathcal{K}_{z}^{\tau} \sum_{i}^{\text{mag}} (\hat{\tau}_{i}^{z})^{2}
        \end{gathered}
        \label{fullH}
    \end{equation}
\end{widetext}
where $\mathcal{J}_{1,ij}^{\sigma}, \mathcal{J}_{1,ij}^{\tau} < 0$ favour staggered spin and orbital arrangement of second-nearest (NN) neighbours, while $\mathcal{J}_{2,ij}^{\sigma}, \mathcal{J}_{2,ij}^{\tau} > 0$ favour aligned spin and orbital arrangement of third-nearest (NNN) neighbours. Additionally, $\mathcal{Q}_{1,ij} > 0$ and $\mathcal{Q}_{2,ij} > 0$ favours the same (ferro- or antiferro-type) arrangement of both spins and orbitals, for NN and NNN neighbours respectively. Here, $\langle\langle i,j \rangle\rangle$ denotes NN neighbours and $\langle\langle\langle i,j \rangle\rangle\rangle$ denotes NNN neighbours.
The orbital isospins are treated on the same footing as the magnetic spins and obey the $\mathfrak{su}(2)$ algebra \cite{Brink2001, Khomskii2014}. Throughout this work, we consider the minimal two active orbital description discussed in Sec.~\ref{model_physics}, such that occupation of either active orbital by a single electron is represented by $\tau^{z} = \pm\tau$ with $\tau = \sfrac{1}{2}$, while the localized magnetic moments are described by $\sigma^{z} = \pm\sigma$ with $\sigma = \sfrac{1}{2}$. The orbital isospin raising and lowering operators, $\tau^{\pm}$, generate quantum fluctuations of the orbital occupation by transferring the electron between the two active orbitals. Since the spins and orbital isospins belong to independent Hilbert spaces, all spin and orbital isospin operators mutually commute.

For notational convenience, Eq.~\eqref{fullH} may be decomposed as $\hat{\mathcal{H}} = \hat{\mathcal{H}}_{\mathrm{NN}} + \hat{\mathcal{H}}_{\mathrm{NNN}} + \hat{\mathcal{H}}_{\mathrm{aniso}}$ where $\hat{\mathcal{H}}_{\mathrm{NN}}$ and $\hat{\mathcal{H}}_{\mathrm{NNN}}$ describe spin, orbital, and biquadratic spin-orbital exchange interactions between second-nearest and third-nearest neighbours respectively; while $\hat{\mathcal{H}}_{\mathrm{aniso}}$ contains the single-ion anisotropy terms. The parameters $\lambda_{1(2)}^{\sigma},\lambda_{1(2)}^{\tau}\in[0,1]$ quantify the uniaxial exchange anisotropy, with $\lambda=0$ as the Ising limit and $\lambda=1$ as the isotropic Heisenberg limit.
The biquadratic exchange $\mathcal{Q}_{1(2),ij}$ couple the spin and orbital sectors and is symmetric with respect to exchange of magnetic sublattices A \& B.
It is important to emphasize that this biquadratic spin-orbital exchange interaction should not be confused with the atomic spin-orbit coupling (SOC) of the form $\hat{\boldsymbol{\sigma}} \cdot \hat{\boldsymbol{\tau}}$ \cite{KugelKhomskii1982}. The SOC term mixes spin and orbital degrees of freedom and gives rise to coupled spin-orbital-wave (SOW) excitations \cite{Kaushal2026}. In the present work we deliberately neglect relativistic SOC in order to isolate the intrinsic non-relativistic physics of altermagnetism. Consequently, any chiral-splitting of the collective excitations originates exclusively from exchange anisotropy and non-magnetic lattice anisotropy rather than from conventional relativistic mechanisms. Furthermore, experimental studies indicate that SW and OW excitations generally have well-separated energy scales \cite{Moussa1996, Inami2003, Coldea2001, Martinelli2024}. Accordingly, magnons and orbitons remain decoupled in our theory and do not hybridize. Nevertheless, the biquadratic spin-orbital exchange interaction produces a strong mutual renormalization of the two excitation spectra, as will be demonstrated in Sec.~\ref{scmft}.

\subsection{Non-Magnetic Lattice Modified Exchange Couplings}
Until now, the effect of non-magnetic lattice has not been taken into account since Eq.~\eqref{fullH} does not contain any term with lattice isospins $\hat{\boldsymbol{\eta}}$.
Now considering that the non-magnetic lattice provides a background interaction that directly modifies the spin and orbital isospin exchange couplings, we define them in the following manner:
\begin{subequations}
    \label{ex_mod}
    \begin{gather}
        \begin{aligned}
            \mathcal{J}_{1,ij}^{\sigma} = J_{1,ij}^{\sigma} + \!\!\!\!\!\!\sideset{}{'}\sum_{(u,v) \rightarrow (i,j)}^{\text{non-mag}}\!\!\!\!\!\! \Phi_{1,ijuv}^{\sigma} (\hat{\boldsymbol{\eta}}_{u} \cdot \hat{\boldsymbol{\eta}}_{v}) \\
            = J_{1,ij}^{\sigma} \bigg[1 + \!\!\!\!\!\!\sideset{}{'}\sum_{(u,v) \rightarrow (i,j)}^{\text{non-mag}}\!\!\!\!\!\! \varphi_{1,uv}^{\sigma} (\hat{\boldsymbol{\eta}}_{u} \cdot \hat{\boldsymbol{\eta}}_{v})\bigg]
        \end{aligned}
        \\[2ex]
        \begin{aligned}
            \mathcal{J}_{1,ij}^{\tau} = J_{1,ij}^{\tau} + \!\!\!\!\!\!\sideset{}{'}\sum_{(u,v) \rightarrow (i,j)}^{\text{non-mag}}\!\!\!\!\!\! \Phi_{1,ijuv}^{\tau} (\hat{\boldsymbol{\eta}}_{u} \cdot \hat{\boldsymbol{\eta}}_{v}) \\
            = J_{1,ij}^{\tau} \bigg[1 + \!\!\!\!\!\!\sideset{}{'}\sum_{(u,v) \rightarrow (i,j)}^{\text{non-mag}}\!\!\!\!\!\! \varphi_{1,uv}^{\tau} (\hat{\boldsymbol{\eta}}_{u} \cdot \hat{\boldsymbol{\eta}}_{v})\bigg]
        \end{aligned}
        \\[2ex]
        \begin{aligned}
            \mathcal{Q}_{1,ij} = Q_{1,ij} + \!\!\!\!\!\!\sideset{}{'}\sum_{(u,v) \rightarrow (i,j)}^{\text{non-mag}}\!\!\!\!\!\! \Phi_{1,ijuv}^{\sigma\tau} (\hat{\boldsymbol{\eta}}_{u} \cdot \hat{\boldsymbol{\eta}}_{v}) \\
            = Q_{1,ij} \bigg[1 + \!\!\!\!\!\!\sideset{}{'}\sum_{(u,v) \rightarrow (i,j)}^{\text{non-mag}}\!\!\!\!\!\! \varphi_{1,uv}^{\sigma\tau} (\hat{\boldsymbol{\eta}}_{u} \cdot \hat{\boldsymbol{\eta}}_{v})\bigg] 
        \end{aligned}
        \\[2ex]
        \begin{aligned}
            \mathcal{J}_{2,ij}^{\sigma} = J_{2,ij}^{\sigma} + \!\!\!\!\sideset{}{'}\sum_{u \rightarrow (i,j)}^{\text{non-mag}}\!\!\!\! (\boldsymbol{\Phi}_{2,iju}^{\sigma} \cdot \hat{\boldsymbol{\eta}}_{u}) \\
            = J_{2,ij}^{\sigma} \bigg[1 + \!\!\!\!\sideset{}{'}\sum_{u \rightarrow (i,j)}^{\text{non-mag}}\!\!\!\! (\boldsymbol{\varphi}_{2,u}^{\sigma} \cdot \hat{\boldsymbol{\eta}}_{u})\bigg]
        \end{aligned}
        \\[2ex]
        \begin{aligned}
            \mathcal{J}_{2,ij}^{\tau} = J_{2,ij}^{\tau} + \!\!\!\!\sideset{}{'}\sum_{u \rightarrow (i,j)}^{\text{non-mag}}\!\!\!\! (\boldsymbol{\Phi}_{2,iju}^{\tau} \cdot \hat{\boldsymbol{\eta}}_{u}) \\
            = J_{2,ij}^{\tau} \bigg[1 + \!\!\!\!\sideset{}{'}\sum_{u \rightarrow (i,j)}^{\text{non-mag}}\!\!\!\! (\boldsymbol{\varphi}_{2,u}^{\tau} \cdot \hat{\boldsymbol{\eta}}_{u})\bigg]
        \end{aligned}
        \\[2ex]
        \begin{aligned}
            \mathcal{Q}_{2,ij} = Q_{2,ij} + \!\!\!\!\sideset{}{'}\sum_{u \rightarrow (i,j)}^{\text{non-mag}}\!\!\!\! (\boldsymbol{\Phi}_{2,iju}^{\sigma\tau} \cdot \hat{\boldsymbol{\eta}}_{u}) \\
            = Q_{2,ij} \bigg[1 + \!\!\!\!\sideset{}{'}\sum_{u \rightarrow (i,j)}^{\text{non-mag}}\!\!\!\! (\boldsymbol{\varphi}_{2,u}^{\sigma\tau} \cdot \hat{\boldsymbol{\eta}}_{u})\bigg]
        \end{aligned}
    \end{gather}
\end{subequations}
Here, the \textit{primed} sum $\sideset{}{'}\sum_{(u,v) \rightarrow (i,j)}$ in the NN couplings is over pairs of \textit{common} non-magnetic nearest neighbours $(u,v)$ connected to a magnetic pair $(i,j)$ and the \textit{primed} sum $\sideset{}{'}\sum_{u \rightarrow (i,j)}$ in the NNN couplings is over non-magnetic nearest neighbours $u$ that are \textit{common} to a pair of magnetic sites $(i,j)$. It should be evident that the non-magnetic sites equidistant from both magnetic sites in a pair $(i,j)$ are \textit{common} to the magnetic pair. Thus, the \textit{primed} sums are actually plaquette sums.
In Eq.~\eqref{ex_mod}, we consider the \textit{effective} longitudinal NN(NNN) couplings $\mathcal{J}_{1(2)}^{\sigma}$, $\mathcal{J}_{1(2)}^{\tau}$, $\mathcal{Q}_{1(2)}$ to be a sum of the \textit{bare} NN(NNN) couplings $J_{1(2)}^{\sigma}$, $J_{1(2)}^{\tau}$, $Q_{1(2)}$ and the non-magnetic lattice NN(NNN) couplings $\Phi_{1}$($\boldsymbol{\Phi}_{2}$)'s with lattice isospin operators $\hat{\boldsymbol{\eta}}$ in both $\sigma$ and $\tau$ sectors, that takes into account the effect of spin-lattice or orbital-lattice interactions. These non-magnetic lattice couplings may further be rewritten in a form where the \textit{bare} NN(NNN) couplings are adjusted by lattice isospin dependent \textit{exchange modifiers} $\varphi_{1}$($\boldsymbol{\varphi}_{2}$)'s in both $\sigma$ and $\tau$ sectors. This dependence on the lattice isospin is due to the possible existence of separate \textit{superexchange} pathways (see bottom right panel of Fig.~\ref{fig1}).
As discussed in Sec.~\ref{model_physics}, we assume a fully ordered non-magnetic lattice with no lattice isospin fluctuations, i.e., $\hat{\boldsymbol{\eta}} = (0,0,\hat{\eta}^{z})$ which we term as the \textit{frozen-lattice approximation}.
Within this approximation, once the lattice order has set in, the non-magnetic lattice provides an average background interaction to the magnetic lattice. The form of \textit{exchange modifiers} become $\sim$$\varphi_{1,uv}\hat{\eta}_{u}^{z}\hat{\eta}_{v}^{z}$ and $\sim$$\varphi_{2,u}^{z}\hat{\eta}_{u}^{z}$ for NN and NNN couplings respectively in both $\sigma$ and $\tau$ sectors.
Under the \textit{frozen-lattice approximation}, the lattice isospin operators may be simply replaced by their maximum and minimum eigenvalues leading to two possible non-magnetic lattice arrangements: i) staggered lattice order with $\hat{\eta}_{u \in \text{C}}^{z} = \pm\eta$, $\hat{\eta}_{v \in \text{D}}^{z} = \mp\eta$ (distinct ligand species), and ii) aligned lattice order with $\hat{\eta}_{u \in \text{C}}^{z} = \pm\eta$, $\hat{\eta}_{v \in \text{D}}^{z} = \pm\eta$ (same ligand species) with $\eta = \sfrac{1}{2}$. Figure~\ref{fig1} shows a staggered lattice order with the convention $+\eta$ denoted by yellow circles and $-\eta$ denoted by green diamonds. Each plaquette consists of a single magnetic pair and one equidistant non-magnetic pair.

Due to discrete lattice translation symmetry, the inter-sublattice NN exchange couplings become site-independent. Furthermore, all the magnetic NN neighbours along the unit cell diagonals have the same equidistant ligands forming similar plaquettes (Fig.~\ref{fig1}). For orbital ordering, which is sensitive to orbital overlaps and hence mutual orientation of magnetic centers, the \textit{superexchange} pathways in every plaquette will be the same. Thus, the \textit{effective} NN couplings will be isotropic and bond-independent (same between any NN $i$-th and $j$-th site), given by: $\mathcal{J}_{1}^{\sigma} = J_{1}^{\sigma} (1 - \varphi_{1,\pm}^{\sigma} \eta^{2})$, $\mathcal{J}_{1}^{\tau} = J_{1}^{\tau} (1 - \varphi_{1,\pm}^{\tau} \eta^{2})$ and $\mathcal{Q}_{1} = Q_{1} (1 - \varphi_{1,\pm}^{\sigma\tau} \eta^{2})$. Notably, the non-magnetic lattice modifies the \textit{effective} NN couplings in an isotropic way (irrespective of the convention used since $\varphi_{1,\pm}=\varphi_{1,\mp}$). This is true even in the case of aligned lattice order, where only the \textit{exchange modifiers} change either to $+\varphi_{1,++}^{\sigma} \eta^{2}$ or $+\varphi_{1,--}^{\sigma} \eta^{2}$ ($\varphi_{1,++}$ and $\varphi_{1,--}$ are unequal in general). 

The situation is different for the intra-sublattice NNN exchange couplings. From Fig.~\ref{fig1}, it is evident that bonds between a magnetic NNN neighbour along $x$- and $y$-directions are inequivalent. There are two distinct reasons for this. Firstly, the staggered lattice order ensures anisotropy along the $x$- and $y$-directions. This is the essential mechanism giving rise to CS-AMs \cite{Brekke2023}. If there is aligned lattice order or very weak lattice coupling, this anisotropy would vanish. Secondly, irrespective of the non-magnetic lattice arrangement, the orbital ordering may induce anisotropy due to different orbital overlaps along the $x$ and $y$ directions. This mechanism gives rise to OO-AMs \cite{Leeb2024, Meier2026, Kaushal2026, Ji2026, Jana2026}. Let us assume that the \textit{bare} NNN couplings for magnetic sublattice A are $J_{2x}$ and $J_{2y}$ along $x$- and $y$-directions respectively as a result of orbital order, then the \textit{effective} couplings between NNN sites of magnetic sublattice A become: $\mathcal{J}_{2,\text{A}x}^{\sigma} = J_{2x}^{\sigma} (1 + \varphi_{1,+}^{\sigma} \eta)$, $\mathcal{J}_{2,\text{A}x}^{\tau} = J_{2x}^{\tau} (1 + \varphi_{1,+}^{\tau} \eta)$, $\mathcal{Q}_{2,\text{A}x} = Q_{2x} (1 + \varphi_{1,+}^{\sigma\tau} \eta)$ along $x$-direction, and $\mathcal{J}_{2,\text{A}y}^{\sigma} = J_{2y}^{\sigma} (1 - \varphi_{2,-}^{\sigma} \eta)$, $\mathcal{J}_{2,\text{A}y}^{\tau} = J_{2y}^{\tau} (1 - \varphi_{2,-}^{\tau} \eta)$, $\mathcal{Q}_{2,\text{A}y} = Q_{2y} (1 - \varphi_{2,-}^{\sigma\tau} \eta)$ along $y$-direction (here the longitudinal component of the $\varphi_{2}$ couplings should be implicitly understood and thus the index $z$ has been dropped). Importantly for NNN sites of magnetic sublattice B, these direction-dependent \textit{effective} couplings must be reversed as a consequence of the 4-fold rotation and diagonal mirror symmetries of the lattice: $\mathcal{J}_{2,\text{B}x}^{\sigma} = J_{2y}^{\sigma} (1 - \varphi_{2,-}^{\sigma} \eta) = \mathcal{J}_{2,\text{A}y}^{\sigma}$, $\mathcal{J}_{2,\text{B}x}^{\tau} = J_{2y}^{\tau} (1 - \varphi_{2,-}^{\tau} \eta) = \mathcal{J}_{2,\text{A}y}^{\tau}$, $\mathcal{Q}_{2,\text{B}x} = Q_{2y} (1 - \varphi_{2,-}^{\sigma\tau} \eta) = \mathcal{Q}_{2,\text{A}y}$ along $x$-direction, and $\mathcal{J}_{2,\text{B}y}^{\sigma} = J_{2x}^{\sigma} (1 + \varphi_{1,+}^{\sigma} \eta) = \mathcal{J}_{2,\text{A}x}^{\sigma}$, $\mathcal{J}_{2,\text{B}y}^{\tau} = J_{2x}^{\tau} (1 + \varphi_{1,+}^{\tau} \eta) = \mathcal{J}_{2,\text{A}x}^{\tau}$, $\mathcal{Q}_{2,\text{B}y} = Q_{2x} (1 + \varphi_{1,+}^{\sigma\tau} \eta) = \mathcal{Q}_{2,\text{A}x}$ along $y$-direction.
Thus, although discrete translation symmetry makes the NNN couplings site-independent within a particular magnetic sublattice, they are now both sublattice-dependent and bond-dependent with different \textit{exchange modifiers} due to staggered lattice order, even in the case of isotropic \textit{bare} couplings. In the case of aligned lattice order, all the \textit{exchange modifiers} become exactly the same. If the \textit{bare} couplings are isotropic in such case, the NNN couplings become both sublattice-independent and bond-independent, i.e., same for both magnetic sublattices.

\subsection{Spin-Wave Orbital-Wave (SW-OW) Theory} \label{swow}
\subsubsection{Holstein-Primakoff Hamiltonian} \label{sectionHP}
Collective spin-wave and orbital-wave excitations are quantized by mapping the spin and orbital isospin operators to Bosonic operators in the spin and orbital sectors respectively using the Holstein-Primakoff (HP) transformations \cite{HolsteinPrimakoff1940} (also see Sec.~S1 of SM \cite{supp}), such that the new HP operators follow the canonical commutation relations of both spins and orbital isospins:
\begin{subequations} \label{HP_comm}
    \begin{align}
        [\annihilatesa[i],\createsa[j]] = \delta_{ij} \;&;\; [\annihilateta[i],\createta[j]] = \delta_{ij} \\
        [\annihilatesb[i],\createsb[j]] = \delta_{ij} \;&;\; [\annihilatetb[i],\createtb[j]] = \delta_{ij}
    \end{align}
\end{subequations}
Any other combination of $a,a^{\dagger},b,b^{\dagger}$ operators commute with each other. The Fourier transforms of HP operators in the $\sigma(\tau)$ sector are given by the following:
\begin{subequations} \label{invFT}
    \begin{align}
        a_{i,\sigma(\tau)} = \sqrt{\frac{2}{N_m}} \sum_{\mathbf{k}} e^{-\mathrm{i}\mathbf{k}\cdot\mathbf{r}_{i}} a_{\mathbf{k},\sigma(\tau)} \\
        a_{i,\sigma(\tau)}^{\dagger} = \sqrt{\frac{2}{N_m}} \sum_{\mathbf{k}} e^{+\mathrm{i}\mathbf{k}\cdot\mathbf{r}_{i}} a_{\mathbf{k},\sigma(\tau)}^{\dagger} \\
        b_{i,\sigma(\tau)} = \sqrt{\frac{2}{N_m}} \sum_{\mathbf{k}} e^{-\mathrm{i}\mathbf{k}\cdot\mathbf{r}_{i}} b_{\mathbf{k},\sigma(\tau)} \\
        b_{i,\sigma(\tau)}^{\dagger} = \sqrt{\frac{2}{N_m}} \sum_{\mathbf{k}} e^{+\mathrm{i}\mathbf{k}\cdot\mathbf{r}_{i}} b_{\mathbf{k},\sigma(\tau)}^{\dagger}
    \end{align}
\end{subequations}
where, $N_m$ is the total number of magnetic sites. The standard procedure now is to substitute the HP operators of Eq.~\eqref{invFT} in Eq.~\eqref{fullH} according to spin/isospin mappings by performing a SW and OW expansion, and consequently obtain the momentum-space Hamiltonian. The expansion is performed under the conditions $\langle \nasigma{i}{i} \rangle , \langle \nbsigma{i}{i} \rangle \ll \sigma$ and $\langle \natau{i}{i} \rangle , \langle \nbtau{i}{i} \rangle \ll \tau$, usually keeping up to terms with quadratic form, which is the linear SW-OW theory. But doing so in the current case completely decouples the spin and orbital sectors despite the presence of interaction terms between spins and orbital isospins in Eq.~\eqref{fullH}. It is evident that any cross-terms between HP operators of spin and orbital sectors should at least be of biquadratic form, forcing us to go beyond linear SW-OW theory by retaining SW-OW expansion terms up to at least second order in spins and isospins, i.e., $\mathcal{O}[\sfrac{1}{\sigma^2}]$, $\mathcal{O}[\sfrac{1}{\tau^2}]$, $\mathcal{O}[\sfrac{1}{\sigma\tau}]$ (also see Sec.~S1 of SM \cite{supp}). Since one finally wishes to obtain the spectrum of collective excitations which requires a diagonalizable Hamiltonian with quadratic form, one then applies standard quantum mean-field approximation via Wick's decoupling procedure \cite{Wick1950}.
Wick's decoupling requires consideration of all possible combinations of operator products and their expectation values (averages). But not all averages will be non-zero in our case because the Hamiltonian in Eq.~\eqref{fullH} has the following two important global symmetries: $[\hat{\mathcal{H}},\hat{\sigma}_{\text{tot}}^{z}] = 0$ and $[\hat{\mathcal{H}},\hat{\tau}_{\text{tot}}^{z}] = 0$ (see Sec.~S2 of SM \cite{supp}), i.e., the total spin $\hat{\sigma}_{\text{tot}}^{z}$ and the total orbital isospin $\hat{\tau}_{\text{tot}}^{z}$ of our spin-orbital system remains conserved. Hence any HP operator or combination of HP operators that change $\hat{\sigma}_{\text{tot}}^{z}$ or $\hat{\tau}_{\text{tot}}^{z}$ of the spin-orbital system must have zero expectation value with respect to eigenstates of $\hat{\mathcal{H}}$. This rules out odd numbered combinations of HP operators and importantly also forbids any quadratic cross-terms averages between HP operators of $\sigma$ and $\tau$ sectors. Only averages of specific quadratic HP operators survive:
\begin{subequations} \label{HPavg}
    \begin{align}
        \langle\fnasigma{\mathbf{k}}{\mathbf{k}'}\rangle = \varrho_{\mathbf{k},\sigma} \delta_{\mathbf{k},\mathbf{k}'} \,&;\,  \langle\fnbsigma{\mathbf{k}}{\mathbf{k}'}\rangle = \varrho_{\mathbf{k},\sigma} \delta_{\mathbf{k},\mathbf{k}'} \\
        \langle\fnatau{\mathbf{k}}{\mathbf{k}'}\rangle = \varrho_{\mathbf{k},\tau} \,\delta_{\mathbf{k},\mathbf{k}'} \,&;\, \langle\fnbtau{\mathbf{k}}{\mathbf{k}'}\rangle = \varrho_{\mathbf{k},\tau} \delta_{\mathbf{k},\mathbf{k}'} \\
        \langle\fannihilatesa[\mathbf{k}]\fannihilatesb[\mathbf{k}']\rangle = \Delta_{\mathbf{k},\sigma} \delta_{\mathbf{k},-\mathbf{k}'} \,&;\, \langle\fcreatesb[\mathbf{k}']\fcreatesa[\mathbf{k}]\rangle = \Delta_{\mathbf{k},\sigma}^{*} \delta_{\mathbf{k},-\mathbf{k}'} \\
        \langle\fannihilateta[\mathbf{k}]\fannihilatetb[\mathbf{k}']\rangle = \Delta_{\mathbf{k},\tau} \delta_{\mathbf{k},-\mathbf{k}'} \,&;\, \langle\fcreatetb[\mathbf{k}']\fcreateta[\mathbf{k}]\rangle = \Delta_{\mathbf{k},\tau}^{*} \delta_{\mathbf{k},-\mathbf{k}'}
    \end{align}
\end{subequations}
with momentum-conservation built into these averages via the Kronecker deltas.
Here in the $\sigma(\tau)$ sector, we have $\varrho_{\mathbf{k},\sigma(\tau)} \in \mathbb{R}$ because $a_{\mathbf{k},\sigma(\tau)}^{\dagger}a_{\mathbf{k},\sigma(\tau)}$ and $b_{\mathbf{k},\sigma(\tau)}^{\dagger}b_{\mathbf{k},\sigma(\tau)}$ are Hermitian operators. But $a_{\mathbf{k},\sigma(\tau)}b_{-\mathbf{k},\sigma(\tau)}$ and $b_{-\mathbf{k},\sigma(\tau)}^{\dagger}a_{\mathbf{k},\sigma(\tau)}^{\dagger}$ are Hermitian conjugates of each other and thus $\Delta_{\mathbf{k},\sigma(\tau)} \in \mathbb{C}$ in general. Also, because we consider collinear $\sigma(\tau)$ arrangement with perfect compensation, we must have $\langle a_{\mathbf{k},\sigma(\tau)}^{\dagger}a_{\mathbf{k},\sigma(\tau)} \rangle = \langle b_{\mathbf{k},\sigma(\tau)}^{\dagger}b_{\mathbf{k},\sigma(\tau)} \rangle$ in accordance with the global symmetries of $\hat{\mathcal{H}}$. For more details of the expectation values, see Sec.~S3 of SM \cite{supp}. Using the relations of Eq.~\eqref{HPavg} and after some lengthy algebra, one arrives at a generic expression for the momentum-space Hamiltonian in terms of HP operators:
\begin{widetext}
    \begin{equation}
        \begin{gathered}
            \hat{\mathcal{H}} = \hat{\mathcal{H}}_{0} + \sum_{\mathbf{k}} \left[\mathcal{E}_{\sigma}^{\text{A}}(\mathbf{k}) \fnasigma{\mathbf{k}}{\mathbf{k}} + \mathcal{E}_{\sigma}^{\text{B}}(\mathbf{k}) \fnbsigma{-\mathbf{k}}{-\mathbf{k}}\right] + \sum_{\mathbf{k}} \left[\mathcal{E}_{\tau}^{\text{A}}(\mathbf{k}) \fnatau{\mathbf{k}}{\mathbf{k}} + \mathcal{E}_{\tau}^{\text{B}}(\mathbf{k}) \fnbtau{-\mathbf{k}}{-\mathbf{k}}\right] \\
            + \sum_{\mathbf{k}} \left[\mathcal{F}_{\sigma}(\mathbf{k}) \fannihilatesa[\mathbf{k}]\fannihilatesb[-\mathbf{k}] + \mathcal{F}_{\sigma}^{*}(\mathbf{k}) \fcreatesb[-\mathbf{k}]\fcreatesa[\mathbf{k}]\right] + \sum_{\mathbf{k}} \left[\mathcal{F}_{\tau}(\mathbf{k}) \fannihilateta[\mathbf{k}]\fannihilatetb[-\mathbf{k}] + \mathcal{F}_{\tau}^{*}(\mathbf{k}) \fcreatetb[-\mathbf{k}]\fcreateta[\mathbf{k}]\right]
        \end{gathered}
        \label{nondiagH}
    \end{equation}
\end{widetext}
Here, the coefficients of HP operators depend on the symmetries of the lattice and the general form is given under Sec.~S4 of SM \cite{supp}. The coefficients for the decorated square lattice considered here will be discussed in Sec.~\ref{scmft}. It is evident from Eq.~\eqref{nondiagH} that $\hat{\mathcal{H}}$ is not diagonal in HP operators and thus the true quantum ground state cannot be the HP-vacuum.

\subsubsection{Bogolyubov Hamiltonian} \label{sectionBogolyubov}
To diagonalize $\hat{\mathcal{H}}$, we invoke the canonical Bogolyubov transformations \cite{Bogolyubov1947, Bogolyubov1958, Valatin1958} on the HP operators, both in the spin and orbital isospin sectors, as follows:
\begin{subequations}
    \label{Bmatrix}
    \begin{align}
        \begin{pmatrix}
            \alpha_{\mathbf{k},\sigma} \\
            \beta_{-\mathbf{k},\sigma}^{\dagger}
        \end{pmatrix}
        &=
        \begin{pmatrix}
            u_{\mathbf{k},\sigma} & -v_{\mathbf{k},\sigma}^{*} \\
            -v_{-\mathbf{k},\sigma}^{*} & u_{-\mathbf{k},\sigma}
        \end{pmatrix}
        \begin{pmatrix}
            a_{\mathbf{k},\sigma} \\
            b_{-\mathbf{k},\sigma}^{\dagger}
        \end{pmatrix}
        \label{Bmatrixsigma} \\
        \begin{pmatrix}
            \alpha_{\mathbf{k},\tau} \\
            \beta_{-\mathbf{k},\tau}^{\dagger}
        \end{pmatrix}
        &=
        \begin{pmatrix}
            u_{\mathbf{k},\tau} & -v_{\mathbf{k},\tau}^{*} \\
            -v_{-\mathbf{k},\tau}^{*} & u_{-\mathbf{k},\tau}
        \end{pmatrix}
        \begin{pmatrix}
            a_{\mathbf{k},\tau} \\
            b_{-\mathbf{k},\tau}^{\dagger}
        \end{pmatrix}
        \label{Bmatrixtau}
    \end{align}
\end{subequations}
Since this is a canonical transformation, the Bogolyubov operators must also follow the same commutation relations of HP operators (see Eq.~\eqref{HP_comm}):
\begin{subequations}
    \label{B_comm}
    \begin{align}
        [\fbannihilatesa[\mathbf{k}],\fbcreatesa[\mathbf{k}']] = \delta_{\mathbf{k},\mathbf{k}'} \;&;\; [\fbannihilatesb[-\mathbf{k}],\fbcreatesb[-\mathbf{k}']] = \delta_{-\mathbf{k},-\mathbf{k}'} \\
        [\fbannihilateta[\mathbf{k}],\fbcreateta[\mathbf{k}']] = \delta_{\mathbf{k},\mathbf{k}'} \;&;\; [\fbannihilatetb[-\mathbf{k}],\fbcreatetb[-\mathbf{k}']] = \delta_{-\mathbf{k},-\mathbf{k}'}
    \end{align}
\end{subequations}
subject to the usual constraints on the Bogolyubov coefficients:
\begin{subequations}
    \label{Bconst}
    \begin{align}
        |u_{\mathbf{k},\sigma}|^{2} - |v_{\mathbf{k},\sigma}|^{2} = 1 \label{Bconst_sigma} \\
        |u_{\mathbf{k},\tau}|^{2} - |v_{\mathbf{k},\tau}|^{2} = 1 \label{Bconst_tau}
    \end{align}
\end{subequations}
Then, $\hat{\mathcal{H}}$ becomes diagonal in Bogolyubov operators:
\begin{equation}
    \begin{gathered}
        \hat{\mathcal{H}} = \hat{\mathcal{H}}_{0} + \sum_{\mathbf{k}} \Big[ \omega_{\mathbf{k},\sigma}^{\alpha} \fbnasigma{\mathbf{k}}{\mathbf{k}} + \omega_{-\mathbf{k},\sigma}^{\beta} \fbnbsigma{-\mathbf{k}}{-\mathbf{k}} \Big] \\
        +\sum_{\mathbf{k}} \Big[ \omega_{\mathbf{k},\tau}^{\alpha} \fbnatau{\mathbf{k}}{\mathbf{k}} + \omega_{-\mathbf{k},\tau}^{\beta} \fbnbtau{-\mathbf{k}}{-\mathbf{k}} \Big]
    \end{gathered}
    \label{diagH}
\end{equation}
Thus, we obtain four possible decoupled Bogolyubov Bosons which are the eigenmodes of our proposed model (also see Sec.~S5 of SM \cite{supp}). The eigenmodes $\alpha_{\mathbf{k},\sigma}$ and $\beta_{-\mathbf{k},\sigma}^{\dagger}$ are the SW (magnon) eigenmodes belonging to the spin sector, while $\alpha_{\mathbf{k},\tau}$ and $\beta_{-\mathbf{k},\tau}^{\dagger}$ are the OW (orbiton) eigenmodes belonging to the orbital sector. These magnons and orbitons are the quasiparticle excitations on top of the true quantum ground state, i.e., the Bogolyubov vacuum. Unlike HP operators, Bogolyubov operators in the $\sigma(\tau)$ sector may comprise of multiple $\sigma(\tau)$ raising or lowering operations that preserve the total $\sigma(\tau)$ symmetries since they are linear combinations of HP operators (see Eq.~\eqref{Bmatrix}).
The diagonal form of $\hat{\mathcal{H}}$ in Eq.~\eqref{diagH} enforces the following relations on the Bogolyubov operators in the $\sigma(\tau)$ sector \cite{Fazekas1999}:
\begin{subequations}
    \label{HEOM}
    \begin{gather}
        [\alpha_{\mathbf{k},\sigma(\tau)}, \hat{\mathcal{H}}] = \omega_{\mathbf{k},\sigma(\tau)}^{\alpha} \alpha_{\mathbf{k},\sigma(\tau)} \\
        [\alpha_{\mathbf{k},\sigma(\tau)}^{\dagger}, \hat{\mathcal{H}}] = -\omega_{\mathbf{k},\sigma(\tau)}^{\alpha} \alpha_{\mathbf{k},\sigma(\tau)}^{\dagger} \\
        [\beta_{-\mathbf{k},\sigma(\tau)}, \hat{\mathcal{H}}] = \omega_{-\mathbf{k},\sigma(\tau)}^{\beta} \beta_{-\mathbf{k},\sigma(\tau)} \\
        [\beta_{-\mathbf{k},\sigma(\tau)}^{\dagger}, \hat{\mathcal{H}}] = -\omega_{-\mathbf{k},\sigma(\tau)}^{\beta} \beta_{-\mathbf{k},\sigma(\tau)}^{\dagger}
    \end{gather}
\end{subequations}
The excitation spectrum for the four decoupled independent collective modes are then obtained by substituting the non-diagonal $\hat{\mathcal{H}}$ of Eq.~\eqref{nondiagH} in Eq.~\eqref{HEOM}, leading to secular equations with the following solutions:
\begin{subequations}
    \begin{align}
        \omega_{\mathbf{k},\sigma}^{\alpha} &= \mathcal{E}_{\sigma}^{-}(\mathbf{k}) + \sqrt{\left(\mathcal{E}_{\sigma}^{+}(\mathbf{k})\right)^{2} - \left|\mathcal{F}_{\sigma}(\mathbf{k})\right|^{2}}
        \label{dispersion1alpha} \\
        \omega_{-\mathbf{k},\sigma}^{\beta} &= -\mathcal{E}_{\sigma}^{-}(\mathbf{k}) + \sqrt{\left(\mathcal{E}_{\sigma}^{+}(\mathbf{k})\right)^{2} - \left|\mathcal{F}_{\sigma}(\mathbf{k})\right|^{2}}
        \label{dispersion1beta} \\
        \omega_{\mathbf{k},\tau}^{\alpha} &= \mathcal{E}_{\tau}^{-}(\mathbf{k}) + \sqrt{\left(\mathcal{E}_{\tau}^{+}(\mathbf{k})\right)^{2} - \left|\mathcal{F}_{\tau}(\mathbf{k})\right|^{2}}
        \label{dispersion2alpha} \\
        \omega_{-\mathbf{k},\tau}^{\beta} &= -\mathcal{E}_{\tau}^{-}(\mathbf{k}) + \sqrt{\left(\mathcal{E}_{\tau}^{+}(\mathbf{k})\right)^{2} - \left|\mathcal{F}_{\tau}(\mathbf{k})\right|^{2}}
        \label{dispersion2beta}
    \end{align}
    \label{dispersion}
\end{subequations}
where, we have defined:
\begin{gather*}
    \mathcal{E}_{\sigma}^{+}(\mathbf{k}) = \frac{\mathcal{E}_{\sigma}^{\text{A}}(\mathbf{k}) + \mathcal{E}_{\sigma}^{\text{B}}(\mathbf{k})}{2} \;;\;
    \mathcal{E}_{\sigma}^{-}(\mathbf{k}) = \frac{\mathcal{E}_{\sigma}^{\text{A}}(\mathbf{k}) - \mathcal{E}_{\sigma}^{\text{B}}(\mathbf{k})}{2} \\
    \mathcal{E}_{\tau}^{+}(\mathbf{k}) = \frac{\mathcal{E}_{\tau}^{\text{A}}(\mathbf{k}) + \mathcal{E}_{\tau}^{\text{B}}(\mathbf{k})}{2} \;;\;
    \mathcal{E}_{\tau}^{-}(\mathbf{k}) = \frac{\mathcal{E}_{\tau}^{\text{A}}(\mathbf{k}) - \mathcal{E}_{\tau}^{\text{B}}(\mathbf{k})}{2}
\end{gather*}
Here, Eqs.~\eqref{dispersion1alpha} and \eqref{dispersion1beta} define the spectrum of SW excitations or magnons and those in Eqs.~\eqref{dispersion2alpha} and \eqref{dispersion2beta} define the spectrum of OW excitations or orbitons. It should be evident from the above equations that the term under square root is the same for both modes within each sector and without the additional first term, both modes would be degenerate. Thus any degeneracy lifting between the magnon and orbiton eigenmodes is a consequence of non-zero $\mathcal{E}_{\sigma}^{-}$ and $\mathcal{E}_{\tau}^{-}$ respectively (see Sec.~S7 of SM \cite{supp} and Sec.~\ref{mag_orb} for detailed discussions).
But finding this spectrum requires the knowledge of Bogolyubov coefficients that enable the canonical transformations to diagonalize $\hat{\mathcal{H}}$. Although the Bogolyubov coefficients in Eqs.~\eqref{Bmatrixsigma} and \eqref{Bmatrixtau} are complex in general, there is an inherent gauge freedom in Eq.~\eqref{diagH} which enables us to choose $u_{\mathbf{k},\sigma},u_{\mathbf{k},\tau} \in \mathbb{R}$ and $v_{\mathbf{k},\sigma},v_{\mathbf{k},\tau} \in \mathbb{C}$ (see Sec.~S6 of SM \cite{supp}). Using Eqs.~\eqref{Bconst} and \eqref{HEOM}, one obtains the following:
\begin{subequations}
    \label{Bcoeff}
    \begin{gather}
        |u_{\mathbf{k},\sigma}| = \sqrt{\frac{\mathcal{E}_{\sigma}^{+}(\mathbf{k})}{2\sqrt{\left(\mathcal{E}_{\sigma}^{+}(\mathbf{k})\right)^{2} - \left|\mathcal{F}_{\sigma}(\mathbf{k})\right|^{2}}} + \frac{1}{2}}
        \label{Bsigma_u} \\
        |v_{\mathbf{k},\sigma}| = \sqrt{\frac{\mathcal{E}_{\sigma}^{+}(\mathbf{k})}{2\sqrt{\left(\mathcal{E}_{\sigma}^{+}(\mathbf{k})\right)^{2} - \left|\mathcal{F}_{\sigma}(\mathbf{k})\right|^{2}}} - \frac{1}{2}}
        \label{Bsigma_v} \\
        u_{\mathbf{k},\sigma}v_{\mathbf{k},\sigma} = -\frac{\mathcal{F}_{\sigma}(\mathbf{k})}{2\sqrt{\left(\mathcal{E}_{\sigma}^{+}(\mathbf{k})\right)^{2} - \left|\mathcal{F}_{\sigma}(\mathbf{k})\right|^{2}}} \\
        |u_{\mathbf{k},\tau}| = \sqrt{\frac{\mathcal{E}_{\tau}^{+}(\mathbf{k})}{2\sqrt{\left(\mathcal{E}_{\tau}^{+}(\mathbf{k})\right)^{2} - \left|\mathcal{F}_{\tau}(\mathbf{k})\right|^{2}}} + \frac{1}{2}}
        \label{Btau_u} \\
        |v_{\mathbf{k},\tau}| = \sqrt{\frac{\mathcal{E}_{\tau}^{+}(\mathbf{k})}{2\sqrt{\left(\mathcal{E}_{\tau}^{+}(\mathbf{k})\right)^{2} - \left|\mathcal{F}_{\tau}(\mathbf{k})\right|^{2}}} - \frac{1}{2}}
        \label{Btau_v} \\
        u_{\mathbf{k},\tau}v_{\mathbf{k},\tau} = -\frac{\mathcal{F}_{\tau}(\mathbf{k})}{2\sqrt{\left(\mathcal{E}_{\tau}^{+}(\mathbf{k})\right)^{2} - \left|\mathcal{F}_{\tau}(\mathbf{k})\right|^{2}}}
    \end{gather}
\end{subequations}
Inverting Eq.~\eqref{Bmatrix} leads to the inverse Bogolyubov transformations:
\begin{subequations}
    \label{invBmatrix}
    \begin{align}
        \begin{pmatrix}
            a_{\mathbf{k},\sigma} \\
            b_{-\mathbf{k},\sigma}^{\dagger}
        \end{pmatrix}
        &=
        \begin{pmatrix}
            u_{\mathbf{k},\sigma} & v_{-\mathbf{k},\sigma} \\
            v_{\mathbf{k},\sigma} & u_{\mathbf{k},\sigma}
        \end{pmatrix}
        \begin{pmatrix}
            \alpha_{\mathbf{k},\sigma} \\
            \beta_{-\mathbf{k},\sigma}^{\dagger}
        \end{pmatrix}
        \label{invBmatrixsigma} \\
        \begin{pmatrix}
            a_{\mathbf{k},\tau} \\
            b_{-\mathbf{k},\tau}^{\dagger}
        \end{pmatrix}
        &=
        \begin{pmatrix}
            u_{\mathbf{k},\tau} & v_{-\mathbf{k},\tau} \\
            v_{\mathbf{k},\tau} & u_{\mathbf{k},\tau}
        \end{pmatrix}
        \begin{pmatrix}
            \alpha_{\mathbf{k},\tau} \\
            \beta_{-\mathbf{k},\tau}^{\dagger}
        \end{pmatrix}
        \label{invBmatrixtau}
    \end{align}
\end{subequations}
where, the relations $u_{\mathbf{k},\sigma} = u_{-\mathbf{k},\sigma}$, $u_{\mathbf{k},\tau} = u_{-\mathbf{k},\tau}$, $v_{-\mathbf{k},\sigma} = v_{\mathbf{k},\sigma}^{*}$ and $v_{-\mathbf{k},\tau} = v_{\mathbf{k},\tau}^{*}$ hold from Eq.~\eqref{Bcoeff}.
The expectation values of HP operators in Eq.~\eqref{HPavg} can now be expressed in terms of the Bogolyubov coefficients (also see Secs.~S7 and S8 of SM \cite{supp}):
\begin{subequations}
    \begin{align}
        \varrho_{\mathbf{k},\sigma} &=  |v_{\mathbf{k},\sigma}|^{2} \\
        \Delta_{\mathbf{k},\sigma} = u_{\mathbf{k},\sigma}v_{-\mathbf{k},\sigma} \;;\;
        &\Delta_{\mathbf{k},\sigma}^{*} = u_{\mathbf{k},\sigma}v_{\mathbf{k},\sigma} = \Delta_{-\mathbf{k},\sigma} \\
        \varrho_{\mathbf{k},\tau} &= |v_{\mathbf{k},\tau}|^{2} \\
        \Delta_{\mathbf{k},\tau} = u_{\mathbf{k},\tau}v_{-\mathbf{k},\tau} \;;\;
        &\Delta_{\mathbf{k},\tau}^{*} = u_{\mathbf{k},\tau}v_{\mathbf{k},\tau} = \Delta_{-\mathbf{k},\tau}
    \end{align}
\end{subequations}
Next, we look at the symmetries of the 2D decorated square lattice which further reduces the expressions derived until now, finally leading to self-consistent mean-field equations.

\subsubsection{Self-consistent mean-field equations} \label{scmft}
The 2D decorated square lattice of Fig.~\ref{fig1} belongs to the wallpaper group $p4mm$ as discussed in Sec.~\ref{model_setup}.
A point $\mathbf{k} = (k_{x},k_{y})$ in the first Brillouin zone will have the following symmetry-equivalent points:
\begin{equation}
    \label{ksym}
    \mathbf{k}' = R\,\mathbf{k} =
    \begin{cases}
        (-k_{x},-k_{y}) & R = C_{2z} = \mathcal{P} \\
        (k_{y},-k_{x}) & R = C_{4z} \\
        (-k_{y},k_{x}) & R = C_{4z}^{3} \\
        (-k_{x},k_{y}) & R = m_{x} \\
        (k_{x},-k_{y}) & R = m_{y} \\
        (k_{y},k_{x}) & R = m_{x+y} \\
        (-k_{y},-k_{x}) & R = m_{x-y} \\
    \end{cases}
    \equiv \mathbf{k}
\end{equation}
This yields reduced forms for the coefficients of HP operators in Eq.~\eqref{nondiagH}:
\begin{subequations}
    \label{HPcoeff}
    \begin{multline}
        \label{E1A}
        \mathcal{E}_{\sigma}^{\text{A}}(\mathbf{k}) = -4J_{1}^{\sigma}(1-\varphi_{1,\pm}^{\sigma}\eta^{2}) (\sigma - \varrho_{\sigma} - \lambda_{1}^{\sigma}\Delta_{\sigma}) \\
        - 4Q_{1}(1-\varphi_{1,\pm}^{\sigma\tau}\eta^{2}) \big[-\tau^{2} (\sigma - \varrho_{\sigma} - \lambda_{1}^{\sigma}\Delta_{\sigma}) \\
        + 2\sigma\tau (\varrho_{\tau} + \lambda_{1}^{\tau}\Delta_{\tau})\big] + 2\mathcal{K}_{z}^{\sigma} (\sigma - 2\varrho_{\sigma}) \\
        + 2J_{2x}^{\sigma} (1+\varphi_{2,+}^{\sigma}\eta) \big[(\sigma - \varrho_{\sigma}) (\,1 - \lambda_{2}^{\sigma}\cos{(2k_{x}a)}\,) \\
        + \mathit{t}_{x,\sigma}^{\text{A}} (\,\lambda_{2}^{\sigma} - \cos{(2k_{x}a)}\,)\big] \\
        + 2J_{2y}^{\sigma} (1-\varphi_{2,-}^{\sigma}\eta) \big[(\sigma - \varrho_{\sigma}) (\,1 - \lambda_{2}^{\sigma}\cos{(2k_{y}a)}\,) \\
        + \mathit{t}_{y,\sigma}^{\text{A}} (\,\lambda_{2}^{\sigma} - \cos{(2k_{y}a)}\,)\big] \\
        + 2Q_{2x} (1+\varphi_{2,+}^{\sigma\tau}\eta) \big[\tau^2 \big\{(\sigma - \varrho_{\sigma}) (\,1 - \lambda_{2}^{\sigma}\cos{(2k_{x} a)}\,)  \\
        + \mathit{t}_{x,\sigma}^{\text{A}} (\,\lambda_{2}^{\sigma} - \cos{(2k_{x} a)}\,)\big\} \\
        - 2\sigma\tau \big\{(\varrho_{\tau} - \lambda_{2}^{\tau} \mathit{t}_{x,\tau}^{\text{A}}) (\,1 - \lambda_{2}^{\sigma} \cos{(2k_{x} a)}\,)\big\}\big] \\
        + 2Q_{2y} (1-\varphi_{2,-}^{\sigma\tau}\eta) \big[\tau^2 \big\{(\sigma - \varrho_{\sigma}) (\,1 - \lambda_{2}^{\sigma}\cos{(2k_{y} a)}\,)  \\
        + \mathit{t}_{y,\sigma}^{\text{A}} (\,\lambda_{2}^{\sigma} - \cos{(2k_{y} a)}\,)\big\} \\
        - 2\sigma\tau \big\{(\varrho_{\tau} - \lambda_{2}^{\tau} \mathit{t}_{y,\tau}^{\text{A}}) (\,1 - \lambda_{2}^{\sigma} \cos{(2k_{y} a)}\,)\big\}\big]
    \end{multline}
    \begin{multline}
        \label{E1B}
        \mathcal{E}_{\sigma}^{\text{B}}(\mathbf{k}) = -4J_{1}^{\sigma}(1-\varphi_{1,\pm}^{\sigma}\eta^{2}) (\sigma - \varrho_{\sigma} - \lambda_{1}^{\sigma}\Delta_{\sigma}) \\
        - 4Q_{1}(1-\varphi_{1,\pm}^{\sigma\tau}\eta^{2}) \big[-\tau^{2} (\sigma - \varrho_{\sigma} - \lambda_{1}^{\sigma}\Delta_{\sigma}) \\
        + 2\sigma\tau (\varrho_{\tau} + \lambda_{1}^{\tau}\Delta_{\tau})\big] + 2\mathcal{K}_{z}^{\sigma} (\sigma - 2\varrho_{\sigma}) \\
        + 2J_{2y}^{\sigma} (1-\varphi_{2,-}^{\sigma}\eta) \big[(\sigma - \varrho_{\sigma}) (\,1 - \lambda_{2}^{\sigma}\cos{(2k_{x}a)}\,) \\
        + \mathit{t}_{x,\sigma}^{\text{B}} (\,\lambda_{2}^{\sigma} - \cos{(2k_{x}a)}\,)\big] \\
        + 2J_{2x}^{\sigma} (1+\varphi_{2,+}^{\sigma}\eta) \big[(\sigma - \varrho_{\sigma}) (\,1 - \lambda_{2}^{\sigma}\cos{(2k_{y}a)}\,) \\
        + \mathit{t}_{y,\sigma}^{\text{B}} (\,\lambda_{2}^{\sigma} - \cos{(2k_{y}a)}\,)\big] \\
        + 2Q_{2y} (1-\varphi_{2,-}^{\sigma\tau}\eta) \big[\tau^2 \big\{(\sigma - \varrho_{\sigma}) (\,1 - \lambda_{2}^{\sigma}\cos{(2k_{x} a)}\,)  \\
        + \mathit{t}_{x,\sigma}^{\text{B}} (\,\lambda_{2}^{\sigma} - \cos{(2k_{x} a)}\,)\big\} \\
        - 2\sigma\tau \big\{(\varrho_{\tau} - \lambda_{2}^{\tau} \mathit{t}_{x,\tau}^{\text{B}}) (\,1 - \lambda_{2}^{\sigma} \cos{(2k_{x} a)}\,)\big\}\big] \\
        + 2Q_{2x} (1+\varphi_{2,+}^{\sigma\tau}\eta) \big[\tau^2 \big\{(\sigma - \varrho_{\sigma}) (\,1 - \lambda_{2}^{\sigma}\cos{(2k_{y} a)}\,)  \\
        + \mathit{t}_{y,\sigma}^{\text{B}} (\,\lambda_{2}^{\sigma} - \cos{(2k_{y} a)}\,)\big\} \\
        - 2\sigma\tau \big\{(\varrho_{\tau} - \lambda_{2}^{\tau} \mathit{t}_{y,\tau}^{\text{B}}) (\,1 - \lambda_{2}^{\sigma} \cos{(2k_{y} a)}\,)\big\}\big]
    \end{multline}
    \begin{multline}
        \label{E2A}
        \mathcal{E}_{\tau}^{\text{A}}(\mathbf{k}) = -4J_{1}^{\tau}(1-\varphi_{1,\pm}^{\tau}\eta^{2}) (\tau - \varrho_{\tau} - \lambda_{1}^{\tau}\Delta_{\tau}) \\
        - 4Q_{1}(1-\varphi_{1,\pm}^{\sigma\tau}\eta^{2}) \big[-\sigma^{2} (\tau - \varrho_{\tau} - \lambda_{1}^{\tau}\Delta_{\tau}) \\
        + 2\sigma\tau (\varrho_{\sigma} + \lambda_{1}^{\sigma}\Delta_{\sigma})\big] + 2\mathcal{K}_{z}^{\tau} (\tau - 2\varrho_{\tau}) \\
        + 2J_{2x}^{\tau} (1+\varphi_{2,+}^{\tau}\eta) \big[(\tau - \varrho_{\tau}) (\,1 - \lambda_{2}^{\tau}\cos{(2k_{x} a)}\,) \\
        + \mathit{t}_{x,\tau}^{\text{A}} (\,\lambda_{2}^{\tau} - \cos{(2k_{x} a)}\,)\big] \\
        + 2J_{2y}^{\tau} (1-\varphi_{2,-}^{\tau}\eta) \big[(\tau - \varrho_{\tau}) (\,1 - \lambda_{2}^{\tau}\cos{(2k_{y} a)}\,) \\
        + \mathit{t}_{y,\tau}^{\text{A}} (\,\lambda_{2}^{\tau} - \cos{(2k_{y} a)}\,)\big] \\
        + 2Q_{2x} (1+\varphi_{2,+}^{\sigma\tau}\eta) \big[\sigma^2 \big\{(\tau - \varrho_{\tau}) (\,1 - \lambda_{2}^{\tau}\cos{(2k_{x} a)}\,)  \\
        + \mathit{t}_{x,\tau}^{\text{A}} (\,\lambda_{2}^{\tau} - \cos{(2k_{x} a)}\,)\big\} \\
        - 2\sigma\tau \big\{(\varrho_{\sigma} - \lambda_{2}^{\sigma}\mathit{t}_{x,\sigma}^{\text{A}}) (\,1 - \lambda_{2}^{\tau} \cos{(2k_{x} a)}\,)\big\}\big] \\
        + 2Q_{2y} (1-\varphi_{2,-}^{\sigma\tau}\eta) \big[\sigma^2 \big\{(\tau - \varrho_{\tau}) (\,1 - \lambda_{2}^{\tau}\cos{(2k_{y} a)}\,)  \\
        + \mathit{t}_{y,\tau}^{\text{A}} (\,\lambda_{2}^{\tau} - \cos{(2k_{y} a)}\,)\big\} \\
        - 2\sigma\tau \big\{(\varrho_{\sigma} - \lambda_{2}^{\sigma}\mathit{t}_{y,\sigma}^{\text{A}}) (\,1 - \lambda_{2}^{\tau} \cos{(2k_{y} a)}\,)\big\}\big]
    \end{multline}
    \begin{multline}
        \label{E2B}
        \mathcal{E}_{\tau}^{\text{B}}(\mathbf{k}) = -4J_{1}^{\tau}(1-\varphi_{1,\pm}^{\tau}\eta^{2}) (\tau - \varrho_{\tau} - \lambda_{1}^{\tau}\Delta_{\tau}) \\
        - 4Q_{1}(1-\varphi_{1,\pm}^{\sigma\tau}\eta^{2}) \big[-\sigma^{2} (\tau - \varrho_{\tau} - \lambda_{1}^{\tau}\Delta_{\tau}) \\
        + 2\sigma\tau (\varrho_{\sigma} + \lambda_{1}^{\sigma}\Delta_{\sigma})\big] + 2\mathcal{K}_{z}^{\tau} (\tau - 2\varrho_{\tau}) \\
        + 2J_{2y}^{\tau} (1-\varphi_{2,-}^{\tau}\eta) \big[(\tau - \varrho_{\tau}) (\,1 - \lambda_{2}^{\tau}\cos{(2k_{x} a)}\,) \\
        + \mathit{t}_{x,\tau}^{\text{B}} (\,\lambda_{2}^{\tau} - \cos{(2k_{x} a)}\,)\big] \\
        + 2J_{2x}^{\tau} (1+\varphi_{2,+}^{\tau}\eta) \big[(\tau - \varrho_{\tau}) (\,1 - \lambda_{2}^{\tau}\cos{(2k_{y} a)}\,) \\
        + \mathit{t}_{y,\tau}^{\text{B}} (\,\lambda_{2}^{\tau} - \cos{(2k_{y} a)}\,)\big] \\
        + 2Q_{2y} (1-\varphi_{2,-}^{\sigma\tau}\eta) \big[\sigma^2 \big\{(\tau - \varrho_{\tau}) (\,1 - \lambda_{2}^{\tau}\cos{(2k_{x} a)}\,)  \\
        + \mathit{t}_{x,\tau}^{\text{B}} (\,\lambda_{2}^{\tau} - \cos{(2k_{x} a)}\,)\big\} \\
        - 2\sigma\tau \big\{(\varrho_{\sigma} - \lambda_{2}^{\sigma}\mathit{t}_{x,\sigma}^{\text{B}}) (\,1 - \lambda_{2}^{\tau} \cos{(2k_{x} a)}\,)\big\}\big] \\
        + 2Q_{2x} (1+\varphi_{2,+}^{\sigma\tau}\eta) \big[\sigma^2 \big\{(\tau - \varrho_{\tau}) (\,1 - \lambda_{2}^{\tau}\cos{(2k_{y} a)}\,)  \\
        + \mathit{t}_{y,\tau}^{\text{B}} (\,\lambda_{2}^{\tau} - \cos{(2k_{y} a)}\,)\big\} \\
        - 2\sigma\tau \big\{(\varrho_{\sigma} - \lambda_{2}^{\sigma}\mathit{t}_{y,\sigma}^{\text{B}}) (\,1 - \lambda_{2}^{\tau} \cos{(2k_{y} a)}\,)\big\}\big]
    \end{multline}
    \begin{multline}
        \label{F1k}
        \mathcal{F}_{\sigma}(\mathbf{k}) = -4J_{1}^{\sigma}(1-\varphi_{1,\pm}^{\sigma}\eta^{2}) \big[\lambda_{1}^{\sigma}(\sigma - \varrho_{\sigma}) - \Delta_{\sigma}\big]\gamma_{\mathbf{k}} \\
        - 4Q_{1}(1-\varphi_{1,\pm}^{\sigma\tau}\eta^{2}) \big[ -\tau^{2} (\lambda_{1}^{\sigma}(\sigma - \varrho_{\sigma}) - \Delta_{\sigma}) \\
        + 2\lambda_{1}^{\sigma}\sigma\tau (\varrho_{\tau} + \lambda_{1}^{\tau}\Delta_{\tau}) \big]\gamma_{\mathbf{k}}
    \end{multline}
    \begin{multline}
        \label{F2k}
        \mathcal{F}_{\tau}(\mathbf{k}) = -4J_{1}^{\tau}(1-\varphi_{1,\pm}^{\tau}\eta^{2}) \big[ \lambda_{1}^{\tau}(\tau - \varrho_{\tau}) - \Delta_{\tau} \big]\gamma_{\mathbf{k}} \\
        -4Q_{1}(1-\varphi_{1,\pm}^{\sigma\tau}\eta^{2}) \big[ -\sigma^{2} (\lambda_{1}^{\tau}(\tau - \varrho_{\tau}) - \Delta_{\tau}) \\
        + 2\lambda_{1}^{\tau}\sigma\tau (\varrho_{\sigma} + \lambda_{1}^{\sigma}\Delta_{\sigma}) \big]\gamma_{\mathbf{k}}
    \end{multline}
\end{subequations}
The NN structure factor has been denoted by $\gamma_{\mathbf{k}}$ and defined as:
\begin{equation}
    \label{struct_fact}
    \gamma_{\mathbf{k}} = \frac{1}{2} [\cos{(k_{x}a + k_{y}a)} + \cos{(k_{x}a - k_{y}a)}] 
\end{equation}
It is clear from Eq.~\eqref{HPcoeff} that the coefficients of HP operators in Eq.~\eqref{nondiagH} for the spin and orbital sectors are interdependent. The coefficients $\mathcal{E}_{\sigma}$ and $\mathcal{F}_{\sigma}$ not only depend on $\sigma$ and $\sigma$-related expectation values but also on $\tau$ and $\tau$-related expectation values. The same is true for $\mathcal{E}_{\tau}$ and $\mathcal{F}_{\tau}$. Thus, the decoupled collective eigenmodes also become interdependent, despite the absence of any hybridized modes (see Eqs.~\eqref{diagH} and \eqref{dispersion}). As mentioned in Sec.~S1 of SM \cite{supp} and Sec.~\ref{model_setup}, this is evidently a consequence of the biquadratic exchange terms. In the absence of such terms, the magnetic and orbital subsystems truly decouple with no dependence on each other.
In Eq.~\eqref{HPcoeff}, $\varrho_{\sigma(\tau)}$ and $\Delta_{\sigma(\tau)}$ are the number density and pairing amplitude of magnons(orbitons) respectively that are usually encountered while going beyond linear SW-OW theory. The additional terms due to considering NNN interactions are $\mathit{t}_{\boldsymbol{\zeta},\sigma(\tau)}^{\text{A}} = \langle a_{i,\sigma(\tau)}^{\dagger} a_{i+\boldsymbol{\zeta},\sigma(\tau)} \rangle$ and $\mathit{t}_{\boldsymbol{\zeta},\sigma(\tau)}^{\text{B}} = \langle b_{i,\sigma(\tau)}^{\dagger} b_{i+\boldsymbol{\zeta},\sigma(\tau)} \rangle$ which are the magnon(orbiton) hopping amplitudes in magnetic sublattices A and B respectively along the four NNN coordination vectors $\boldsymbol{\zeta} = (\pm 2a,0),(0,\pm 2a)$. All of these are defined as follows (also see Secs.~S3, S8 and S9 of SM \cite{supp}):
\begin{subequations}
    \label{sc_eq}
    \begin{gather}
        \varrho_{\sigma} = \frac{2}{N_m} \sum_{\mathbf{k}} \varrho_{\mathbf{k},\sigma} = \frac{2}{N_m} \sum_{\mathbf{k}} v_{\mathbf{k},\sigma}^{2} \\
        \Delta_{\sigma} = \frac{2}{N_m} \sum_{\mathbf{k}} \gamma_{\mathbf{k}} \Delta_{\mathbf{k},\sigma} = \frac{2}{N_m} \sum_{\mathbf{k}} \gamma_{\mathbf{k}} u_{\mathbf{k},\sigma}v_{\mathbf{k},\sigma} \\
        \mathit{t}_{\boldsymbol{\zeta},\sigma}^{\text{A}} = \frac{2}{N_m} \!\sum_{\mathbf{k}} e^{\mathrm{i} \mathbf{k} \cdot \boldsymbol{\zeta}} \varrho_{\mathbf{k},\sigma} = \frac{2}{N_m} \!\sum_{\mathbf{k}} e^{\mathrm{i} \mathbf{k} \cdot \boldsymbol{\zeta}} v_{\mathbf{k},\sigma}^{2} = \mathit{t}_{\boldsymbol{\zeta},\sigma}^{\text{B}} \\
        \varrho_{\tau} = \frac{2}{N_m} \sum_{\mathbf{k}} \varrho_{\mathbf{k},\tau} = \frac{2}{N_m} \sum_{\mathbf{k}} v_{\mathbf{k},\tau}^{2} \\
        \Delta_{\tau} = \frac{2}{N_m} \sum_{\mathbf{k}} \gamma_{\mathbf{k}} \Delta_{\mathbf{k},\tau} = \frac{2}{N_m} \sum_{\mathbf{k}} \gamma_{\mathbf{k}} u_{\mathbf{k},\tau}v_{\mathbf{k},\tau} \\
        \mathit{t}_{\boldsymbol{\zeta},\tau}^{\text{A}} = \frac{2}{N_m} \!\sum_{\mathbf{k}} e^{\mathrm{i} \mathbf{k} \cdot \boldsymbol{\zeta}} \varrho_{\mathbf{k},\tau} = \frac{2}{N_m} \!\sum_{\mathbf{k}} e^{\mathrm{i} \mathbf{k} \cdot \boldsymbol{\zeta}} v_{\mathbf{k},\tau}^{2} = \mathit{t}_{\boldsymbol{\zeta},\tau}^{\text{B}}
    \end{gather}
\end{subequations}
where, we have used the fact $v_{\mathbf{k},\sigma}, v_{\mathbf{k},\tau} \in \mathbb{R} \Rightarrow \Delta_{\mathbf{k},\sigma}, \Delta_{\mathbf{k},\tau} \in \mathbb{R}$ as is evident from Eqs.~\eqref{Bcoeff}, \eqref{ksym}, \eqref{F1k}, \eqref{F2k} and \eqref{struct_fact}. Also, $\mathit{t}_{\boldsymbol{\zeta},\sigma(\tau)}^{\text{A}},\mathit{t}_{\boldsymbol{\zeta},\sigma(\tau)}^{\text{B}} \in \mathbb{R}$ since forward and backward hoppings are equivalent and due to the lattice symmetries of Eq.~\eqref{ksym}, $\mathit{t}_{\boldsymbol{\zeta},\sigma(\tau)}^{\text{A}} = \mathit{t}_{\boldsymbol{\zeta},\sigma(\tau)}^{\text{B}}$.
It is worth mentioning that both sides of Eq.~\eqref{sc_eq} contain unknowns that are dependent on each other, yielding a set of equations where they have to be evaluated self-consistently given an initial guess value. Thus, Eq.~\eqref{sc_eq} defines the self-consistent mean-field (SCMF) equations for our minimal spin-orbital model,

\begin{figure*}[!htb]
    \centering
    \includegraphics[width=\textwidth]{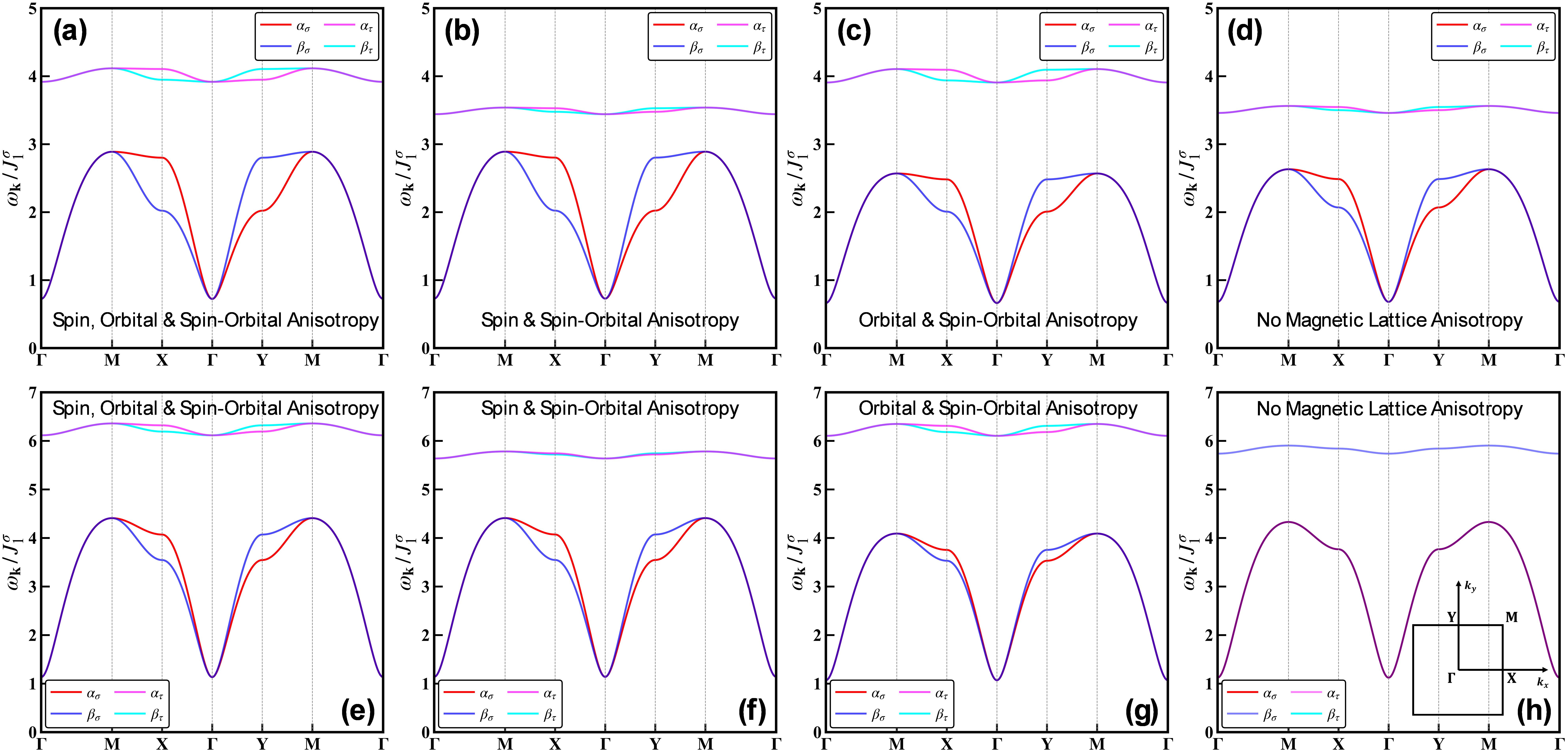}
    \caption{Spectrum of collective spin-wave and orbital-wave excitations on a 2D decorated square lattice. Top row (a-d) corresponds to staggered lattice order and bottom row (e-h) corresponds to aligned lattice order. Each column corresponds to different combinations of anisotropy. The two magnon bands corresponding to magnon eigenmodes $\alpha_{\mathbf{k},\sigma}$ and $\beta_{-\mathbf{k},\sigma}^{\dagger}$ have been denoted by $\alpha_{\sigma}$ and $\beta_{\sigma}$ respectively. Similarly, the two orbiton bands corresponding to orbiton eigenmodes $\alpha_{\mathbf{k},\tau}$ and $\beta_{-\mathbf{k},\tau}^{\dagger}$ have been denoted by $\alpha_{\tau}$ and $\beta_{\tau}$ respectively. For more details and the base model parameters used in all the dispersions, refer Sec.~\ref{mag_orb}. The model parameters chosen for \textit{bare} spin, orbital and spin-orbital anisotropy are $\sfrac{J_{2x}^{\sigma}}{J_{1}^{\sigma}} = 0.2$, $\sfrac{J_{2y}^{\sigma}}{J_{1}^{\sigma}} = 0.1$; $\sfrac{J_{2x}^{\tau}}{J_{1}^{\sigma}} = 0.4$, $\sfrac{J_{2y}^{\tau}}{J_{1}^{\sigma}} = 0.1$; and $\sfrac{Q_{2x}}{J_{1}^{\sigma}} = -0.35$, $\sfrac{Q_{2y}}{J_{1}^{\sigma}} = -0.05$ respectively where $J_{1}^{\sigma} < 0$. Inset of (h) shows the 2D Brillouin zone.}
    \label{fig2}
\end{figure*}

\section{Results and Discussions}
\subsection{Dispersion of Magnons and Orbitons} \label{mag_orb}

The spectrum of collective SW and OW excitations is obtained by solving the SCMF equations within the Brillouin zone of a 2D decorated square lattice. Usually, the OW excitations occur at a higher energy scale and are quite separated from SW excitations \cite{Brink1998}, as observed in real systems such as LaMnO$_{3}$ \cite{Moussa1996, Inami2003} and La$_{2}$CuO$_{4}$ \cite{Coldea2001, Martinelli2024}. Further, the orbital subsystem is highly directional due to orbital structure compared to the magnetic subsystem \cite{Khomskii2014}. This motivates the following physically relevant choice of base parameters for our minimal spin-orbital model (taking $J_{1}^{\sigma} < 0$): $\sfrac{\mathcal{J}_{1}^{\tau}}{\mathcal{J}_{1}^{\sigma}} = \sfrac{J_{1}^{\tau}}{J_{1}^{\sigma}} = 2.0$, $\sfrac{\mathcal{Q}_{1}}{\mathcal{J}_{1}^{\sigma}} = \sfrac{Q_{1}}{J_{1}^{\sigma}} = -0.5$, $\sfrac{\mathcal{K}_{z}^{\sigma}}{\mathcal{J}_{1}^{\sigma}} = \sfrac{\mathcal{K}_{z}^{\tau}}{\mathcal{J}_{1}^{\sigma}} = 0$ (single-ion anisotropy neglected for simplicity), $\lambda_{1}^{\sigma} = \lambda_{2}^{\sigma} = 0.9$ (isotropic Heisnberg-like spins), $\lambda_{1}^{\tau} = \lambda_{2}^{\tau} = 0.1$ (Ising-like orbital isospins), $\varphi_{1,\pm}^{\sigma} = \varphi_{1,\pm}^{\tau} = \varphi_{1,\pm}^{\sigma\tau} = 1.0$ (\textit{exchange modifier} for effective plaquette coupling, see bottom right panel of Fig.~\ref{fig1}), $\varphi_{2,+}^{\sigma} = \varphi_{2,+}^{\tau} = \varphi_{2,+}^{\sigma\tau} = 1.5$ (stronger \textit{exchange modifier}, denoted by solid arrows in bottom right panel of Fig.~\ref{fig1}), $\varphi_{2,-}^{\sigma} = \varphi_{2,-}^{\tau} = \varphi_{2,-}^{\sigma\tau} = 1.1$ (weaker \textit{exchange modifier}, denoted by dashed arrows in bottom right panel of Fig.~\ref{fig1}). The results after fixing the above parameters are shown in Figure~\ref{fig2} (\textit{bare} NNN exchange couplings provided in caption). The solid red and blue branches correspond to $\alpha_{\sigma}$ and $\beta_{\sigma}$ magnon modes respectively, whereas the solid magenta and cyan branches correspond to $\alpha_{\tau}$ and $\beta_{\tau}$ orbiton modes respectively. Further, the orbiton branches have noticeably less dispersion compared to magnon branches because of the chosen Ising-like exchange ($\lambda_{1}^{\tau} = \lambda_{2}^{\tau} = 0.1$). There are also differences due to staggered lattice order and aligned lattice order, shown in top and bottom rows respectively of Fig.~\ref{fig2} for the same set of parameters. Here, \textit{bare} spin, orbital and spin-orbital anisotropy refers to $J_{2x}^{\sigma} \neq J_{2y}^{\sigma}$, $J_{2x}^{\tau} \neq J_{2y}^{\tau}$ and $Q_{2x} \neq Q_{2y}$ respectively. Before discussing the dispersions, we need to look at chiral-splitting in more detail.

From Eq.~\eqref{HEOM}, which has the form of Heisenberg equation of motion, it can be shown that the eigenmodes $\alpha_{\mathbf{k},\sigma}$ and $\beta_{-\mathbf{k},\sigma}^{\dagger}$ evolve in time $\mathrm{t}$ as $\sim$$e^{-\mathrm{i}\omega_{\mathbf{k},\sigma}^{\alpha}\mathrm{t}}$ and $\sim$$e^{\mathrm{i}\omega_{-\mathbf{k},\sigma}^{\beta}\mathrm{t}}$ respectively. Hence, $\alpha_{\mathbf{k},\sigma}$ and $\beta_{-\mathbf{k},\sigma}^{\dagger}$ may be considered as rotations in the opposite sense leading to a handedness or \textit{chirality}. Similarly, eigenmodes $\alpha_{\mathbf{k},\tau}$ and $\beta_{-\mathbf{k},\tau}^{\dagger}$ evolve in time $\mathrm{t}$ as $\sim$$e^{-\mathrm{i}\omega_{\mathbf{k},\tau}^{\alpha}\mathrm{t}}$ and $\sim$$e^{\mathrm{i}\omega_{-\mathbf{k},\tau}^{\beta}\mathrm{t}}$ respectively and hence are chiral (see Sec.~S5 of SM \cite{supp}). Further, the Bogolyubov operators in the $\sigma$ and $\tau$ sectors are associated with definite spin and orbital isospin since $\langle \alpha_{\mathbf{k},\sigma} \,\hat{\sigma}_{\text{tot}}^{z}\, \alpha_{\mathbf{k},\sigma}^{\dagger} \rangle = -1$, $\langle \beta_{\mathbf{k},\sigma} \,\hat{\sigma}_{\text{tot}}^{z}\, \beta_{\mathbf{k},\sigma}^{\dagger} \rangle = +1$ and $\langle \alpha_{\mathbf{k},\tau} \,\hat{\tau}_{\text{tot}}^{z}\, \alpha_{\mathbf{k},\tau}^{\dagger} \rangle = -1$, $\langle \beta_{\mathbf{k},\tau} \,\hat{\tau}_{\text{tot}}^{z}\, \beta_{\mathbf{k},\tau}^{\dagger} \rangle = +1$ respectively. Thus, $\alpha_{\mathbf{k},\sigma}$ and $\beta_{-\mathbf{k},\sigma}^{\dagger}$ are two chiral magnon modes that generate opposite spin precessions in the $\sigma$ sector. Analogously in the $\tau$ sector, $\alpha_{\mathbf{k},\tau}$ and $\beta_{-\mathbf{k},\tau}^{\dagger}$ are two chiral orbiton modes that generate opposite orbital isospin precessions. It was mentioned in Sec.~\ref{sectionBogolyubov} that non-zero $\mathcal{E}_{\sigma}^{-}$ and $\mathcal{E}_{\tau}^{-}$ leads to degeneracy lifting in the $\sigma$ and $\tau$ sectors respectively. We can now obtain an expression for this chiral-splitting, denoted by $\Omega_{\sigma}$ for magnons and $\Omega_{\tau}$ for orbitons, from Eqs.~\eqref{dispersion}, \eqref{E1A}, \eqref{E1B}, \eqref{E2A} and \eqref{E2B}:
\begin{subequations}
    \label{splitting}
    \begin{gather}
        \label{mag_split}
        \begin{gathered}
            \Omega_{\sigma} = \omega_{\mathbf{k},\sigma}^{\alpha} - \omega_{\mathbf{k},\sigma}^{\beta} = 2\mathcal{E}_{\sigma}^{-}(\mathbf{k}) = \mathcal{E}_{\sigma}^{\text{A}}(\mathbf{k}) - \mathcal{E}_{\sigma}^{\text{B}}(\mathbf{k}) \\
            \begin{multlined}
                = 2\Big[\big\{\lambda_{2}^{\sigma}(\sigma-\varrho_{\sigma}) \big[(J_{2x}^{\sigma} - J_{2y}^{\sigma}) + \eta(J_{2x}^{\sigma}\varphi_{2,+}^{\sigma} + J_{2y}^{\sigma}\varphi_{2,-}^{\sigma})\big] \\
                + \big[(J_{2x}^{\sigma}\mathit{t}_{x,\sigma}^{\text{A}} - J_{2y}^{\sigma}\mathit{t}_{y,\sigma}^{\text{A}}) + \eta(J_{2x}^{\sigma}\varphi_{2,+}^{\sigma}\mathit{t}_{x,\sigma}^{\text{A}} + J_{2y}^{\sigma}\varphi_{2,-}^{\sigma}\mathit{t}_{y,\sigma}^{\text{A}})\big]\big\} \\
                + \big\{\lambda_{2}^{\sigma} \big[\tau^{2}(\sigma-\varrho_{\sigma}) - 2\sigma\tau\varrho_{\tau}\big] \\
                \big[(Q_{2x} - Q_{2y}) + \eta(Q_{2x}\varphi_{2,+}^{\sigma\tau} + Q_{2y}\varphi_{2,-}^{\sigma\tau})\big] \\
                + \tau^{2} \big[(Q_{2x}\mathit{t}_{x,\sigma}^{\text{A}} - Q_{2y}\mathit{t}_{y,\sigma}^{\text{A}}) + \eta(Q_{2x}\varphi_{2,+}^{\sigma\tau}\mathit{t}_{x,\sigma}^{\text{A}} + Q_{2y}\varphi_{2,-}^{\sigma\tau}\mathit{t}_{y,\sigma}^{\text{A}})\big] \\
                - 2\sigma\tau \lambda_{2}^{\tau} \big[(Q_{2x}\mathit{t}_{x,\tau}^{\text{A}} - Q_{2y}\mathit{t}_{y,\tau}^{\text{A}}) \\+ \eta(Q_{2x}\varphi_{2,+}^{\sigma\tau}\mathit{t}_{x,\tau}^{\text{A}} + Q_{2y}\varphi_{2,-}^{\sigma\tau}\mathit{t}_{y,\tau}^{\text{A}})\big]\big\}\Big] \\
                \big[-\cos{(2k_{x}a)} + \cos{(2k_{y}a)}\big]
            \end{multlined} \\
            \Rightarrow \Omega_{\sigma} = \{\overline{\Omega}_{J,\sigma} + \Omega_{J,\sigma}^{\eta}\} + \{\overline{\Omega}_{Q,\sigma} + \Omega_{Q,\sigma}^{\eta}\} \\
            \therefore \boxed{\Omega_{\sigma} = \overline{\Omega}_{\sigma} + \Omega_{\sigma}^{\eta}}
        \end{gathered}
        \\[2ex]
        \begin{gathered}
            \label{orb_split}
            \Omega_{\tau} = \omega_{\mathbf{k},\tau}^{\alpha} - \omega_{\mathbf{k},\tau}^{\beta} = 2\mathcal{E}_{\tau}^{-}(\mathbf{k}) = \mathcal{E}_{\tau}^{\text{A}}(\mathbf{k}) - \mathcal{E}_{\tau}^{\text{B}}(\mathbf{k}) \\
            \begin{multlined}
                = 2\Big[\big\{\lambda_{2}^{\tau}(\tau-\varrho_{\tau}) \big[(J_{2x}^{\tau} - J_{2y}^{\tau}) + \eta(J_{2x}^{\tau}\varphi_{2,+}^{\tau} + J_{2y}^{\tau}\varphi_{2,-}^{\tau})\big] \\
                + \big[(J_{2x}^{\tau}\mathit{t}_{x,\tau}^{\text{A}} - J_{2y}^{\tau}\mathit{t}_{y,\tau}^{\text{A}}) + \eta(J_{2x}^{\tau}\varphi_{2,+}^{\tau}\mathit{t}_{x,\tau}^{\text{A}} + J_{2y}^{\tau}\varphi_{2,-}^{\tau}\mathit{t}_{y,\tau}^{\text{A}})\big]\big\} \\
                + \big\{\lambda_{2}^{\tau} \big[\sigma^{2}(\tau-\varrho_{\tau}) - 2\sigma\tau\varrho_{\sigma}\big] \\
                \big[(Q_{2x} - Q_{2y}) + \eta(Q_{2x}\varphi_{2,+}^{\sigma\tau} + Q_{2y}\varphi_{2,-}^{\sigma\tau})\big] \\
                + \sigma^{2} \big[(Q_{2x}\mathit{t}_{x,\tau}^{\text{A}} - Q_{2y}\mathit{t}_{y,\tau}^{\text{A}}) + \eta(Q_{2x}\varphi_{2,+}^{\sigma\tau}\mathit{t}_{x,\tau}^{\text{A}} + Q_{2y}\varphi_{2,-}^{\sigma\tau}\mathit{t}_{y,\tau}^{\text{A}})\big] \\
                - 2\sigma\tau \lambda_{2}^{\sigma} \big[(Q_{2x}\mathit{t}_{x,\sigma}^{\text{A}} - Q_{2y}\mathit{t}_{y,\sigma}^{\text{A}}) \\+ \eta(Q_{2x}\varphi_{2,+}^{\sigma\tau}\mathit{t}_{x,\sigma}^{\text{A}} + Q_{2y}\varphi_{2,-}^{\sigma\tau}\mathit{t}_{y,\sigma}^{\text{A}})\big]\big\}\Big] \\
                \big[-\cos{(2k_{x}a)} + \cos{(2k_{y}a)}\big]
            \end{multlined} \\
            \Rightarrow \Omega_{\tau} = \{\overline{\Omega}_{J,\tau} + \Omega_{J,\tau}^{\eta}\} + \{\overline{\Omega}_{Q,\tau} + \Omega_{Q,\tau}^{\eta}\} \\
            \therefore \boxed{\Omega_{\tau} = \overline{\Omega}_{\tau} + \Omega_{\tau}^{\eta}}
        \end{gathered}
    \end{gather}
\end{subequations}
where we have used $\mathit{t}_{x,\sigma(\tau)}^{\text{A}} = \mathit{t}_{y,\sigma(\tau)}^{\text{B}}$ and $\mathit{t}_{x,\sigma(\tau)}^{\text{B}} = \mathit{t}_{y,\sigma(\tau)}^{\text{A}}$ due to model symmetries (Eq.~\eqref{ksym}).
The structure of Eq.~\eqref{HPcoeff} makes it clear that only the NNN couplings can enter Eq.~\eqref{splitting} due to anisotropic $k$-dependence. This is in fact a general feature of effective spin models on bipartite lattices such as our proposed model in Eq.~\eqref{fullH} on a 2D decorated square lattice, where the intra-sublattice couplings (NNN couplings in our case) must have an anisotropic structure to generate chiral-splitting of collective excitation modes, as also pointed out in Ref.~\cite{Xie2026}. It is seen that in $\Omega_{\sigma}$($\Omega_{\tau}$), there are terms directly dependent on the non-magnetic lattice arrangement, denoted by $\Omega_{\sigma}^{\eta}$($\Omega_{\tau}^{\eta}$), and terms independent of $\eta$, denoted by $\overline{\Omega}_{\sigma}$($\overline{\Omega}_{\tau}$). Thus, Eqs.~\eqref{mag_split} and \eqref{orb_split} define the chiral-splitting of magnons and orbitons respectively for staggered lattice order. In Fig.~\ref{fig2}(a), none of the NNN couplings are isotropic resulting in chiral-splitting of both magnon and orbiton modes. In Fig.~\ref{fig2}(b), the magnon modes split because of spin anisotropy but there is also a small splitting in orbiton modes despite \textit{bare} NNN isospin couplings being isotropic ($J_{2x}^{\tau} = J_{2y}^{\tau}$). 
Similarly in Fig.~\ref{fig2}(c), the orbiton modes split due to orbital anisotropy but the magnon modes are also split despite \textit{bare} spin isotropy ($J_{2x}^{\sigma} = J_{2y}^{\sigma}$), although the magnitude is smaller compared to Fig.~\ref{fig2}(a). Finally in Fig.~\ref{fig2}(d), where all \textit{bare} NNN couplings are isotropic ($J_{2x}^{\sigma} = J_{2y}^{\sigma}$, $J_{2x}^{\tau} = J_{2y}^{\tau}$ and $Q_{2x} = Q_{2y}$), one still sees (reduced) chiral-splitting of both magnon and orbiton modes. Now, it is important to note that the biquadratic spin-orbital exchange interaction that couples the magnetic and orbital subsystems is usually not entirely independent of the quadratic exchange of spins and orbitals in the Kugel\textquotesingle-Khomski\u{\i} model \cite{KugelKhomskii1982, Khomskii2014}.
Thus it is natural to assume that anisotropy in quadratic exchange of either $\sigma$ or $\tau$ sector should automatically lead to anisotropy in the biquadratic exchange that couples these two sectors. In other words, presence of \textit{bare} spin or orbital anisotropy will lead to \textit{bare} spin-orbital anisotropy. It will only be isotropic if both \textit{bare} spin and orbital exchanges are isotropic. Following this principle, it is now seen that in Figs.~\ref{fig2}(b) and \ref{fig2}(c), the chiral-splitting may be attributed to two sources. First is the presence of \textit{bare} spin-orbital anisotropy because of \textit{bare} orbital and spin anisotropy respectively. But here interestingly, had we considered \textit{bare} spin-orbital isotropy ($Q_{2x} = Q_{2y}$), there would still be chiral-splitting of both magnon and orbiton modes, especially visible in Fig.~\ref{fig2}(d) where all \textit{bare} NNN couplings are isotropic. This is because for staggered lattice order, although $\overline{\Omega}_{\sigma}$ and $\overline{\Omega}_{\tau}$ may be zero for isotropic \textit{bare} NNN couplings, $\Omega_{\sigma}$ and $\Omega_{\tau}$ will still be non-zero because of finite $\Omega_{\sigma}^{\eta}$ and $\Omega_{\tau}^{\eta}$ respectively resulting in small but finite chiral-splitting, as shown in Figs.~\ref{fig2}(b-d). Thus, $\Omega_{\sigma}^{\eta}$ and $\Omega_{\tau}^{\eta}$ is the second source of chiral-splitting in magnon and orbital modes respectively that can only be zero if the lattice couplings are very weak ($\varphi_{2}^{\sigma/\tau/\sigma\tau} \rightarrow 0$) or the \textit{superexchange} pathways favour ferro-type order along one direction and antiferro-type along the other with the same \textit{exchange modifier} ($\varphi_{2,+}^{\sigma/\tau/\sigma\tau} = -\varphi_{2,-}^{\sigma/\tau/\sigma\tau}$). Therefore, Figs.~\ref{fig2}(a-d) correspond to collective excitations in CS-AMs.

It should not be hard to see from Eq.~\eqref{splitting} (and Eq.~\eqref{HPcoeff}) that for aligned lattice order, since there can only be one type of \textit{exchange modifier}, they can be globally factored out resulting in $\Omega_{\sigma(\tau)} \sim 2(1+\varphi_{2,+}^{\sigma(\tau)}\eta)[\cdots] + 2(1+\varphi_{2,+}^{\sigma\tau}\eta)[\cdots]$ or $\Omega_{\sigma(\tau)} \sim 2(1-\varphi_{2,-}^{\sigma(\tau)}\eta)[\cdots] + 2(1-\varphi_{2,-}^{\sigma\tau}\eta)[\cdots]$. Thus for aligned lattice order, the \textit{exchange modifiers} simply get reduced to scaling factors and there are no $\Omega_{\sigma}^{\eta}$ and $\Omega_{\tau}^{\eta}$ terms separate from $\overline{\Omega}_{\sigma}$ and $\overline{\Omega}_{\tau}$.
This leads to a overall reduction in the chiral-splitting of magnons and orbitons despite all other model parameters being the same, as seen in Fig.~\ref{fig2}(e-g). In Fig.~\ref{fig2}(e), chiral-splitting of magnon and orbiton modes is from the usual \textit{bare} spin, orbital and spin-orbital anisotropy. In Fig.~\ref{fig2}(f) the only source of chiral-splitting of orbitons, in spite of \textit{bare} orbital isotropy, is the \textit{bare} spin-orbital anisotropy. The same is true for chiral-splitting of magnons in Fig.~\ref{fig2}(g). The ultimate consequence of aligned lattice order is seen in Fig.~\ref{fig2}(h) where full \textit{bare} NNN isotropy leads to degenerate magnon and orbiton branches throughout the Brillouin zone. This corresponds to collective excitations in conventional antiferromagnets. In contrast, Figs.~\ref{fig2}(e-g) correspond to collective excitations in OO-AMs.

\begin{figure*}[!htb]
    \centering
    \includegraphics[width=\textwidth]{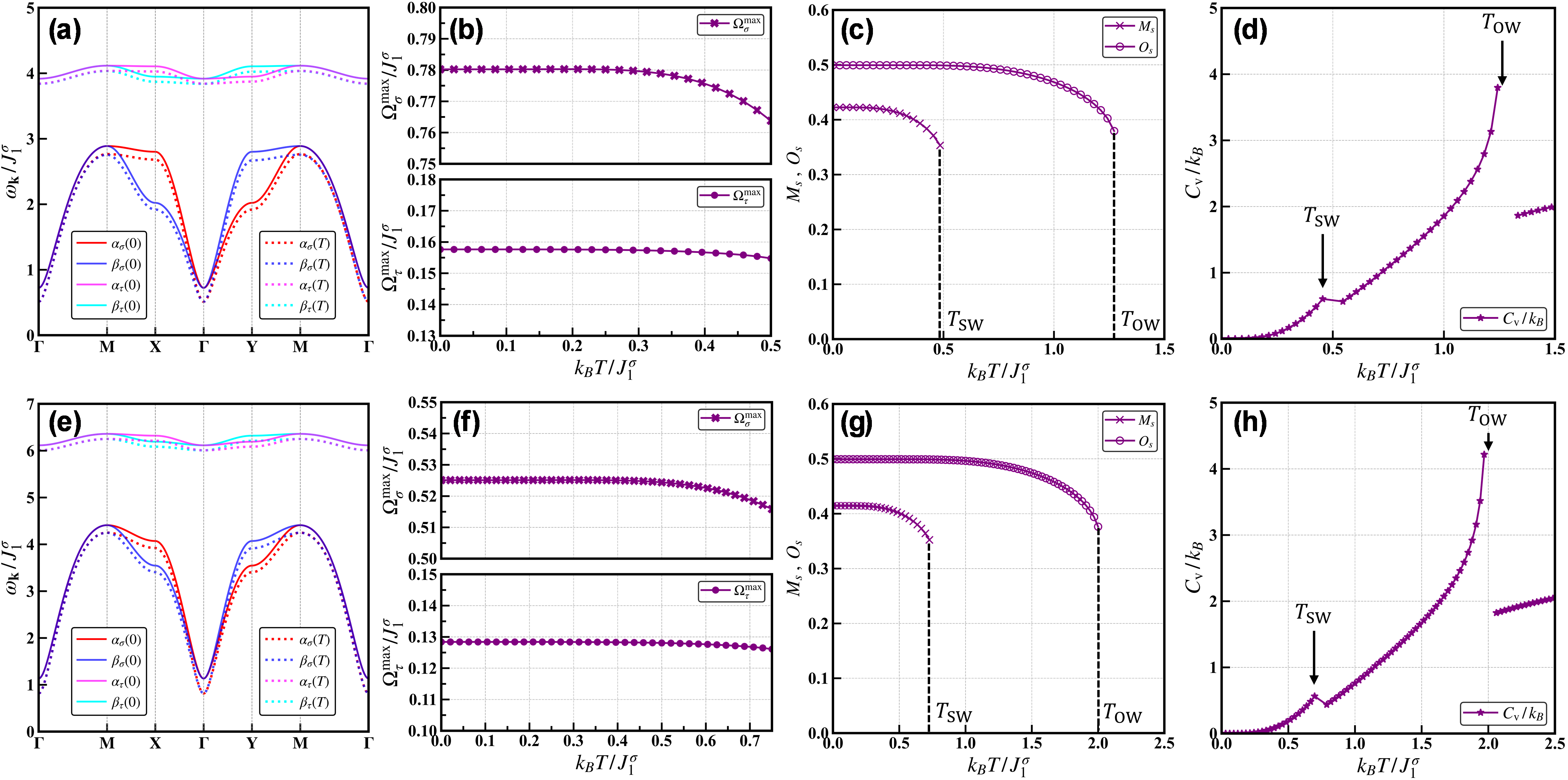}
    \caption{Temperature dependence of the (a,e) magnon spectrum with solid lines at $T=0$, dotted lines at $T=0.50\left(\sfrac{J_{1}^{\sigma}}{k_B}\right)$; and orbiton spectrum with solid lines at $T=0$, dotted lines at $T=0.75\left(\sfrac{J_{1}^{\sigma}}{k_B}\right)$, (b,f) chiral-splitting of magnon and orbiton modes, (c,g) magnetization ($M_s$) and orbitization ($O_s$) order parameters and (d,h) specific heat ($C_{\text{v}}$) in dimensionless units. Top row (a-d) corresponds to staggered lattice order while bottom row (e-h) corresponds to aligned lattice order. For more details, refer Sec.~\ref{TDSWOW}.}
    \label{fig3}
\end{figure*}

\begin{figure}
    \centering
    \includegraphics[width=\columnwidth]{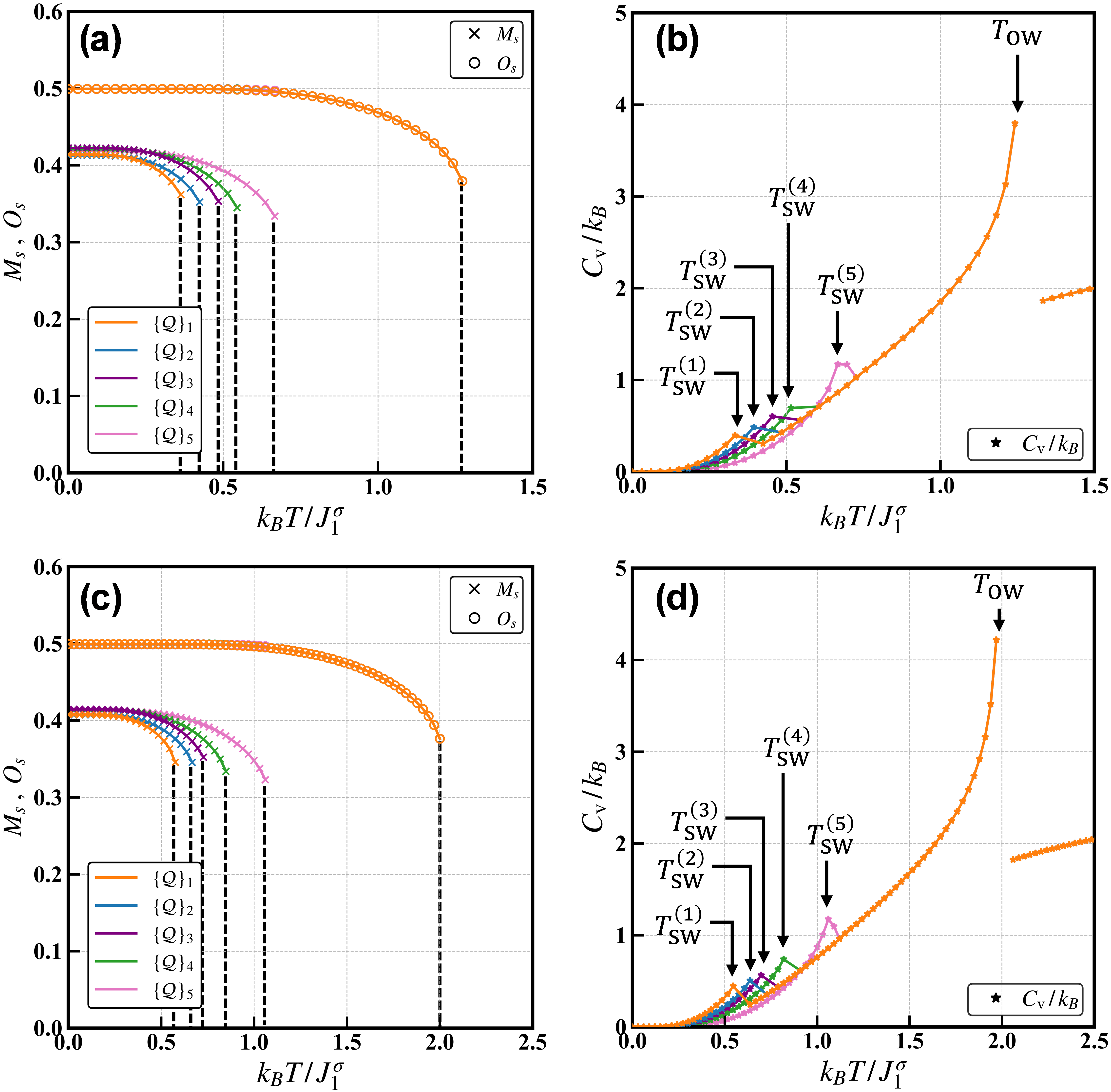}
    \caption{Temperature dependence of the (a,c) order parameters  $M_s$ and $O_s$ and (b,d) specific heat ($C_{\text{v}}$) in dimensionless units for different sets of biquadratic exchange couplings $\{\mathcal{Q}\}_{i} (i=1,\ldots,5)$ (refer Sec.~\ref{TDSWOW}). Top row (a,b) corresponds to staggered lattice order and bottom row (c,d) corresponds to aligned lattice order.}
    \label{fig4}
\end{figure}

\subsection{Effect of Finite Temperature} \label{TDSWOW}
Till now, we have focused on zero temperature results of our proposed minimal spin-orbital model with only quantum fluctuations present. We now investigate how thermal fluctuations at finite temperatures affect the collective SW and OW excitations. At zero temperature ($T=0$) in the absence of any thermal fluctuations, our spin-orbital system is expected to be fully ordered with zero magnon populations $\langle \alpha_{\mathbf{k},\sigma}^{\dagger}\alpha_{\mathbf{k},\sigma} \rangle$ \& $\langle \beta_{\mathbf{k},\sigma}^{\dagger}\beta_{\mathbf{k},\sigma} \rangle$, and orbiton populations $\langle \alpha_{\mathbf{k},\tau}^{\dagger}\alpha_{\mathbf{k},\tau} \rangle$ \& $\langle \beta_{\mathbf{k},\tau}^{\dagger}\beta_{\mathbf{k},\tau} \rangle$ in the ground state. As $T$ rises, the populations increase following the Bose-Einstein distribution \cite{Bose1924, Fazekas1999}: $f(\omega_{\mathbf{k}},T) = [\exp{(\frac{\omega_{\mathbf{k}}}{k_{B}T})-1}]^{-1}$ and the SCMF equations at $T=0$ of Eq.~\eqref{sc_eq} have to be modified to incorporate finite $T$ effects:
\begin{subequations}
    \label{sc_eq_T}
    \begin{gather}
        \begin{multlined}
            \varrho_{\sigma}(T) = \frac{2}{N_m} \sum_{\mathbf{k}} \big[f(\omega_{\mathbf{k},\sigma}^{\alpha},T) \\
            + v_{\mathbf{k},\sigma}^{2} \{f(\omega_{\mathbf{k},\sigma}^{\alpha},T) + f(\omega_{\mathbf{k},\sigma}^{\beta},T)\}\big]
        \end{multlined}
        \\[2ex]
        \Delta_{\sigma}(T) = \frac{2}{N_m} \sum_{\mathbf{k}} \gamma_{\mathbf{k}} u_{\mathbf{k},\sigma}v_{\mathbf{k},\sigma} [1 + f(\omega_{\mathbf{k},\sigma}^{\alpha}) + f(\omega_{\mathbf{k},\sigma}^{\beta})] \\
        \begin{multlined}
            \mathit{t}_{\boldsymbol{\zeta},\sigma}^{\text{A}}(T) = \frac{2}{N_m} \sum_{\mathbf{k}} e^{\mathrm{i} \mathbf{k} \cdot \boldsymbol{\zeta}} \big[f(\omega_{\mathbf{k},\sigma}^{\alpha},T) \\
            + v_{\mathbf{k},\sigma}^{2} \{f(\omega_{\mathbf{k},\sigma}^{\alpha},T) + f(\omega_{\mathbf{k},\sigma}^{\beta},T)\}\big]
        \end{multlined}
        \\[2ex]
        \begin{multlined}
            \mathit{t}_{\boldsymbol{\zeta},\sigma}^{\text{B}}(T) = \frac{2}{N_m} \sum_{\mathbf{k}} e^{\mathrm{i} \mathbf{k} \cdot \boldsymbol{\zeta}} \big[f(\omega_{\mathbf{k},\sigma}^{\beta},T) \\
            + v_{\mathbf{k},\sigma}^{2} \{f(\omega_{\mathbf{k},\sigma}^{\alpha},T) + f(\omega_{\mathbf{k},\sigma}^{\beta},T)\}\big]
        \end{multlined}
        \\[2ex]
        \begin{multlined}
            \varrho_{\tau}(T) = \frac{2}{N_m} \sum_{\mathbf{k}} \big[f(\omega_{\mathbf{k},\tau}^{\alpha},T) \\
            + v_{\mathbf{k},\tau}^{2} \{f(\omega_{\mathbf{k},\tau}^{\alpha},T) + f(\omega_{\mathbf{k},\tau}^{\beta},T)\}\big]
        \end{multlined}
        \\[2ex]
        \Delta_{\tau}(T) = \frac{2}{N_m} \sum_{\mathbf{k}} \gamma_{\mathbf{k}} u_{\mathbf{k},\tau}v_{\mathbf{k},\tau} [1 + f(\omega_{\mathbf{k},\tau}^{\alpha}) + f(\omega_{\mathbf{k},\tau}^{\beta})] \\
        \begin{multlined}
            \mathit{t}_{\boldsymbol{\zeta},\tau}^{\text{A}}(T) = \frac{2}{N_m} \sum_{\mathbf{k}} e^{\mathrm{i} \mathbf{k} \cdot \boldsymbol{\zeta}} \big[f(\omega_{\mathbf{k},\tau}^{\alpha},T) \\
            + v_{\mathbf{k},\tau}^{2} \{f(\omega_{\mathbf{k},\tau}^{\alpha},T) + f(\omega_{\mathbf{k},\tau}^{\beta},T)\}\big]
        \end{multlined}
        \\[2ex]
        \begin{multlined}
            \mathit{t}_{\boldsymbol{\zeta},\tau}^{\text{B}}(T) = \frac{2}{N_m} \sum_{\mathbf{k}} e^{\mathrm{i} \mathbf{k} \cdot \boldsymbol{\zeta}} \big[f(\omega_{\mathbf{k},\tau}^{\beta},T) \\
            + v_{\mathbf{k},\tau}^{2} \{f(\omega_{\mathbf{k},\tau}^{\alpha},T) + f(\omega_{\mathbf{k},\tau}^{\beta},T)\}\big]
        \end{multlined}
    \end{gather}
\end{subequations}
These are the SCMF equations at finite $T$. At $T=0$ for collinear $\sigma(\tau)$ arrangement with perfect compensation, we had $\langle a_{\mathbf{k},\sigma(\tau)}^{\dagger}a_{\mathbf{k},\sigma(\tau)} \rangle = \langle b_{\mathbf{k},\sigma(\tau)}^{\dagger}b_{\mathbf{k},\sigma(\tau)} \rangle$ as mentioned in Sec.~\ref{sectionHP}. At finite $T$, $\langle a_{\mathbf{k},\sigma(\tau)}^{\dagger}a_{\mathbf{k},\sigma(\tau)} \rangle = f(\omega_{\mathbf{k},\sigma(\tau)}^{\alpha}) + v_{\mathbf{k},\sigma(\tau)}^{2} [f(\omega_{\mathbf{k},\sigma(\tau)}^{\alpha}) + f(\omega_{\mathbf{k},\sigma(\tau)}^{\beta})]$ and $\langle b_{\mathbf{k},\sigma(\tau)}^{\dagger}b_{\mathbf{k},\sigma(\tau)} \rangle = f(\omega_{\mathbf{k},\sigma(\tau)}^{\beta}) + v_{\mathbf{k},\sigma(\tau)}^{2} [f(\omega_{\mathbf{k},\sigma(\tau)}^{\alpha}) + f(\omega_{\mathbf{k},\sigma(\tau)}^{\beta})]$. Since for general $\mathbf{k}$, $\omega_{\mathbf{k},\sigma(\tau)}^{\alpha} \neq \omega_{\mathbf{k},\sigma(\tau)}^{\beta}$, we must also have $\langle a_{\mathbf{k},\sigma(\tau)}^{\dagger}a_{\mathbf{k},\sigma(\tau)} \rangle \neq \langle b_{\mathbf{k},\sigma(\tau)}^{\dagger}b_{\mathbf{k},\sigma(\tau)} \rangle$ locally. But when summed over the Brillouin zone, these expectation values will become equal as a result of model symmetries and hence we need only consider $\varrho_{\sigma(\tau)}$ globally. The magnon hopping terms will also be locally distinct within a magnetic sublattice because of the extra $e^{\mathrm{i} \mathbf{k} \cdot \boldsymbol{\zeta}}$ factor, i.e., $\mathit{t}_{x,\sigma(\tau)}^{\text{A}} \neq \mathit{t}_{y,\sigma(\tau)}^{\text{A}}$ and $\mathit{t}_{x,\sigma(\tau)}^{\text{B}} \neq \mathit{t}_{y,\sigma(\tau)}^{\text{B}}$ (also see Sec.~S10 of SM \cite{supp}). Effect of thermal fluctuations enter Eq.~\eqref{dispersion} via the $T$-dependent number densities $\varrho_{\sigma}(T)$ \& $\varrho_{\tau}(T)$, pairing amplitudes $\Delta_{\sigma}(T)$ \& $\Delta_{\tau}(T)$, and hopping amplitudes $\mathit{t}_{\sigma}(T)$ \& $\mathit{t}_{\tau}(T)$ defined in Eq.~\eqref{sc_eq_T}, renormalizing the magnon and orbiton spectrums as shown in Fig.~\ref{fig3}. In Fig.~\ref{fig3}(a) and \ref{fig3}(e), where the same model parameters of Fig.~\ref{fig2}(a) and \ref{fig2}(e) have been used respectively, an overall downward shift of magnon and orbiton bands is observed. The solid spectrum is at $T = 0$ and the dotted spectrum is at finite $T = 0.50 \left(\frac{J_{1}^{\sigma}}{k_{B}}\right)$ (corresponding Fig.~\ref{fig3}(a)) and $T = 0.75 \left(\frac{J_{1}^{\sigma}}{k_{B}}\right)$ (corresponding Fig.~\ref{fig3}(e)). This is because at higher $T$, which indicates availability of higher thermal energy to the system, the excitation spectrum becomes more accessible to the ground state. Further, the chiral-splitting of magnons and orbitons also decrease with increasing $T$ as shown in Figs.~\ref{fig3}(b) and \ref{fig3}(f). Here, the maximum value of chiral-splitting magnitude of magnons and orbitons, denoted by $\Omega_{\sigma}^{\mathrm{max}}$ and $\Omega_{\tau}^{\mathrm{max}}$ respectively, have been shown that were evaluated using Eq.~\eqref{splitting}. The decrease in chiral-splitting is less for orbiton branches since they have less dispersion compared to the magnon branches. All the above effects are the results of spectrum renormalization by thermal fluctuations and these trends remain same irrespective of staggered or aligned lattice order.

The results discussed until now have been derived within the framework of SW-OW theory which is essentially a perturbative expansion of spin-waves and orbital-waves using the HP transformations \cite{HolsteinPrimakoff1940, Oguchi1960} of spin and isospin operators, resting upon certain assumptions also mentioned in Sec.~\ref{sectionHP}. For the expansion to be valid, one must satisfy the following conditions: $\varrho_{\sigma} = \langle \nasigma{i}{i} \rangle = \langle \nbsigma{i}{i} \rangle \ll \sigma$ and $\varrho_{\tau} = \langle \natau{i}{i} \rangle = \langle \nbtau{i}{i} \rangle \ll \tau$. At $T=0$, magnons or orbitons may only be created by quantum fluctuations yielding finite $\varrho_{\sigma}$ and $\varrho_{\tau}$. With increasing $T$, $\varrho_{\sigma}$ and $\varrho_{\tau}$ also increases due to thermal fluctuations until there are no physically meaningful self-consistent solutions possible for Eq.~\eqref{sc_eq_T}. This usually occurs close to the critical temperature \cite{Liu1966}, signalling a classical order-disorder phase transition driven by thermal fluctuations and leading to the breakdown of SW-OW theory as the system effectively `melts'. There are alternate approaches to theoretically investigate collective excitations that also work at finite T, such as Schwinger Boson representations \cite{Schwinger1952, Arovas1988, Sarker1989} or Dyson-Maleev transformations \cite{Dyson1956a, Dyson1956b, Maleev1958, Takahashi1989}, which is beyond the scope of current work and shall not be pursued here. Using HP transformations, the results of SW-OW theory are only reliable up to certain temperatures $T_{\mathrm{SW}}$ and $T_{\mathrm{OW}}$, which are close to the actual magnetic ordering (Néel) temperature $T_{\mathrm{N}}$ and orbital ordering temperature $T_{\mathrm{OO}}$, for the magnetic and orbital subsystems respectively. Figures~\ref{fig3}(c) and \ref{fig3}(g) show the staggered magnetization $M_s$ and staggered orbitization $O_s$ for staggered and aligned lattice arrangements respectively. They are defined as:
\begin{subequations}
    \begin{gather}
        M_s = \frac{1}{2} (M_{\mathrm{A}} - M_{\mathrm{B}}) = \sigma - \varrho_{\sigma} \\
        O_s = \frac{1}{2} (O_{\mathrm{A}} - O_{\mathrm{B}}) = \tau - \varrho_{\tau}
    \end{gather}
\end{subequations}
Since we have chosen $\sfrac{\mathcal{J}_{1}^{\tau}}{\mathcal{J}_{1}^{\sigma}} > 1$, it is natural to expect $T_{\mathrm{OO}} > T_{\mathrm{N}}$ meaning $T_{\mathrm{OW}} > T_{\mathrm{SW}}$ as seen in Figs.~\ref{fig3}(c) and \ref{fig3}(g), while choosing $\sfrac{\mathcal{J}_{1}^{\tau}}{\mathcal{J}_{1}^{\sigma}} < 1$ would lead to $T_{\mathrm{SW}} > T_{\mathrm{OW}}$. But there remains a caveat. Since the magnetic and orbital subsystems are coupled via $\mathcal{Q}_{1}$ and $\mathcal{Q}_{2}$, if the SW-OW expansion breaks down for one subsystem with no further self-consistent solutions of Eq.~\eqref{sc_eq_T}, so it does for the other subsystem. Hence, one would always get a single temperature at which our spin-orbital system `melts'. But numerically, one can trick the system into not recognizing the breakdown of one of the subsystems by decoupling them beyond $T_{\mathrm{SW}}$ or $T_{\mathrm{OW}}$, i.e., setting $\mathcal{Q}_{1}=\mathcal{Q}_{2}=0$. We call this high-temperature decoupling (HTD) and it is simply a numerical procedure enabling us to peek into the regime where SW-OW expansion for the fully coupled system has failed. The reason we see two distinct temperatures $T_{\mathrm{SW}}$ and $T_{\mathrm{OW}}$, denoted by black dashed lines in Figs.~\ref{fig3}(c) and \ref{fig3}(g) respectively, is because we have invoked the HTD procedure. This at least shows us qualitatively that there should exist two distinct ordering temperatures for the two different subsystems, as expected of a spin-orbital system. At low $T$, well within the regime of SW-OW theory, it is seen that orbital isospins are fully ordered ($O_s \approx 0.5$ at $T=0$) but the magnetization of spins is diminished ($M_s \approx 0.422$ at $T=0$). This is a consequence of Heisenberg-like ($\lambda_{1}^{\sigma}=0.9$) magnetic subsystem with enhanced quantum spin fluctuations $\sim$($\hat{\sigma}^{+}\hat{\sigma}^{-}+\hat{\sigma}^{-}\hat{\sigma}^{+}$) and Ising-like ($\lambda_{1}^{\tau}=0.9$) orbital subsystem with highly suppressed quantum orbital isospin fluctuations $\sim$($\hat{\tau}^{+}\hat{\tau}^{-}+\hat{\tau}^{-}\hat{\tau}^{+}$). As $T$ increases, both $\varrho_{\sigma}$ and $\varrho_{\tau}$ increase and hence $M_s$ and $O_s$ decrease. Beyond $T_{\mathrm{SW}}$, $\varrho_{\sigma}$ abruptly overshoots $\sigma$ without reaching $M_s=0$ and no physically meaningful self-consistent $\varrho_{\sigma}$ is possible. Performing HTD enables us to look beyond $T_{\mathrm{SW}}$ and we finally reach $T_{\mathrm{OW}}$ where $\varrho_{\tau}$ overshoots $\tau$ without reaching $O_s=0$. Although the magnetic and orbital transitions here seem to be of first-order, one must keep in mind that SW-OW theory has already broken down and obtaining order parameters around the critical temperatures will require a framework more suited at high $T$ as we shall see in Sec.~\ref{MC}. But the SW-OW framework importantly captures the quantum reduction of order parameters and the correct trend of decreasing $M_s$ and $O_s$ with increasing $T$.

The specific heat of a system can also track the transitions occurring in the individual subsystems and is obtained from $C_{\mathrm{v}} = \frac{\partial \langle\hat{\mathcal{H}}\rangle}{\partial T}$ where $\langle\hat{\mathcal{H}}\rangle = \langle\hat{\mathcal{H}}_{0}\rangle + \sum_{\mathbf{k}} [\omega_{\mathbf{k},\sigma}^{\alpha} f(\omega_{\mathbf{k},\sigma}^{\alpha},T) + \omega_{\mathbf{k},\sigma}^{\beta} f(\omega_{\mathbf{k},\sigma}^{\beta},T) + \omega_{\mathbf{k},\tau}^{\alpha} f(\omega_{\mathbf{k},\tau}^{\alpha},T) + \omega_{\mathbf{k},\tau}^{\beta} f(\omega_{\mathbf{k},\tau}^{\beta},T)]$ is the internal energy of the system obtained from Eq.~\eqref{diagH}.
Figures~\ref{fig3}(d) and \ref{fig3}(h) show the variation of specific heat for staggered and aligned lattice arrangement respectively by invoking the HTD procedure. Two discontinuity peaks corresponding to $T_{\mathrm{SW}}$ and $T_{\mathrm{OW}}$ is observed in both figures.

When the coupling strengths of spins and orbital isospins are comparable, physically one would expect $T_{\mathrm{N}} \approx T_{\mathrm{OO}}$, i.e., the spin-orbital system would show a single transition with simultaneous ordering of spins and orbitals. But these ordering temperatures will be different in general for disparate coupling strengths of spins and orbitals like we have considered in our minimal spin-orbital model. Within SW-OW theory, this manifests as distinct $T_{\mathrm{SW}}$ and $T_{\mathrm{OW}}$. Then the natural question to ask is how $T_{\mathrm{SW}}$ and $T_{\mathrm{OW}}$ may be affected by the coupling strengths, specifically the biquadratic exchange couplings since these couple the magnetic and orbital subsystems. Figure~\ref{fig4} shows the variation of $T_{\mathrm{SW}}$ and $T_{\mathrm{OW}}$ with different sets of biquadratic exchange couplings $\{\mathcal{Q}\}_{i}$ $(i = 1, \ldots, 5)$. The sets are as follows:
\begin{enumerate}
    \item $\{\mathcal{Q}\}_{1}$ = \{$\sfrac{Q_{1}}{J_{1}^{\sigma}} = 0.00$, $\sfrac{Q_{2x}}{J_{1}^{\sigma}} = 0.00$, $\sfrac{Q_{2y}}{J_{1}^{\sigma}} = 0.00$\}
    \item $\{\mathcal{Q}\}_{2}$ = \{$\sfrac{Q_{1}}{J_{1}^{\sigma}} = -0.50$, $\sfrac{Q_{2x}}{J_{1}^{\sigma}} = 0.00$, $\sfrac{Q_{2y}}{J_{1}^{\sigma}} = 0.00$\}
    \item $\{\mathcal{Q}\}_{3}$ = \{$\sfrac{Q_{1}}{J_{1}^{\sigma}} = -0.50$, $\sfrac{Q_{2x}}{J_{1}^{\sigma}} = -0.35$, $\sfrac{Q_{2y}}{J_{1}^{\sigma}} = -0.05$\}
    \item $\{\mathcal{Q}\}_{4}$ = \{$\sfrac{Q_{1}}{J_{1}^{\sigma}} = -1.00$, $\sfrac{Q_{2x}}{J_{1}^{\sigma}} = -0.35$, $\sfrac{Q_{2y}}{J_{1}^{\sigma}} = -0.05$\}
    \item $\{\mathcal{Q}\}_{5}$ = \{$\sfrac{Q_{1}}{J_{1}^{\sigma}} = -2.00$, $\sfrac{Q_{2x}}{J_{1}^{\sigma}} = -0.35$, $\sfrac{Q_{2y}}{J_{1}^{\sigma}} = -0.05$\}
\end{enumerate}
with the appropriate \textit{exchange modifiers} for the corresponding \textit{bare} couplings mentioned in Sec.~\ref{mag_orb}. The purple curve in Figs.~\ref{fig4}(a) and \ref{fig4}(b) has the same model parameters as in Figs.~\ref{fig3}(c) and \ref{fig3}(d) with $\{\mathcal{Q}\}_{3}$ and acts as a reference. Similarly, the reference purple curve in Figs.~\ref{fig4}(c) and \ref{fig4}(d) has the same model parameters as in Figs.~\ref{fig3}(g) and \ref{fig3}(h) with $\{\mathcal{Q}\}_{3}$. As mentioned earlier, $T_{\mathrm{OW}} > T_{\mathrm{SW}}$ because $\sfrac{\mathcal{J}_{1}^{\tau}}{\mathcal{J}_{1}^{\sigma}} > 1$. Since we have used the HTD procedure here, $T_{\mathrm{OW}}$ effectively becomes fixed with only varying $T_{\mathrm{SW}}$ corresponding to different sets of biquadratic exchange. There are five such $T_{\mathrm{SW}}^{(i)}$'s, one for each set, beyond which the orbital subsystem is decoupled numerically and has no effect on $T_{\mathrm{OW}}$, i.e., orbital ordering. It should be evident that the situation would be reversed for $\sfrac{\mathcal{J}_{1}^{\tau}}{\mathcal{J}_{1}^{\sigma}} < 1$ within the HTD procedure since $\hat{\mathcal{H}}$ is symmetric under the swapping of $\sigma$ and $\tau$ (see Eq.~\eqref{fullH}). Thus, it should be understood from Fig.~\ref{fig4} that increasing the NN or NNN biquadratic exchange strength increases $T_{\mathrm{SW}}$ or $T_{\mathrm{OW}}$ depending on whichever subsystem has stronger exchange coupling, i.e., the subsystem with stronger coupling strength drives the ordering of the subsystem with weaker coupling strength. It is further seen that finite temperature behaviour also depends on the type of non-magnetic lattice arrangement and affects the magnetic and orbital ordering temperatures. 

\subsection{Classical Monte-Carlo Simulations} \label{MC}

\begin{figure}[!htb]
    \centering
    \includegraphics[width=\columnwidth]{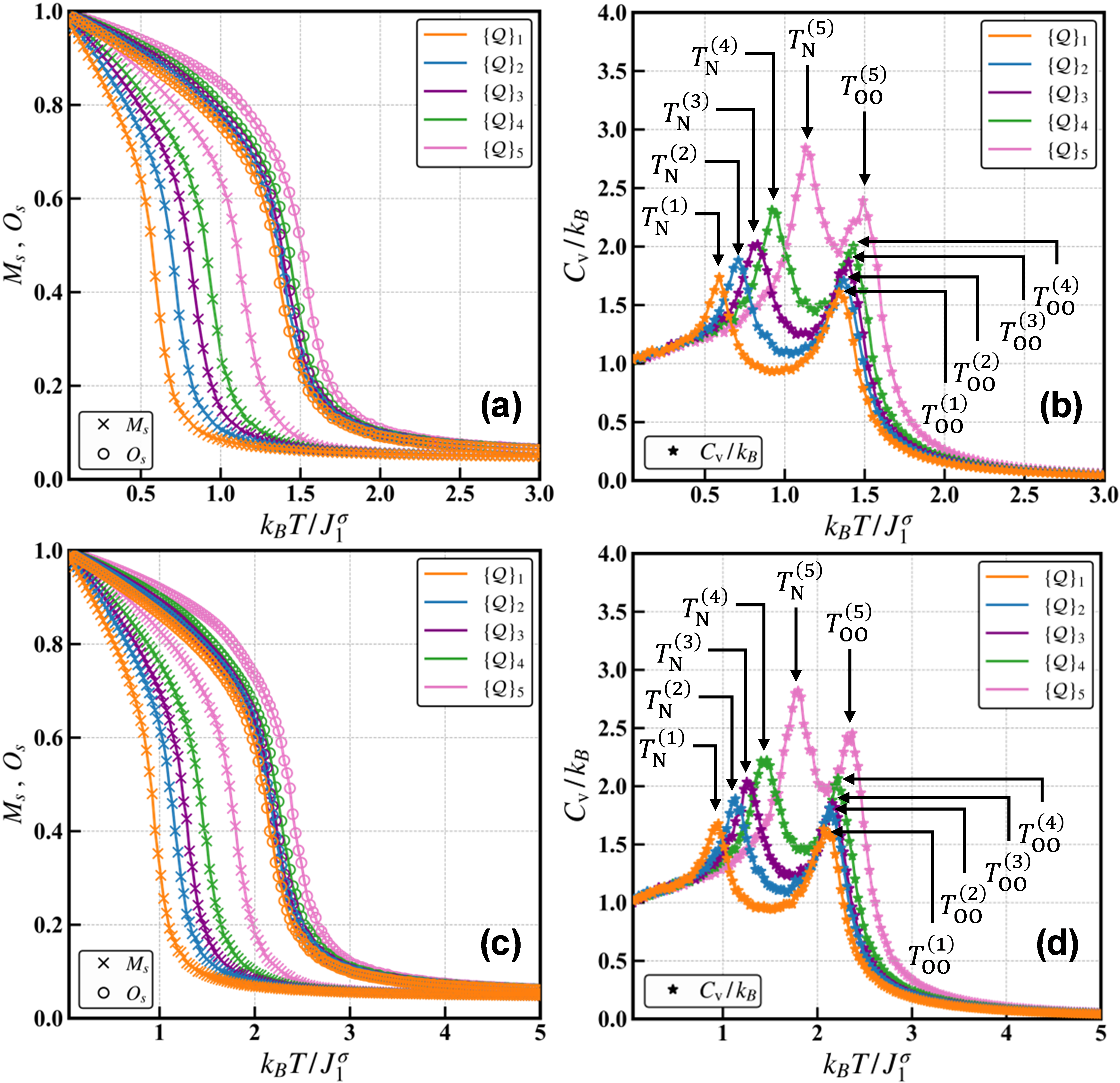}
    \caption{Monte-Carlo simulated results for the (a,c) order parameters $M_s$ and  $O_s$, (b,d) specific heat ($C_{\text{v}}$) vs. temperature in dimensionless units for different sets of model parameters $\{\mathcal{Q}\}_{i} (i=1,\ldots,5)$ as in Fig.~\ref{fig4}. Top row (a,b) corresponds to staggered lattice order and bottom row (c,d) corresponds to aligned lattice order.}
    \label{fig5}
\end{figure}

\begin{table*}[!htb]
    \caption {Comparison of transition temperatures between self-consistent mean-field spin-wave orbital-wave (SCMF SW-OW) theory and classical Monte-Carlo (MC) simulation. Here, spin-wave and orbital-wave transition temperatures have been denoted by $T_{\mathrm{SW}}$ and $T_{\mathrm{OW}}$ respectively while Néel-ordering and orbital-ordering temperatures have been denoted by $T_{\mathrm{N}}$ and $T_{\mathrm{OO}}$ respectively. The temperatures shown here are in units of $\left(\sfrac{J_{1}^{\sigma}}{k_{B}}\right)$.}
    \label{tab1}
    \begin{ruledtabular}
        \begin{tabular}{cccccc}
            \multirow{2}{*}{\textbf{\shortstack{Parameter \\ Set}}} & \multirow{2}{*}{\textbf{\shortstack{Non-magnetic \\ Lattice}}} & \multicolumn{2}{c}{\textbf{SCMF SW-OW Theory}} & \multicolumn{2}{c}{\textbf{Classical MC Simulation}} \\
            \cline{3-6}
            & & $T_{\mathrm{SW}}$ & $T_{\mathrm{OW}}$ & $T_{\mathrm{N}}$ & $T_{\mathrm{OO}}$ \\
            \hline\hline
            \multirow{2}{*}{$\{\mathcal{Q}\}_{1}$} & staggered & 0.36 & 1.26 & 0.59 & 1.34 \\
            & aligned & 0.57 & 2.01 & 0.95 & 2.06 \\
            \hline
            \multirow{2}{*}{$\{\mathcal{Q}\}_{2}$} & staggered & 0.42 & 1.26 & 0.71 & 1.37 \\
            & aligned & 0.66 & 2.01 & 1.13 & 2.12 \\
            \hline
            \multirow{2}{*}{$\{\mathcal{Q}\}_{3}$} & staggered & 0.48 & 1.26 & 0.83 & 1.40 \\
            & aligned & 0.72 & 2.01 & 1.25 & 2.15 \\
            \hline
            \multirow{2}{*}{$\{\mathcal{Q}\}_{4}$} & staggered & 0.54 & 1.26 & 0.92 & 1.43 \\
            & aligned & 0.84 & 2.01 & 1.46 & 2.21 \\
            \hline
            \multirow{2}{*}{$\{\mathcal{Q}\}_{5}$} & staggered & 0.66 & 1.26 & 1.13 & 1.49 \\
            & aligned & 1.05 & 2.01 & 1.79 & 2.36 \\
        \end{tabular}
    \end{ruledtabular}
\end{table*}

As mentioned in Sec.~\ref{TDSWOW}, the framework of SW-OW theory is valid only at low temperatures when the conditions $\varrho_{\sigma} \ll \sigma$ and $\varrho_{\tau} \ll \tau$ are satisfied. But while approaching the actual ordering temperatures, these conditions are violated since there is a proliferation of magnons and orbitons in the spin-orbital system due to increased thermal fluctuations. Since at high temperatures the quantum fluctuations become negligibly small compared to the thermal fluctuations \cite{Rosengaard1997, Cuccoli2000, Pajda2001, Torelli2019}, one may then consider the classical limit of Eq.~\eqref{fullH}, where quantum spins and isospins lose their operator structure becoming classical 3D vectors with fixed magnitude, usually normalized to unity, whose components commute with each other. The simplest method to probe such situations is the classical Monte-Carlo method where the system is considered to be in contact with a heat-bath providing thermal energy to the system that cause deviations of spins or isospins from their ordered directions. The details of the method are available in Sec.~S11 of SM \cite{supp}. The results of the classical Monte-Carlo simulations for the five different sets of biquadratic exchange $\{\mathcal{Q}\}_{i}$ $(i = 1,\dots,5)$ provided in Sec.~\ref{TDSWOW} are shown in Figure~\ref{fig5}. The top and bottom rows correspond to staggered and aligned lattice orders respectively.
In Figs.~\ref{fig5}(a) and \ref{fig5}(c), both $M_s$ and $O_s$ (normalized to unity) show a continuous change from finite ordered values towards zero, indicating a second-order magnetic and orbital phase transition for each $\{\mathcal{Q}\}_{i}$. The specific heat $C_{\text{v}}$ in Figs.~\ref{fig5}(b) and \ref{fig5}(d) corroborates this and shows two peaks, corresponding to ordering of spins and orbital isospins for each $\{\mathcal{Q}\}_{i}$. The magnetic ordering temperature $T_{\mathrm{N}}^{(i)}$ follows the same trend as $T_{\mathrm{SW}}^{(i)}$ in Fig.~\ref{fig4} and increases with increasing biquadratic exchange coupling strength. Importantly, $T_{\mathrm{OO}}^{(i)}$ also increases with increasing biquadratic exchange coupling strength unlike $T_{\mathrm{OW}}$ in Fig.~\ref{fig4} where HTD procedure was used, fixing it at a single value.
The classical nature of the model becomes apparent at low temperatures with the absence of any quantum reduction of $M_s$ and $O_s$ and finite $C_{\text{v}}$.
Nevertheless the conclusions of Sec.~\ref{TDSWOW}, namely that the subsystem with stronger coupling strength drives the ordering of the subsystem with weaker coupling strength is validated by the classical limit of our minimal model. Furthermore, the explicit modification of exchange couplings by the non-magnetic lattice provides an effective route for tuning the magnetic and orbital ordering temperatures. A summary of the transition temperatures obtained within self-consistent mean-field spin-wave orbital-wave theory (corresponding Fig.~\ref{fig4}) and classical Monte-Carlo simulation (corresponding Fig.~\ref{fig5}) is given in Table~\ref{tab1}. In all parameter cases, $T_{\mathrm{SW}} < T_{\mathrm{N}}$ and $T_{\mathrm{OW}} < T_{\mathrm{OO}}$ as expected since SW-OW expansions become invalid beyond $T_{\mathrm{SW}}$ and $T_{\mathrm{OW}}$.

\section{Conclusions}

We have developed a unified microscopic framework for describing collective excitations in altermagnets by proposing an extended Kugel\textquotesingle-Khomski\u{\i} spin-orbital model that explicitly incorporates both inequivalent non-magnetic lattice environments and correlation-driven orbital ordering. Within a self-consistent mean-field spin-wave orbital-wave formalism, the model supports two distinct types of mutually unhybridized collective excitations, magnons and orbitons, which nevertheless remain intrinsically coupled via biquadratic spin-orbital exchange interactions.

A central finding of this work is that both magnon and orbiton spectra exhibit characteristic chiral-splitting governed by two physically distinct microscopic mechanisms. The first originates directly from inequivalent non-magnetic lattice environments and therefore represents the collective excitation fingerprint of crystalline symmetry-mediated altermagnetism. The second arises from exchange anisotropy induced by orbital order and survives even in crystallographically equivalent magnetic lattices, providing the corresponding signature of orbitally ordered altermagnetism. Our results therefore establish, within a single theoretical framework, how these two seemingly different microscopic routes towards altermagnetism manifest themselves in the collective excitation spectrum.

Finite temperature effects leads to renormalization of the magnon and orbiton spectrums, further demonstrating the interplay between the spin and orbital sectors, and reveals the presence of two distinct phase transitions in the spin-orbital system driven by thermal fluctuations. It is further seen that the subsystem with stronger coupling strength can enhance the ordering of the subsystem with weaker coupling strength, increasing the ordering temperatures of the overall spin-orbital system.
While the self-consistent mean-field SW-OW theory predicts abrupt quenching of the order parameters near the transition temperatures, signalling the breakdown of SW-OW approximations through spurious first-order transitions, complementary classical Monte Carlo simulations recover the expected continuous second-order phase transitions in the thermodynamic limit. Furthermore, the explicit modulation of exchange interactions by the non-magnetic lattice provides an effective mechanism for tuning the ordering temperatures of both magnetic and orbital subsystems.

Beyond providing a minimal description of collective excitations in altermagnets, the present work bridges two recently proposed microscopic origins of altermagnetism within a common spin-orbital framework. We anticipate that our results will motivate future experimental investigations of magnons and orbitons in candidate altermagnetic materials and provide a theoretical foundation for exploiting collective spin-wave and orbital-wave excitations in spintronic and orbitronic devices based on altermagnets.

\begin{acknowledgments}
    BD thanks Industrial Research and Consultancy Centre (IRCC) and Department of Physics, IIT Bombay, for financial support. BD and CKB also thank Surajit Sarkar for valuable suggestions.
\end{acknowledgments}
\bibliography{main}
\clearpage

\resumetoc

\title{Supplementary Material for `A Unified Theory of Collective Magnon and Orbiton Excitations in Altermagnets'}

\maketitle
\onecolumngrid
\tableofcontents

\setcounter{section}{0}
\setcounter{subsection}{0}
\setcounter{equation}{0}
\setcounter{figure}{0}
\setcounter{table}{0}
\makeatletter
\renewcommand{\thesection}{S\arabic{section}}
\renewcommand{\thesubsection}{\Alph{subsection}}
\renewcommand{\thefigure}{S\arabic{figure}}
\renewcommand{\theequation}{\thesection.\arabic{equation}}
\renewcommand{\thetable}{S\arabic{table}}

\begin{bibunit}[apsrev4-2]

\bigskip
\noindent Here, we provide general results of the self-consistent mean-field spin-wave orbital-wave theory using Holstein-Primakoff and Bogolyubov transformations and computational details of classical Monte-Carlo simulation.
\clearpage

\section{Bosonization of Spins and Orbital Isospins}

Collective spin-wave (SW) and orbital-wave (OW) excitations are quantized by mapping the spin and orbital isospin operators to Bosonic creation and annihilation operators in the spin and orbital sectors respectively using the Holstein-Primakoff (HP) transformations \cite{HolsteinPrimakoff1940_sm, Fazekas1999_sm}:
\begin{subequations} \label{sigA}
    \begin{gather}
        \hat{\sigma}_{i}^{z} = \sigma - \nasigma{i}{i} \label{saz} \\
        \hat{\sigma}_{i}^{+} = \sqrt{2\sigma} \sqrt{1 - \frac{\nasigma{i}{i}}{2\sigma}} \annihilatesa \simeq \sqrt{2\sigma} \left(1 - \frac{\nasigma{i}{i}}{4\sigma}\right) \annihilatesa \label{sa+} \\
        \hat{\sigma}_{i}^{-} = \sqrt{2\sigma} \createsa \sqrt{1 - \frac{\nasigma{i}{i}}{2\sigma}} \simeq \sqrt{2\sigma} \createsa \left(1 - \frac{\nasigma{i}{i}}{4\sigma}\right) \label{sa-}
    \end{gather}
\end{subequations}
\begin{subequations} \label{sigB}
    \begin{gather}
        \hat{\sigma}_{j}^{z} = -\sigma + \nbsigma{j}{j} \label{sbz} \\
        \hat{\sigma}_{j}^{+} = \sqrt{2\sigma} \createsb \sqrt{1 - \frac{\nbsigma{j}{j}}{2\sigma}} \simeq \sqrt{2\sigma} \createsb \left(1 - \frac{\nbsigma{j}{j}}{4\sigma}\right) \label{sb+} \\
        \hat{\sigma}_{j}^{-} = \sqrt{2\sigma} \sqrt{1 - \frac{\nbsigma{j}{j}}{2\sigma}} \annihilatesb \simeq \sqrt{2\sigma} \left(1 - \frac{\nbsigma{j}{j}}{4\sigma}\right) \annihilatesb \label{sb-}
    \end{gather}
\end{subequations}
\begin{subequations} \label{tauA}
    \begin{gather}
        \hat{\tau}_{i}^{z} = \tau - \natau{i}{i} \label{taz} \\
        \hat{\tau}_{i}^{+} = \sqrt{2\tau}\; \sqrt{1 - \frac{\natau{i}{i}}{2\tau}} \annihilateta \simeq \sqrt{2\tau} \left(1 - \frac{\natau{i}{i}}{4\tau}\right) \annihilateta \label{ta+} \\
        \hat{\tau}_{i}^{-} = \sqrt{2\tau} \createta \sqrt{1 - \frac{\natau{i}{i}}{2\tau}} \simeq \sqrt{2\tau} \createta \left(1 - \frac{\natau{i}{i}}{4\tau}\right) \label{ta-}
    \end{gather}
\end{subequations}
\begin{subequations}
    \label{tauB}
    \begin{gather}
        \hat{\tau}_{j}^{z} = -\tau + \nbtau{j}{j} \label{tbz} \\
        \hat{\tau}_{j}^{+} = \sqrt{2\tau} \createtb \sqrt{1 - \frac{\nbtau{j}{j}}{2\tau}} \simeq \sqrt{2\tau} \createtb \left(1 - \frac{\nbtau{j}{j}}{4\tau}\right) \label{tb+} \\
        \hat{\tau}_{j}^{-} = \sqrt{2\tau} \sqrt{1 - \frac{\nbtau{j}{j}}{2\tau}} \annihilatetb \simeq \sqrt{2\tau} \left(1 - \frac{\nbtau{j}{j}}{4\tau}\right) \annihilatetb \label{tb-}
    \end{gather}
\end{subequations}
Here, $i \in$ A and $j \in$ B have been assumed with magnetic sublattice A consisting of up-spins and magnetic sublattice B consisting of down-spins. For $i \in$ B and $j \in$ A, the HP-Bosons $a,a^{\dagger}$ and $b,b^{\dagger}$ will be interchanged. The above mapping relations can be substituted in the following spin and orbital isospin Hamiltonians:
\begin{subequations}
    \begin{equation}
        \hat{\mathcal{H}}_{\sigma} = -\sum_{\langle\langle i,j \rangle\rangle}^{\text{mag}} \mathcal{J}_{1,ij}^{\sigma} \left[ \hat{\sigma}_{i}^{z} \hat{\sigma}_{j}^{z} + \frac{\lambda_{1}^{\sigma}}{2} \left(\hat{\sigma}_{i}^{+} \hat{\sigma}_{j}^{-} + \hat{\sigma}_{i}^{-} \hat{\sigma}_{j}^{+}\right) \right] -\sum_{\langle\langle\langle i,j \rangle\rangle\rangle}^{\text{mag}} \mathcal{J}_{2,ij}^{\sigma} \left[ \hat{\sigma}_{i}^{z} \hat{\sigma}_{j}^{z} + \frac{\lambda_{2}^{\sigma}}{2} \left(\hat{\sigma}_{i}^{+} \hat{\sigma}_{j}^{-} + \hat{\sigma}_{i}^{-} \hat{\sigma}_{j}^{+}\right) \right] - \mathcal{K}_{z}^{\sigma} \sum_{i}^{\text{mag}} (\hat{\sigma}_{i}^{z})^{2}
    \end{equation}
    \begin{equation}
        \hat{\mathcal{H}}_{\tau} = -\sum_{\langle\langle i,j \rangle\rangle}^{\text{mag}} \mathcal{J}_{1,ij}^{\tau} \left[ \hat{\tau}_{i}^{z}\hat{\tau}_{j}^{z} + \frac{\lambda_{1}^{\tau}}{2} \left(\hat{\tau}_{i}^{+}\hat{\tau}_{j}^{-} + \hat{\tau}_{i}^{-}\hat{\tau}_{j}^{+}\right) \right] -\sum_{\langle\langle\langle i,j \rangle\rangle\rangle}^{\text{mag}} \mathcal{J}_{2,ij}^{\tau} \left[ \hat{\tau}_{i}^{z}\hat{\tau}_{j}^{z} + \frac{\lambda_{2}^{\tau}}{2} \left(\hat{\tau}_{i}^{+}\hat{\tau}_{j}^{-} + \hat{\tau}_{i}^{-}\hat{\tau}_{j}^{+}\right) \right] - \mathcal{K}_{z}^{\tau} \sum_{i}^{\text{mag}} (\hat{\tau}_{i}^{z})^{2}
    \end{equation}
\end{subequations}
yielding terms of the following forms:
\begin{enumerate}[label=(\roman*)]
    \item
    $\begin{aligned}[t]
        \hat{\sigma}_{i \in \text{A}}^{z} \hat{\sigma}_{j \in \text{B}}^{z} = (\sigma - \nasigma{i}{i})(-\sigma + \nbsigma{j}{j}) = -\sigma^2 \Bigg[1 + \left(\frac{1}{\sigma}\right) \underbrace{(\nasigma{i}{i} + \nbsigma{j}{j})}_{\text{2-operator combinations}} - \left(\frac{1}{\sigma^2}\right) \underbrace{\nasigma{i}{i}\nbsigma{j}{j}}_{\substack{\text{4-operator} \\ \text{combination}}}\Bigg]
    \end{aligned}$

    \item
    $\begin{alignedat}[t]{2}
        &\frac{1}{2}\left(\hat{\sigma}_{i \in \text{A}}^{+} \hat{\sigma}_{j \in \text{B}}^{-} + \hat{\sigma}_{i \in \text{A}}^{-} \hat{\sigma}_{j \in \text{B}}^{+}\right)& \\
        &= \sigma^2 \Bigg\{ \left(\frac{1}{\sigma}\right) \underbrace{[\annihilatesa\annihilatesb + \createsa\createsb]}_{\text{2-operator combinations}}
        &- \left(\frac{1}{4\sigma^2}\right) \underbrace{[\nasigma{i}{i}\annihilatesa\annihilatesb + \annihilatesa\nbsigma{j}{j}\annihilatesb + \createsa\nasigma{i}{i}\createsb + \createsa\createsb\nbsigma{j}{j}]}_{\text{4-operator combinations}} \\
        &&+ \left(\frac{1}{16\sigma^4}\right) [\nasigma{i}{i}\annihilatesa\nbsigma{j}{j}\annihilatesb + \createsa\nasigma{i}{i}\createsb\nbsigma{j}{j}] \Bigg\}
    \end{alignedat}$

    \item 
    $\begin{aligned}[t]
        (\hat{\sigma}_{i \in \text{A}}^{z})^2 = (\sigma - \nasigma{i}{i})(\sigma - \nasigma{i}{i}) = \sigma^2 \Bigg[1 - \left(\frac{2}{\sigma}\right) \underbrace{\nasigma{i}{i}}_{\substack{\text{2-operator} \\ \text{combination}}} + \left(\frac{1}{\sigma^2}\right) \underbrace{\nasigma{i}{i}\nasigma{i}{i}}_{\substack{\text{4-operator} \\ \text{combination}}}\Bigg]
    \end{aligned}$

    \item 
    $\begin{aligned}[t]
        (\hat{\sigma}_{j \in \text{B}}^{z})^2 = (-\sigma + \nbsigma{j}{j})(-\sigma + \nbsigma{j}{j}) = \sigma^2 \Bigg[1 - \left(\frac{2}{\sigma}\right) \underbrace{\nbsigma{j}{j}}_{\substack{\text{2-operator} \\ \text{combination}}} + \left(\frac{1}{\sigma^2}\right) \underbrace{\nbsigma{j}{j}\nbsigma{j}{j}}_{\substack{\text{4-operator} \\ \text{combination}}}\Bigg]
    \end{aligned}$

    \item 
    $\begin{aligned}[t]
        \hat{\tau}_{i \in \text{A}}^{z} \hat{\tau}_{j \in \text{B}}^{z} = (\tau - \natau{i}{i})(-\tau + \nbtau{j}{j}) = -\tau^2 \Bigg[1 + \left(\frac{1}{\tau}\right) \underbrace{(\natau{i}{i} + \nbtau{j}{j})}_{\text{2-operator combinations}} - \left(\frac{1}{\tau^2}\right) \underbrace{\natau{i}{i}\nbtau{j}{j}}_{\substack{\text{4-operator} \\ \text{combination}}}\Bigg]
    \end{aligned}$

    \item 
    $\begin{alignedat}[t]{2}
        &\frac{1}{2}\left(\hat{\tau}_{i \in \text{A}}^{+} \hat{\tau}_{j \in \text{B}}^{-} + \hat{\tau}_{i \in \text{A}}^{-} \hat{\tau}_{j \in \text{B}}^{+}\right)& \\
        &= \tau^2 \Bigg\{ \left(\frac{1}{\tau}\right) \underbrace{[\annihilateta\annihilatetb + \createta\createtb]}_{\text{2-operator combinations}} 
        &- \left(\frac{1}{4\tau^2}\right) \underbrace{[\natau{i}{i}\annihilateta\annihilatetb + \annihilateta\nbtau{j}{j}\annihilatetb + \createta\natau{i}{i}\createtb + \createta\createtb\nbtau{j}{j}]}_{\text{4-operator combinations}} \\
        &&+ \left(\frac{1}{16\tau^4}\right) [\natau{i}{i}\annihilateta\nbtau{j}{j}\annihilatetb + \createta\natau{i}{i}\createtb\nbtau{j}{j}] \Bigg\}
    \end{alignedat}$

    \item 
    $\begin{aligned}[t]
        (\hat{\tau}_{i \in \text{A}}^{z})^2 = (\tau - \natau{i}{i})(\tau - \natau{i}{i}) = \tau^2 \Bigg[1 - \left(\frac{2}{\tau}\right) \underbrace{\natau{i}{i}}_{\substack{\text{2-operator} \\ \text{combination}}} + \left(\frac{1}{\tau^2}\right) \underbrace{\natau{i}{i}\natau{i}{i}}_{\substack{\text{4-operator} \\ \text{combination}}}\Bigg]
    \end{aligned}$

    \item 
    $\begin{aligned}[t]
        (\hat{\tau}_{j \in \text{B}}^{z})^2 = (-\tau + \nbtau{j}{j})(-\tau + \nbtau{j}{j}) = \tau^2 \Bigg[1 - \left(\frac{2}{\tau}\right) \underbrace{\nbtau{j}{j}}_{\substack{\text{2-operator} \\ \text{combination}}} + \left(\frac{1}{\tau^2}\right) \underbrace{\nbtau{j}{j}\nbtau{j}{j}}_{\substack{\text{4-operator} \\ \text{combination}}}\Bigg]
    \end{aligned}$
\end{enumerate}
Within linear SW and OW theory, the square root expansions in Eqs.~\eqref{sigA}, \eqref{sigB}, \eqref{tauA} and \eqref{tauB} are truncated to only unity such that $\hat{\mathcal{H}}_{\sigma}$ and $\hat{\mathcal{H}}_{\tau}$ exclusively contain up to terms linear in $\sfrac{1}{\sigma}$ and $\sfrac{1}{\tau}$ respectively. From the above expressions, it is clear that only terms containing 2-operator combinations (quadratic form) of HP operators in both spin and orbital sectors are of order $\mathcal{O}[\sfrac{1}{\sigma}]$ and $\mathcal{O}[\sfrac{1}{\tau}]$ respectively. The resulting $\hat{\mathcal{H}}_{\sigma}$ and $\hat{\mathcal{H}}_{\tau}$, after linear SW and OW expansions, will thus have a quadratic form. Now let us consider the full spin-orbital Hamiltonian $\hat{\mathcal{H}}$ (given by \textcolor{blue}{Eq.~\eqref{fullH} of main manuscript}), which may be rewritten as $\hat{\mathcal{H}} = \hat{\mathcal{H}}_{\sigma} + \hat{\mathcal{H}}_{\tau} + \hat{\mathcal{H}}_{\sigma\tau}$, where $\hat{\mathcal{H}}_{\sigma\tau}$ is the interaction between spins and orbital isospins given by:
\begin{multline}
    \hat{\mathcal{H}}_{\sigma\tau} = -\sum_{\langle\langle i,j \rangle\rangle}^{\text{mag}} \mathcal{Q}_{1,ij} \left[ \hat{\sigma}_{i}^{z} \hat{\sigma}_{j}^{z} + \frac{\lambda_{1}^{\sigma}}{2} \left(\hat{\sigma}_{i}^{+} \hat{\sigma}_{j}^{-} + \hat{\sigma}_{i}^{-} \hat{\sigma}_{j}^{+}\right) \right]\!\! \left[ \hat{\tau}_{i}^{z}\hat{\tau}_{j}^{z} + \frac{\lambda_{1}^{\tau}}{2} \left(\hat{\tau}_{i}^{+}\hat{\tau}_{j}^{-} + \hat{\tau}_{i}^{-}\hat{\tau}_{j}^{+}\right) \right] \\
    -\sum_{\langle\langle\langle i,j \rangle\rangle\rangle}^{\text{mag}} \mathcal{Q}_{2,ij} \left[ \hat{\sigma}_{i}^{z} \hat{\sigma}_{j}^{z} + \frac{\lambda_{2}^{\sigma}}{2} \left(\hat{\sigma}_{i}^{+} \hat{\sigma}_{j}^{-} + \hat{\sigma}_{i}^{-} \hat{\sigma}_{j}^{+}\right) \right]\!\! \left[ \hat{\tau}_{i}^{z}\hat{\tau}_{j}^{z} + \frac{\lambda_{2}^{\tau}}{2} \left(\hat{\tau}_{i}^{+}\hat{\tau}_{j}^{-} + \hat{\tau}_{i}^{-}\hat{\tau}_{j}^{+}\right) \right]
\end{multline}
Because $\hat{\mathcal{H}}_{\sigma\tau}$ contains product of expressions (i)+(ii) and (v)+(vi) listed above, there will be cross-terms between HP operators of spin and orbital sectors as shown below:
\begin{enumerate}[resume,label=(\roman*)]
    \item 
    $\begin{aligned}[t]
        \hat{\sigma}_{i \in A}^{z}\hat{\sigma}_{j \in B}^{z}\hat{\tau}_{i \in A}^{z}\hat{\tau}_{j \in B}^{z}
    \end{aligned}$ \\
    $\begin{multlined}[t]
        = \sigma^2\tau^2 \Bigg[ 1 + \left(\frac{1}{\sigma}\right) \underbrace{(\nasigma{i}{i} + \nbsigma{j}{j})}_{\text{2-operator combinations}} + \left(\frac{1}{\sigma^2}\right) \underbrace{\nasigma{i}{i}\nbsigma{j}{j}}_{\substack{\text{4-operator} \\ \text{combination}}}- \left(\frac{1}{\tau}\right) \underbrace{(\natau{i}{i} + \nbtau{j}{j})}_{\text{2-operator combinations}} + \left(\frac{1}{\tau^2}\right) \underbrace{\natau{i}{i}\nbtau{j}{j}}_{\substack{\text{4-operator} \\ \text{combination}}}  \\
        + \left(\frac{1}{\sigma\tau}\right) \underbrace{(\nasigma{i}{i}\natau{i}{i} + \nasigma{i}{i}\nbtau{j}{j} + \nbsigma{j}{j}\natau{i}{i} + \nbsigma{j}{j}\nbtau{j}{j})}_{\text{4-operator combinations (cross-terms)}} \\
        - \left(\frac{1}{\sigma\tau^2}\right) (\nasigma{i}{i}\natau{i}{i}\nbtau{j}{j} + \nbsigma{j}{j}\natau{i}{i}\nbtau{j}{j}) - \left(\frac{1}{\sigma^2\tau}\right) (\nasigma{i}{i}\nbsigma{j}{j}\natau{i}{i} + \nasigma{i}{i}\nbsigma{j}{j}\nbtau{j}{j}) \\
        + \left(\frac{1}{\sigma^2\tau^2}\right) \nasigma{i}{i}\nbsigma{j}{j}\natau{i}{i}\nbtau{j}{j} \Bigg]
    \end{multlined}$

    \item 
    $\begin{aligned}[t]
        \frac{1}{2} \left( \hat{\sigma}_{i \in A}^{+}\hat{\sigma}_{j \in B}^{-} + \hat{\sigma}_{i \in A}^{-}\hat{\sigma}_{j \in B}^{+} \right) \hat{\tau}_{i \in A}^{z}\hat{\tau}_{j \in B}^{z}
    \end{aligned}$ \\
    $\begin{multlined}[t]
        = \sigma^2\tau^2 \Bigg[ -\left(\frac{1}{\sigma}\right) \underbrace{(\annihilatesa\annihilatesb + \createsa\createsb)}_{\text{2-operator combinations}} + \left(\frac{1}{4\sigma^2}\right) \underbrace{(\nasigma{i}{i}\annihilatesa\annihilatesb + \annihilatesa\nbsigma{j}{j}\annihilatesb + \createsa\nasigma{i}{i}\createsb + \createsa\createsb\nbsigma{j}{j})}_{\text{4-operator combinations}} \\
        + \left(\frac{1}{\sigma\tau}\right) \underbrace{(\annihilatesa\annihilatesb\natau{i}{i} + \annihilatesa\annihilatesb\nbtau{j}{j} + \createsa\createsb\natau{i}{i} + \createsa\createsb\nbtau{j}{j})}_{\text{4-operator combinations (cross-terms)}} \\
        - \left(\frac{1}{4\sigma^2\tau}\right) (\annihilatesa\nbsigma{j}{j}\annihilatesb\natau{i}{i} + \nasigma{i}{i}\annihilatesa\annihilatesb\natau{i}{i} + \createsa\nasigma{i}{i}\createsb\natau{i}{i} + \createsa\createsb\nbsigma{j}{j}\natau{i}{i} \\
        + \annihilatesa\nbsigma{j}{j}\annihilatesb\nbtau{j}{j} + \nasigma{i}{i}\annihilatesa\annihilatesb\nbtau{j}{j} + \createsa\nasigma{i}{i}\createsb\nbtau{j}{j} + \createsa\createsb\nbsigma{j}{j}\nbtau{j}{j}) \\
        - \left(\frac{1}{\sigma\tau^2}\right) (\annihilatesa\annihilatesb\natau{i}{i}\nbtau{j}{j} + \createsa\createsb\natau{i}{i}\nbtau{j}{j}) \\
        + \left(\frac{1}{4\sigma^2\tau^2}\right) (\annihilatesa\nbsigma{j}{j}\annihilatesb + \nasigma{i}{i}\annihilatesa\annihilatesb + \createsa\nasigma{i}{i}\createsb + \createsa\createsb\nbsigma{j}{j})\natau{i}{i}\nbtau{j}{j} \Bigg]
    \end{multlined}$

    \item 
    $\begin{aligned}[t]
        \frac{1}{2} \hat{\sigma}_{i \in B}^{z}\hat{\sigma}_{j \in A}^{z} \left( \hat{\tau}_{i \in B}^{+}\hat{\tau}_{j \in A}^{-} + \hat{\tau}_{i \in B}^{-}\hat{\tau}_{j \in A}^{+} \right)
    \end{aligned}$ \\
    $\begin{multlined}[t]
        = \sigma^2\tau^2 \Bigg[ - \left(\frac{1}{\tau}\right) \underbrace{(\createtb[i]\createta[j] + \annihilatetb[i]\annihilateta[j])}_{\text{2-operator combinations}} + \left(\frac{1}{4\tau^2}\right) \underbrace{(\createtb[i]\nbtau{i}{i}\createta[j] + \createtb[i]\createta[j]\natau{j}{j} + \nbtau{i}{i}\annihilatetb[i]\annihilateta[j] + \annihilatetb[i]\natau{j}{j}\annihilateta[j])}_{\text{4-operator combinations}} \\
        + \left(\frac{1}{\sigma\tau}\right) \underbrace{(\nbsigma{i}{i}\createtb[i]\createta[j] + \nasigma{j}{j}\createtb[i]\createta[j] + \nbsigma{i}{i}\annihilatetb[i]\annihilateta[j] + \nasigma{j}{j}\annihilatetb[i]\annihilateta[j])}_{\text{4-operator combinations (cross-terms)}} \\
        - \left(\frac{1}{4\sigma\tau^2}\right) (\nasigma{i}{i}\natau{i}{i}\annihilateta\annihilatetb + \nasigma{i}{i}\annihilateta\nbtau{j}{j}\annihilatetb + \nasigma{i}{i}\createta\natau{i}{i}\createtb + \nasigma{i}{i}\createta\createtb\nbtau{j}{j} \\
        + \nbsigma{j}{j}\natau{i}{i}\annihilateta\annihilatetb + \nbsigma{j}{j}\annihilateta\nbtau{j}{j}\annihilatetb + \nbsigma{j}{j}\createta\natau{i}{i}\createtb + \nbsigma{j}{j}\createta\createtb\nbtau{j}{j}) \\
        - \left(\frac{1}{\sigma^2\tau}\right) (\nasigma{i}{i}\nbsigma{j}{j}\annihilateta\annihilatetb + \nasigma{i}{i}\nbsigma{j}{j}\createta\createtb) \\
        + \left(\frac{1}{4\sigma^2\tau^2}\right) \nasigma{i}{i}\nbsigma{j}{j}(\natau{i}{i}\annihilateta\annihilatetb + \annihilateta\nbtau{j}{j}\annihilatetb + \createta\natau{i}{i}\createtb + \createta\createtb\nbtau{j}{j}) \Bigg]
    \end{multlined}$
    
    \item 
    $\begin{aligned}[t]
        \frac{1}{4} \left( \hat{\sigma}_{i \in A}^{+}\hat{\sigma}_{j \in B}^{-} + \hat{\sigma}_{i \in A}^{-}\hat{\sigma}_{j \in B}^{+} \right) \left( \hat{\tau}_{i \in A}^{+}\hat{\tau}_{j \in B}^{-} + \hat{\tau}_{i \in A}^{-}\hat{\tau}_{j \in B}^{+} \right)
    \end{aligned}$ \\
    $\begin{multlined}[t]
        = \sigma^2\tau^2 \Bigg[ \left(\frac{1}{\sigma\tau}\right) \underbrace{(\annihilatesa\annihilatesb\annihilateta\annihilatetb + \annihilatesa\annihilatesb\createta\createtb + \createsa\createsb\annihilateta\annihilatetb + \createsa\createsb\createta\createtb)}_{\text{4-operator combinations (cross-terms)}} \\
        - \left(\frac{1}{4\sigma^2\tau}\right) (\nasigma{i}{i}\annihilatesa\annihilatesb\annihilateta\annihilatetb + \annihilatesa\nbsigma{j}{j}\annihilatesb\annihilateta\annihilatetb + \createsa\nasigma{i}{i}\createsb\annihilateta\annihilatetb + \createsa\createsb\nbsigma{j}{j}\annihilateta\annihilatetb \\
        + \nasigma{i}{i}\annihilatesa\annihilatesb\createta\createtb + \annihilatesa\nbsigma{j}{j}\annihilatesb\createta\createtb + \createsa\nasigma{i}{i}\createsb\createta\createtb + \createsa\createsb\nbsigma{j}{j}\createta\createtb) \\
        - \left(\frac{1}{4\sigma\tau^2}\right) (\annihilatesa\annihilatesb\natau{i}{i}\annihilateta\annihilatetb + \annihilatesa\annihilatesb\annihilateta\nbtau{j}{j}\annihilatetb + \annihilatesa\annihilatesb\createta\natau{i}{i}\createtb + \annihilatesa\annihilatesb\createta\createtb\nbtau{j}{j} \\
        + \createsa\createsb\natau{i}{i}\annihilateta\annihilatetb + \createsa\createsb\annihilateta\nbtau{j}{j}\annihilatetb + \createsa\createsb\createta\natau{i}{i}\createtb + \createsa\createsb\createta\createtb\nbtau{j}{j}) \\
        + \left(\frac{1}{16\sigma^2\tau^2}\right) (\nasigma{i}{i}\annihilatesa\annihilatesb + \annihilatesa\nbsigma{j}{j}\annihilatesb + \createsa\nasigma{i}{i}\createsb + \createsa\createsb\nbsigma{j}{j}) \\
        (\natau{i}{i}\annihilateta\annihilatetb + \annihilateta\nbtau{j}{j}\annihilatetb + \createta\natau{i}{i}\createtb + \createta\createtb\nbtau{j}{j}) \Bigg]
    \end{multlined}$
\end{enumerate}
It is now evident that the cross-terms will at least contain 4-operator combinations (biquadratic form) which are second order in spins and isospins, i.e., $\mathcal{O}[\sfrac{1}{\sigma\tau}]$.
As a result, if one performs linear SW and OW expansion and truncates $\hat{\mathcal{H}}$ at $\mathcal{O}[\sfrac{1}{\sigma}]$ and $\mathcal{O}[\sfrac{1}{\tau}]$ (or equivalently the HP operator square root expansions to unity), the spin and orbital sectors completely decouple. Thus to include interactions between spins and orbital isospins, one needs to at least consider biquadratic form of HP operators with orders $\mathcal{O}[\sfrac{1}{\sigma^2}]$, $\mathcal{O}[\sfrac{1}{\tau^2}]$ and $\mathcal{O}[\sfrac{1}{\sigma\tau}]$.

\section{Conservation of Total Spin and Orbital Isospin}

In Quantum Mechanics, quantities that commute with the Hamiltonian of a system are conserved. To conclude that the total spin ($\hat{\sigma}_{\text{tot}}^{z}$) and orbital isospin ($\hat{\tau}_{\text{tot}}^{z}$) is conserved, one needs to show $[\hat{\mathcal{H}},\hat{\sigma}_{\text{tot}}^{z}] = 0$ and $[\hat{\mathcal{H}},\hat{\tau}_{\text{tot}}^{z}] = 0$, where $\hat{\mathcal{H}}$ is the full Hamiltonian in \textcolor{blue}{Eq.~\eqref{fullH} of main manuscript}. The spin and orbital isospin operators follow $\mathfrak{su}(2)$ algebra obeying the canonical commutation relations:
\begin{equation}
    \big[\,\hat{\sigma}_{i}^{\mu},\hat{\sigma}_{j}^{\nu}\,\big] = \mathrm{i}\epsilon_{\mu\nu\vartheta}\,\hat{\sigma}_{i}^{\vartheta}\delta_{ij} \;;\; \big[\,\hat{\tau}_{i}^{\mu},\hat{\tau}_{j}^{\nu}\,\big] = \mathrm{i}\epsilon_{\mu\nu\vartheta}\,\hat{\tau}_{i}^{\vartheta}\delta_{ij}
    \label{comm}
\end{equation}
where, $i,j$ are site indices, $\delta_{ij}$ is Kronecker delta and $\epsilon_{\mu\nu\vartheta}$ is the Levi-Civita symbol and $\mu,\nu,\vartheta = \{x,y,z\}$. Further, $\big[\,\hat{\sigma}_{i}^{\mu},\hat{\tau}_{j}^{\nu}\,\big] = 0$ since spin and orbital isospin operators belong to distinct vector spaces. The spin raising and lowering operators are defined by $\hat{\sigma}_{i}^{+} = \hat{\sigma}_{i}^{x} + \mathrm{i}\hat{\sigma}_{i}^{y}$ and $\hat{\sigma}_{i}^{-} = \hat{\sigma}_{i}^{x} - \mathrm{i}\hat{\sigma}_{i}^{y}$ respectively. Similarly, the orbital isospin raising and lowering operators are defined by $\hat{\tau}_{i}^{+} = \hat{\tau}_{i}^{x} + \mathrm{i}\hat{\tau}_{i}^{y}$ and $\hat{\tau}_{i}^{-} = \hat{\tau}_{i}^{x} - \mathrm{i}\hat{\tau}_{i}^{y}$ respectively. Using these relations and Eq.~\eqref{comm}, the following commutators are obtained:
\begin{subequations}
    \label{raise_lower_comm}
    \begin{gather}
        \big[\,\hat{\sigma}_{i}^{+},\hat{\sigma}_{j}^{z}\,\big] = \big[\,(\hat{\sigma}_{i}^{x} + \mathrm{i}\hat{\sigma}_{i}^{y}),\hat{\sigma}_{j}^{z}\,\big] = \big[\,\hat{\sigma}_{i}^{x},\hat{\sigma}_{j}^{z}\,\big] + \mathrm{i} \big[\,\hat{\sigma}_{i}^{y},\hat{\sigma}_{j}^{z}\,\big] = (-\mathrm{i} \hat{\sigma}_{i}^{y} - \hat{\sigma}_{i}^{x}) \delta_{ij} = -\hat{\sigma}_{i}^{+}\delta_{ij} \\
        \big[\,\hat{\sigma}_{i}^{-},\hat{\sigma}_{j}^{z}\,\big] = \big[\,(\hat{\sigma}_{i}^{x} - \mathrm{i}\hat{\sigma}_{i}^{y}),\hat{\sigma}_{j}^{z}\,\big] = \big[\,\hat{\sigma}_{i}^{x},\hat{\sigma}_{j}^{z}\,\big] - \mathrm{i} \big[\,\hat{\sigma}_{i}^{y},\hat{\sigma}_{j}^{z}\,\big] = (-\mathrm{i} \hat{\sigma}_{i}^{y} + \hat{\sigma}_{i}^{x}) \delta_{ij} = \hat{\sigma}_{i}^{-}\delta_{ij} \\
        \big[\,\hat{\tau}_{i}^{+},\hat{\tau}_{j}^{z}\,\big] = \big[\,(\hat{\tau}_{i}^{x} + \mathrm{i}\hat{\tau}_{i}^{y}),\hat{\tau}_{j}^{z}\,\big] = \big[\,\hat{\tau}_{i}^{x},\hat{\tau}_{j}^{z}\,\big] + \mathrm{i} \big[\,\hat{\tau}_{i}^{y},\hat{\tau}_{j}^{z}\,\big] = (-\mathrm{i} \hat{\tau}_{i}^{y} - \hat{\tau}_{i}^{x}) \delta_{ij} = -\hat{\tau}_{i}^{+}\delta_{ij} \\
        \big[\,\hat{\tau}_{i}^{-},\hat{\tau}_{j}^{z}\,\big] = \big[\,(\hat{\tau}_{i}^{x} - \mathrm{i}\hat{\tau}_{i}^{y}),\hat{\tau}_{j}^{z}\,\big] = \big[\,\hat{\tau}_{i}^{x},\hat{\tau}_{j}^{z}\,\big] - \mathrm{i} \big[\,\hat{\tau}_{i}^{y},\hat{\tau}_{j}^{z}\,\big] = (-\mathrm{i} \hat{\tau}_{i}^{y} + \hat{\tau}_{i}^{x}) \delta_{ij} = \hat{\tau}_{i}^{-}\delta_{ij}
    \end{gather}
\end{subequations}
The total spin and orbital isospin is given by $\hat{\sigma}_{\text{tot}}^{z} = \sum_{i}^{\text{mag}} \hat{\sigma}_{i}^{z}$ and $\hat{\tau}_{\text{tot}}^{z} = \sum_{i}^{\text{mag}} \hat{\tau}_{i}^{z}$ respectively where $i$ runs over both magnetic sublattices A and B. We now consider the following commutators:
\begin{enumerate}[label=(\roman*)]
    \item 
    $\begin{aligned}[t]
        \left[\,\hat{\sigma}_{i}^{z} \hat{\sigma}_{j}^{z}\,,\,\hat{\sigma}_{\text{tot}}^{z}\,\right] = \sum_{l}^{\text{mag}}\left[\,\hat{\sigma}_{i}^{z} \hat{\sigma}_{j}^{z}\,,\,\hat{\sigma}_{l}^{z}\,\right] = 0
    \end{aligned}$
    
    \item 
    $\begin{aligned}[t]
        &\left[\,(\hat{\sigma}_{i}^{+} \hat{\sigma}_{j}^{-} + \hat{\sigma}_{i}^{-} \hat{\sigma}_{j}^{+})\,,\,\hat{\sigma}_{\text{tot}}^{z}\,\right] = \sum_{l}^{\text{mag}}\left[\,(\hat{\sigma}_{i}^{+} \hat{\sigma}_{j}^{-} + \hat{\sigma}_{i}^{-} \hat{\sigma}_{j}^{+})\,,\,\hat{\sigma}_{l}^{z}\,\right] = \sum_{l}^{\text{mag}}\Big\{\left[\,\hat{\sigma}_{i}^{+} \hat{\sigma}_{j}^{-}\,,\,\hat{\sigma}_{l}^{z}\,\right] + \left[\,\hat{\sigma}_{i}^{-} \hat{\sigma}_{j}^{+}\;,\;\hat{\sigma}_{l}^{z}\,\right]\Big\} \\
        &= \sum_{l}^{\text{mag}}\Big\{\hat{\sigma}_{i}^{+}\big[\,\hat{\sigma}_{j}^{-},\hat{\sigma}_{l}^{z}\,\big] + \big[\,\hat{\sigma}_{i}^{+} ,\hat{\sigma}_{l}^{z}\,\big]\hat{\sigma}_{j}^{-} + \hat{\sigma}_{i}^{-}\big[\, \hat{\sigma}_{j}^{+},\hat{\sigma}_{l}^{z}\,\big] + \big[\,\hat{\sigma}_{i}^{-} ,\hat{\sigma}_{l}^{z}\,\big]\hat{\sigma}_{j}^{+}\Big\} \\
        &= \sum_{l}^{\text{mag}}\Big\{\hat{\sigma}_{i}^{+}\hat{\sigma}_{j}^{-}\delta_{jl} - \hat{\sigma}_{i}^{+}\delta_{il}\hat{\sigma}_{j}^{-} - \hat{\sigma}_{i}^{-}\hat{\sigma}_{j}^{+}\delta_{jl} + \hat{\sigma}_{i}^{-}\delta_{il}\hat{\sigma}_{j}^{+}\Big\} = \hat{\sigma}_{i}^{+}\hat{\sigma}_{j}^{-} - \hat{\sigma}_{i}^{+}\hat{\sigma}_{j}^{-} - \hat{\sigma}_{i}^{-}\hat{\sigma}_{j}^{+} + \hat{\sigma}_{i}^{-}\hat{\sigma}_{j}^{+} = 0
    \end{aligned}$

    \item 
    $\begin{aligned}[t]
        \left[\,\hat{\tau}_{i}^{z} \hat{\tau}_{j}^{z}\,,\,\hat{\tau}_{\text{tot}}^{z}\,\right] = \sum_{l}^{\text{mag}}\left[\,\hat{\tau}_{i}^{z} \hat{\tau}_{j}^{z}\,,\,\hat{\tau}_{l}^{z}\,\right] = 0
    \end{aligned}$

    \item 
    $\begin{aligned}[t]
        &\left[\,(\hat{\tau}_{i}^{+} \hat{\tau}_{j}^{-} + \hat{\tau}_{i}^{-} \hat{\tau}_{j}^{+})\,,\,\hat{\tau}_{\text{tot}}^{z}\,\right] = \sum_{l}^{\text{mag}}\left[\,(\hat{\tau}_{i}^{+} \hat{\tau}_{j}^{-} + \hat{\tau}_{i}^{-} \hat{\tau}_{j}^{+})\,,\,\hat{\tau}_{l}^{z}\,\right] = \sum_{l}^{\text{mag}}\Big\{\left[\,\hat{\tau}_{i}^{+} \hat{\tau}_{j}^{-}\,,\,\hat{\tau}_{l}^{z}\,\right] + \left[\,\hat{\tau}_{i}^{-} \hat{\tau}_{j}^{+}\;,\;\hat{\tau}_{l}^{z}\,\right]\Big\} \\
        &= \sum_{l}^{\text{mag}}\Big\{\hat{\tau}_{i}^{+}\big[\,\hat{\tau}_{j}^{-},\hat{\tau}_{l}^{z}\,\big] + \big[\,\hat{\tau}_{i}^{+} ,\hat{\tau}_{l}^{z}\,\big]\hat{\tau}_{j}^{-} + \hat{\tau}_{i}^{-}\big[\, \hat{\tau}_{j}^{+},\hat{\tau}_{l}^{z}\,\big] + \big[\,\hat{\tau}_{i}^{-} ,\hat{\tau}_{l}^{z}\,\big]\hat{\tau}_{j}^{+}\Big\} \\
        &= \sum_{l}^{\text{mag}}\Big\{\hat{\tau}_{i}^{+}\hat{\tau}_{j}^{-}\delta_{jl} - \hat{\tau}_{i}^{+}\delta_{il}\hat{\tau}_{j}^{-} - \hat{\tau}_{i}^{-}\hat{\tau}_{j}^{+}\delta_{jl} + \hat{\tau}_{i}^{-}\delta_{il}\hat{\tau}_{j}^{+}\Big\} = \hat{\tau}_{i}^{+}\hat{\tau}_{j}^{-} - \hat{\tau}_{i}^{+}\hat{\tau}_{j}^{-} - \hat{\tau}_{i}^{-}\hat{\tau}_{j}^{+} + \hat{\tau}_{i}^{-}\hat{\tau}_{j}^{+} = 0
    \end{aligned}$
\end{enumerate}
where Eq.~\eqref{raise_lower_comm} has been used.
Since $\hat{\mathcal{H}}$ contains terms of the form $\hat{\sigma}_{i}^{z}\hat{\sigma}_{j}^{z}, (\hat{\sigma}_{i}^{+}\hat{\sigma}_{j}^{-} + \hat{\sigma}_{i}^{-}\hat{\sigma}_{j}^{+}), \hat{\tau}_{i}^{z}\hat{\tau}_{j}^{z}, (\hat{\tau}_{i}^{+}\hat{\tau}_{j}^{-} + \hat{\tau}_{i}^{-}\hat{\tau}_{j}^{+})$, we can conclude using the listed commutators that $[\hat{\mathcal{H}},\hat{\sigma}_{\text{tot}}^{z}] = 0$ and $[\hat{\mathcal{H}},\hat{\tau}_{\text{tot}}^{z}] = 0$. Hence, the extended Heisenberg Hamiltonian in \textcolor{blue}{Eq.~\eqref{fullH} of main manuscript} conserves total spin and orbital isospin globally.

\section{Expectation Values of Allowed HP Operators with Quadratic Form} \label{expct_val}

As mentioned in \textcolor{blue}{Sec.~\ref{sectionHP} of main manuscript}, only specific combinations of quadratic form HP operators will not change total spin and orbital isospin, i.e., $\hat{\sigma}_{\text{tot}}^{z}$ and $\hat{\tau}_{\text{tot}}^{z}$ remains conserved. Restricting to HP operators with quadratic form, the allowed combinations are defined below. Here, we consider the total number of magnetic sites to be $N_m$ such that A and B sublattices each consist of $\frac{N_m}{2}$ sites.
\begin{enumerate}
    \item Number density: \\
    \begin{enumerate}[label=(\roman*), leftmargin=4pt, topsep=0pt]
        \item $\varrho_{i,\sigma}^{\text{A}} = \langle\nasigma{i}{i}\rangle = \Big\langle \sqrt{\frac{2}{N_m}} \sum_{\mathbf{k}} e^{+\mathrm{i}\mathbf{k}\cdot\mathbf{r}_{i}}\, \fcreatesa[\mathbf{k}] \sqrt{\frac{2}{N_m}} \sum_{\mathbf{k}'} e^{-\mathrm{i}\mathbf{k}'\cdot\mathbf{r}_{i}}\, \fannihilatesa[\mathbf{k}'] \Big\rangle = \frac{2}{N_m} \sum_{\mathbf{k},\mathbf{k}'} e^{\mathrm{i}(\mathbf{k} - \mathbf{k}')\cdot\mathbf{r}_{i}} \langle\fcreatesa[\mathbf{k}]\fannihilatesa[\mathbf{k}']\rangle = \frac{2}{N_m} \sum_{\mathbf{k},\mathbf{k}'} e^{\mathrm{i}(\mathbf{k} - \mathbf{k}')\cdot\mathbf{r}_{i}} \varrho_{\mathbf{k},\sigma}^{\text{A}} \,\delta_{\mathbf{k},\mathbf{k}'} \Rightarrow \varrho_{i,\sigma}^{\text{A}} = \frac{2}{N_m} \sum_{\mathbf{k}} \varrho_{\mathbf{k},\sigma}^{\text{A}} \Rightarrow \varrho_{\sigma}^{\text{A}} = \frac{2}{N_m} \sum_{\mathbf{k}} \varrho_{\mathbf{k},\sigma}^{\text{A}}$
        \item $\varrho_{i,\sigma}^{\text{B}} = \langle\nbsigma{i}{i}\rangle = \Big\langle \sqrt{\frac{2}{N_m}} \sum_{\mathbf{k}} e^{+\mathrm{i}\mathbf{k}\cdot\mathbf{r}_{i}}\, \fcreatesb[\mathbf{k}] \sqrt{\frac{2}{N_m}} \sum_{\mathbf{k}'} e^{-\mathrm{i}\mathbf{k}'\cdot\mathbf{r}_{i}}\, \fannihilatesb[\mathbf{k}'] \Big\rangle = \frac{2}{N_m} \sum_{\mathbf{k},\mathbf{k}'} e^{\mathrm{i}(\mathbf{k} - \mathbf{k}')\cdot\mathbf{r}_{i}} \langle\fcreatesb[\mathbf{k}]\fannihilatesb[\mathbf{k}']\rangle = \frac{2}{N_m} \sum_{\mathbf{k},\mathbf{k}'} e^{\mathrm{i}(\mathbf{k} - \mathbf{k}')\cdot\mathbf{r}_{i}} \varrho_{\mathbf{k},\sigma}^{\text{B}} \,\delta_{\mathbf{k},\mathbf{k}'} \Rightarrow \varrho_{i,\sigma}^{\text{B}} = \frac{2}{N_m} \sum_{\mathbf{k}} \varrho_{\mathbf{k},\sigma}^{\text{B}} \Rightarrow \varrho_{\sigma}^{\text{B}} = \frac{2}{N_m} \sum_{\mathbf{k}} \varrho_{\mathbf{k},\sigma}^{\text{B}}$
    \end{enumerate}
    \noindent Physically, for a N\'{e}el ordered state, number density of sublattices A and B must be same. Thus, $\varrho_{\mathbf{k},\sigma}^{\text{A}} = \varrho_{\mathbf{k},\sigma}^{\text{B}}$ $\Rightarrow \varrho_{\sigma}^{\text{A}} = \varrho_{\sigma}^{\text{B}} = \varrho_{\sigma}$. The same is true for both $\langle\natau{i}{i}\rangle$ and $\langle\nbtau{i}{i}\rangle$. Together one obtains:
    \begin{subequations}
        \begin{gather}
            \boxed{\varrho_{\sigma} = \frac{2}{N_m} \sum_{\mathbf{k}} \varrho_{\mathbf{k},\sigma}} \\
            \boxed{\varrho_{\tau} = \frac{2}{N_m} \sum_{\mathbf{k}} \varrho_{\mathbf{k},\tau}}
        \end{gather}
    \end{subequations}
    By definition, $\varrho_{\sigma}, \varrho_{\tau} \in \mathbb{R}$.

    \item Pairing amplitude: \\
    We define $\mathbf{r}_{j} = \mathbf{r}_{i} + \boldsymbol{\xi}$ where $\boldsymbol{\xi}$ is the next-nearest (NN) neighbour coordination vector.
    \begin{subequations}
        \begin{enumerate}[label=(\roman*), leftmargin=4pt, topsep=0pt]
            \item $\Delta_{ij,\sigma} = \langle\annihilatesa\annihilatesb\rangle = \Big\langle \sqrt{\frac{2}{N_m}} \sum_{\mathbf{k}} e^{-\mathrm{i}\mathbf{k}\cdot\mathbf{r}_{i}}\, \fannihilatesa[\mathbf{k}] \sqrt{\frac{2}{N_m}} \sum_{\mathbf{k}'} e^{-\mathrm{i}\mathbf{k}'\cdot\mathbf{r}_{j}}\, \fannihilatesb[\mathbf{k}'] \Big\rangle = \frac{2}{N_m} \sum_{\mathbf{k},\mathbf{k}'} e^{-\mathrm{i}(\mathbf{k} + \mathbf{k}')\cdot\mathbf{r}_{i}} e^{-\mathrm{i}\mathbf{k}' \cdot \boldsymbol{\xi}} \langle\fannihilatesa[\mathbf{k}]\fannihilatesb[\mathbf{k}']\rangle = \frac{2}{N_m} \sum_{\mathbf{k},\mathbf{k}'} e^{-\mathrm{i}(\mathbf{k} + \mathbf{k}')\cdot\mathbf{r}_{i}} e^{-\mathrm{i}\mathbf{k}' \cdot \boldsymbol{\xi}} \Delta_{\mathbf{k},\sigma} \,\delta_{\mathbf{k},-\mathbf{k}'} \Rightarrow \Delta_{ij,\sigma} = \frac{2}{N_m} \sum_{\mathbf{k}} e^{+\mathrm{i}\mathbf{k} \cdot \boldsymbol{\xi}} \Delta_{\mathbf{k},\sigma} = \Delta_{\boldsymbol{\xi},\sigma}$
            \begin{equation}
                \implies \boxed{\Delta_{\boldsymbol{\xi},\sigma} = \frac{2}{N_m} \sum_{\mathbf{k}} e^{+\mathrm{i}\mathbf{k} \cdot \boldsymbol{\xi}} \Delta_{\mathbf{k},\sigma}}
            \end{equation}
            Finally, averaging over all nearest neighbours $\mathit{n}$:
            \begin{equation}
                \begin{gathered}
                    \Delta_{\sigma} = \frac{1}{\mathit{n}} \sum_{\boldsymbol{\xi}} \Delta_{\boldsymbol{\xi},\sigma} = \frac{2}{N_m} \sum_{\mathbf{k}} \bigg(\frac{1}{\mathit{n}}\sum_{\boldsymbol{\xi}} e^{+\mathrm{i}\mathbf{k} \cdot \boldsymbol{\xi}}\bigg) \Delta_{\mathbf{k},\sigma} \\
                    \implies \boxed{\Delta_{\sigma} = \frac{2}{N_m} \sum_{\mathbf{k}} \gamma_{\mathbf{k}} \Delta_{\mathbf{k},\sigma}}
                \end{gathered}
            \end{equation}
            Similarly, for $\langle\annihilateta\annihilatetb\rangle$:
            \begin{gather}
                \boxed{\Delta_{\boldsymbol{\xi},\tau} = \frac{2}{N_m} \sum_{\mathbf{k}} e^{+\mathrm{i}\mathbf{k} \cdot \boldsymbol{\xi}} \Delta_{\mathbf{k},\tau}} \\
                \boxed{\Delta_{\tau} = \frac{2}{N_m} \sum_{\mathbf{k}} \gamma_{\mathbf{k}} \Delta_{\mathbf{k},\tau}}
            \end{gather}

            \item $\Delta_{ij,\sigma}^{*} = \langle\createsa\createsb\rangle = \Big\langle \sqrt{\frac{2}{N_m}} \sum_{\mathbf{k}} e^{+\mathrm{i}\mathbf{k}\cdot\mathbf{r}_{i}}\, \fcreatesa[\mathbf{k}] \sqrt{\frac{2}{N_m}} \sum_{\mathbf{k}'} e^{+\mathrm{i}\mathbf{k}'\cdot\mathbf{r}_{j}}\, \fcreatesb[\mathbf{k}'] \Big\rangle = \frac{2}{N_m} \sum_{\mathbf{k},\mathbf{k}'} e^{+\mathrm{i}(\mathbf{k} + \mathbf{k}')\cdot\mathbf{r}_{i}} e^{+\mathrm{i}\mathbf{k}' \cdot \boldsymbol{\xi}} \langle\fcreatesa[\mathbf{k}]\fcreatesb[\mathbf{k}']\rangle = \frac{2}{N_m} \sum_{\mathbf{k},\mathbf{k}'} e^{+\mathrm{i}(\mathbf{k} + \mathbf{k}')\cdot\mathbf{r}_{i}} e^{+\mathrm{i}\mathbf{k}' \cdot \boldsymbol{\xi}} \Delta_{\mathbf{k},\sigma}^{*} \,\delta_{\mathbf{k},-\mathbf{k}'} \Rightarrow \Delta_{ij,\sigma}^{*} = \frac{2}{N_m} \sum_{\mathbf{k}} e^{-\mathrm{i}\mathbf{k} \cdot \boldsymbol{\xi}} \Delta_{\mathbf{k},\sigma}^{*} = \Delta_{\boldsymbol{\xi},\sigma}^{*}$
            \begin{equation}
                \implies \boxed{\Delta_{\boldsymbol{\xi},\sigma}^{*} = \frac{2}{N_m} \sum_{\mathbf{k}} e^{-\mathrm{i}\mathbf{k} \cdot \boldsymbol{\xi}} \Delta_{\mathbf{k},\sigma}^{*}}
            \end{equation}
            Finally, averaging over all nearest neighbours $\mathit{n}$:
            \begin{equation}
                \begin{gathered}
                    \Delta_{\sigma}^{*} = \frac{1}{\mathit{n}} \sum_{\boldsymbol{\xi}} \Delta_{\boldsymbol{\xi},\sigma}^{*} = \frac{2}{N_m} \sum_{\mathbf{k}} \Big(\frac{1}{\mathit{n}}\sum_{\boldsymbol{\xi}} e^{-\mathrm{i}\mathbf{k} \cdot \boldsymbol{\xi}}\Big) \Delta_{\mathbf{k},\sigma}^{*} \\
                    \implies \boxed{\Delta_{\sigma}^{*} = \frac{2}{N_m} \sum_{\mathbf{k}} \gamma_{\mathbf{k}}^{*} \Delta_{\mathbf{k},\sigma}^{*}}
                \end{gathered}
            \end{equation}
            Similarly for $\langle\createta\createtb\rangle$:
            \begin{gather}
                \boxed{\Delta_{\boldsymbol{\xi},\tau}^{*} = \frac{2}{N_m} \sum_{\mathbf{k}} e^{-\mathrm{i}\mathbf{k} \cdot \boldsymbol{\xi}} \Delta_{\mathbf{k},\tau}^{*}} \\
                \boxed{\Delta_{\tau}^{*} = \frac{2}{N_m} \sum_{\mathbf{k}} \gamma_{\mathbf{k}}^{*} \Delta_{\mathbf{k},\tau}^{*}}
            \end{gather}
        \end{enumerate}
    \end{subequations}
    For 2D decorated square lattice, the NN coordination vectors are $\boldsymbol{\xi} = (\pm a, \pm a),(\pm a, \mp a)$ and $\mathit{n} = 4$.

    \item Hopping amplitude: \\
    We define $\mathbf{r}_{j} = \mathbf{r}_{i} + \boldsymbol{\zeta}$ where $\boldsymbol{\zeta}$ is the next-to-next-nearest (NNN) neighbour coordination vector.
    \begin{subequations}
        \begin{enumerate}[label=(\roman*), leftmargin=4pt, topsep=0pt]
            \item $\mathit{t}_{ij,\sigma} = \langle\createsa[j]\annihilatesa[i]\rangle = \Big\langle \sqrt{\frac{2}{N_m}} \sum_{\mathbf{k}} e^{+\mathrm{i}\mathbf{k}\cdot\mathbf{r}_{j}}\, \fcreatesa[\mathbf{k}] \sqrt{\frac{2}{N_m}} \sum_{\mathbf{k}'} e^{-\mathrm{i}\mathbf{k}'\cdot\mathbf{r}_{i}}\, \fannihilatesa[\mathbf{k}'] \Big\rangle = \frac{2}{N_m} \sum_{\mathbf{k},\mathbf{k}'} e^{+\mathrm{i}\mathbf{k}'\cdot\boldsymbol{\zeta}} e^{\mathrm{i}(\mathbf{k} - \mathbf{k}')\cdot\mathbf{r}_{i}} \langle\fcreatesa[\mathbf{k}]\fannihilatesa[\mathbf{k}']\rangle = \frac{2}{N_m} \sum_{\mathbf{k},\mathbf{k}'} e^{+\mathrm{i}\mathbf{k}'\cdot\boldsymbol{\zeta}} e^{\mathrm{i}(\mathbf{k} - \mathbf{k}')\cdot\mathbf{r}_{i}} \varrho_{\mathbf{k},\sigma} \delta_{\mathbf{k},\mathbf{k}'} \Rightarrow \mathit{t}_{ij,\sigma} =\frac{2}{N_m} \sum_{\mathbf{k}} e^{+\mathrm{i} \mathbf{k} \cdot \boldsymbol{\zeta}} \varrho_{\mathbf{k},\sigma}$
            \begin{equation}
                \implies \boxed{\mathit{t}_{\boldsymbol{\zeta},\sigma} = \frac{2}{N_m} \sum_{\mathbf{k}} e^{+\mathrm{i} \mathbf{k} \cdot \boldsymbol{\zeta}} \varrho_{\mathbf{k},\sigma}}
            \end{equation}
            Similarly for $\langle\createta[j]\annihilateta[i]\rangle$:
            \begin{equation}
                \boxed{\mathit{t}_{\boldsymbol{\zeta},\tau} = \frac{2}{N_m} \sum_{\mathbf{k}} e^{+\mathrm{i} \mathbf{k} \cdot \boldsymbol{\zeta}} \varrho_{\mathbf{k},\tau}}
            \end{equation}
            
            \item $\mathit{t}_{ij,\sigma}^{*} = \langle\createsa[i]\annihilatesa[j]\rangle = \Big\langle \sqrt{\frac{2}{N_m}} \sum_{\mathbf{k}} e^{+\mathrm{i}\mathbf{k}\cdot\mathbf{r}_{i}}\, \fcreatesa[\mathbf{k}] \sqrt{\frac{2}{N_m}} \sum_{\mathbf{k}'} e^{-\mathrm{i}\mathbf{k}'\cdot\mathbf{r}_{j}}\, \fannihilatesa[\mathbf{k}'] \Big\rangle = \frac{2}{N_m} \sum_{\mathbf{k},\mathbf{k}'} e^{\mathrm{i}(\mathbf{k} - \mathbf{k}')\cdot\mathbf{r}_{i}} e^{-\mathrm{i}\mathbf{k}'\cdot\boldsymbol{\zeta}} \langle\fcreatesa[\mathbf{k}]\fannihilatesa[\mathbf{k}']\rangle = \frac{2}{N_m} \sum_{\mathbf{k},\mathbf{k}'} e^{\mathrm{i}(\mathbf{k} - \mathbf{k}')\cdot\mathbf{r}_{i}} e^{-\mathrm{i}\mathbf{k}'\cdot\boldsymbol{\zeta}} \varrho_{\mathbf{k},\sigma} \delta_{\mathbf{k},\mathbf{k}'} \Rightarrow \mathit{t}_{ij,\sigma}^{*} = \frac{2}{N_m} \sum_{\mathbf{k}} e^{-\mathrm{i} \mathbf{k} \cdot \boldsymbol{\zeta}} \varrho_{\mathbf{k},\sigma}$
            \begin{equation}
                \implies \boxed{\mathit{t}_{\boldsymbol{\zeta},\sigma}^{*} = \frac{2}{N_m} \sum_{\mathbf{k}} e^{-\mathrm{i} \mathbf{k} \cdot \boldsymbol{\zeta}} \varrho_{\mathbf{k},\sigma}}
            \end{equation}
            Similarly for $\langle\createta[i]\annihilateta[j]\rangle$:
            \begin{equation}
                \boxed{\mathit{t}_{\boldsymbol{\zeta},\tau}^{*} = \frac{2}{N_m} \sum_{\mathbf{k}} e^{-\mathrm{i} \mathbf{k} \cdot \boldsymbol{\zeta}} \varrho_{\mathbf{k},\tau}}
            \end{equation}
        \end{enumerate}
        
        Physically, $\mathcal{J}_{2,ij}^{\sigma} = \mathcal{J}_{2,ji}^{\sigma}$ and $\mathcal{J}_{2,ij}^{\tau} = \mathcal{J}_{2,ji}^{\tau}$. Thus, $\langle\createsa[i]\annihilatesa[j]\rangle = \langle\createsa[j]\annihilatesa[i]\rangle \Rightarrow \mathit{t}_{ij,\sigma} \in \mathbb{R}$ and $\langle\createta[i]\annihilateta[j]\rangle = \langle\createta[j]\annihilateta[i]\rangle \Rightarrow \mathit{t}_{ij,\tau} \in \mathbb{R}$. The results for sublattice A is equally valid for sublattice B and because for a N\'{e}el ordered state, the number density of sublattices A and B are equal, their hopping density of sublattices must also be same by definition. The hopping amplitude appears only because NNN neighbour interactions have been taken into account giving rise to an intra-sublattice hopping of collective excitations. For 2D decorated square lattice, the NNN coordination vectors are $\boldsymbol{\zeta} = (\pm 2a, 0),(0, \pm 2a)$.
    \end{subequations}
\end{enumerate}

\section{General Coefficients of HP Operators with Quadratic Form} \label{gen_coeff}

Considering the spin-orbital Hamiltonian given by \textcolor{blue}{Eq.~\eqref{fullH} of main manuscript} for a generic bipartite lattice, the coefficients of quadratic form HP operators in the mean-field Hamiltonian given by \textcolor{blue}{Eq.~\eqref{nondiagH} of main manuscript} are as follows:
\begin{subequations}
    \begin{multline}
            \mathcal{E}_{\sigma}^{\text{A}}(\mathbf{k}) = -\mathit{n} \mathcal{J}_{1}^{\sigma} (\,\sigma - \varrho_{\sigma} - \lambda_{1}^{\sigma}\Re(\Delta_{\sigma})\,) - \mathit{n} \mathcal{Q}_{1} \big[-\tau^{2} (\,\sigma - \varrho_{\sigma} - \lambda_{1}^{\sigma}\Re(\Delta_{\sigma})\,) + 2\sigma\tau (\,\varrho_{\tau} + \lambda_{1}^{\tau}\Re(\Delta_{\tau})\,)\big] 
            + 2\mathcal{K}_{z}^{\sigma} (\sigma - 2\varrho_{\sigma}) \\
            + \textstyle{\sum_{\boldsymbol{\zeta}}} \mathcal{J}_{2,\text{A}\boldsymbol{\zeta}}^{\sigma} \big[(\sigma - \varrho_{\sigma}) (\,1 - \lambda_{2}^{\sigma}\cos{(\mathbf{k} \cdot \boldsymbol{\zeta})}\,) + \mathit{t}_{\boldsymbol{\zeta},\sigma}^{\text{A}} (\,\lambda_{2}^{\sigma} - \cos{(\mathbf{k} \cdot \boldsymbol{\zeta})}\,)\big] \\
            + \textstyle{\sum_{\boldsymbol{\zeta}}} \mathcal{Q}_{2,\text{A}\boldsymbol{\zeta}} \big[\tau^2 \big\{(\sigma - \varrho_{\sigma}) (\,1 - \lambda_{2}^{\sigma}\cos{(\mathbf{k} \cdot \boldsymbol{\zeta})}\,) + \mathit{t}_{\boldsymbol{\zeta},\sigma}^{\text{A}} (\,\lambda_{2}^{\sigma} - \cos{(\mathbf{k} \cdot \boldsymbol{\zeta})}\,)\big\} \\
            - 2\sigma\tau \big\{(\varrho_{\tau} - \lambda_{2}^{\tau} \mathit{t}_{\boldsymbol{\zeta},\tau}^{\text{A}}) (\,1 - \lambda_{2}^{\sigma} \cos{(\mathbf{k} \cdot \boldsymbol{\zeta})}\,)\big\}\big]
        \end{multline}
        \begin{multline}
            \mathcal{E}_{\sigma}^{\text{B}}(\mathbf{k}) = -\mathit{n} \mathcal{J}_{1}^{\sigma} (\,\sigma - \varrho_{\sigma} - \lambda_{1}^{\sigma}\Re(\Delta_{\sigma})\,) - \mathit{n} \mathcal{Q}_{1} \big[-\tau^{2} (\,\sigma - \varrho_{\sigma} - \lambda_{1}^{\sigma}\Re(\Delta_{\sigma})\,) + 2\sigma\tau (\,\varrho_{\tau} + \lambda_{1}^{\tau}\Re(\Delta_{\tau})\,)\big] + 2\mathcal{K}_{z}^{\sigma} (\sigma - 2\varrho_{\sigma}) \\
            + \textstyle{\sum_{\boldsymbol{\zeta}}} \mathcal{J}_{2,\text{B}\boldsymbol{\zeta}}^{\sigma} \big[(\sigma - \varrho_{\sigma}) (\,1 - \lambda_{2}^{\sigma}\cos{(\mathbf{k} \cdot \boldsymbol{\zeta})}\,) + \mathit{t}_{\boldsymbol{\zeta},\sigma}^{\text{B}} (\,\lambda_{2}^{\sigma} - \cos{(\mathbf{k} \cdot \boldsymbol{\zeta})}\,)\big] \\
            + \textstyle{\sum_{\boldsymbol{\zeta}}} \mathcal{Q}_{2,\text{B}\boldsymbol{\zeta}} \big[\tau^2 \big\{(\sigma - \varrho_{\sigma}) (\,1 - \lambda_{2}^{\sigma}\cos{(\mathbf{k} \cdot \boldsymbol{\zeta})}\,) + \mathit{t}_{\boldsymbol{\zeta},\sigma}^{\text{B}} (\,\lambda_{2}^{\sigma} - \cos{(\mathbf{k} \cdot \boldsymbol{\zeta})}\,)\big\} \\
            - 2\sigma\tau \big\{(\varrho_{\tau} - \lambda_{2}^{\tau} \mathit{t}_{\boldsymbol{\zeta},\tau}^{\text{B}}) (\,1 - \lambda_{2}^{\sigma} \cos{(\mathbf{k} \cdot \boldsymbol{\zeta})}\,)\big\}\big]
        \end{multline}
        \begin{multline}
            \mathcal{E}_{\tau}^{\text{A}}(\mathbf{k}) = -\mathit{n} \mathcal{J}_{1}^{\tau} (\,\tau - \varrho_{\tau} - \lambda_{1}^{\tau}\Re(\Delta_{\tau})\,) - \mathit{n} \mathcal{Q}_{1} \big[-\sigma^{2} (\,\tau - \varrho_{\tau} - \lambda_{1}^{\tau}\Re(\Delta_{\tau})\,) + 2\sigma\tau (\,\varrho_{\sigma} + \lambda_{1}^{\sigma}\Re(\Delta_{\sigma})\,)\big] + 2\mathcal{K}_{z}^{\tau} (\tau - 2\varrho_{\tau}) \\
            + \textstyle{\sum_{\boldsymbol{\zeta}}} \mathcal{J}_{2,\text{A}\boldsymbol{\zeta}}^{\tau} \big[(\tau - \varrho_{\tau}) (\,1 - \lambda_{2}^{\tau}\cos{(\mathbf{k} \cdot \boldsymbol{\zeta})}\,) + \mathit{t}_{\boldsymbol{\zeta},\tau}^{\text{A}} (\,\lambda_{2}^{\tau} - \cos{(\mathbf{k} \cdot \boldsymbol{\zeta})}\,)\big] \\
            + \textstyle{\sum_{\boldsymbol{\zeta}}} \mathcal{Q}_{2,\text{A}\boldsymbol{\zeta}} \big[\sigma^2 \big\{(\tau - \varrho_{\tau}) (\,1 - \lambda_{2}^{\tau}\cos{(\mathbf{k} \cdot \boldsymbol{\zeta})}\,) + \mathit{t}_{\boldsymbol{\zeta},\tau}^{\text{A}} (\,\lambda_{2}^{\tau} - \cos{(\mathbf{k} \cdot \boldsymbol{\zeta})}\,)\big\} \\
            - 2\sigma\tau \big\{(\varrho_{\sigma} - \lambda_{2}^{\sigma}\mathit{t}_{\boldsymbol{\zeta},\sigma}^{\text{A}}) (\,1 - \lambda_{2}^{\tau} \cos{(\mathbf{k} \cdot \boldsymbol{\zeta})}\,)\big\}\big]
        \end{multline}
        \begin{multline}
            \mathcal{E}_{\tau}^{\text{B}}(\mathbf{k}) = -\mathit{n} \mathcal{J}_{1}^{\tau} (\,\tau - \varrho_{\tau} - \lambda_{1}^{\tau}\Re(\Delta_{\tau})\,) - \mathit{n} \mathcal{Q}_{1} \big[-\sigma^{2} (\,\tau - \varrho_{\tau} - \lambda_{1}^{\tau}\Re(\Delta_{\tau})\,) + 2\sigma\tau (\,\varrho_{\sigma} + \lambda_{1}^{\sigma}\Re(\Delta_{\sigma})\,)\big] + 2\mathcal{K}_{z}^{\tau} (\tau - 2\varrho_{\tau}) \\
            + \textstyle{\sum_{\boldsymbol{\zeta}}} \mathcal{J}_{2,\text{B}\boldsymbol{\zeta}}^{\tau} \big[(\tau - \varrho_{\tau}) (\,1 - \lambda_{2}^{\tau}\cos{(\mathbf{k} \cdot \boldsymbol{\zeta})}\,) + \mathit{t}_{\boldsymbol{\zeta},\tau}^{\text{B}} (\,\lambda_{2}^{\tau} - \cos{(\mathbf{k} \cdot \boldsymbol{\zeta})}\,)\big] \\
            + \textstyle{\sum_{\boldsymbol{\zeta}}} \mathcal{Q}_{2,\text{B}\boldsymbol{\zeta}} \big[\sigma^2 \big\{(\tau - \varrho_{\tau}) (\,1 - \lambda_{2}^{\tau}\cos{(\mathbf{k} \cdot \boldsymbol{\zeta})}\,) + \mathit{t}_{\boldsymbol{\zeta},\tau}^{\text{B}} (\,\lambda_{2}^{\tau} - \cos{(\mathbf{k} \cdot \boldsymbol{\zeta})}\,)\big\} \\
            - 2\sigma\tau \big\{(\varrho_{\sigma} - \lambda_{2}^{\sigma}\mathit{t}_{\boldsymbol{\zeta},\sigma}^{\text{B}}) (\,1 - \lambda_{2}^{\tau} \cos{(\mathbf{k} \cdot \boldsymbol{\zeta})}\,)\big\}\big]
        \end{multline}
        \begin{multline}
            \mathcal{F}_{\sigma}(\mathbf{k}) = -\mathit{n} \mathcal{J}_{1}^{\sigma} \big[\lambda_{1}^{\sigma}(\sigma - \varrho_{\sigma})\gamma_{\mathbf{k}} - \gamma_{\mathbf{k},\sigma}'\big] \\
            - \mathit{n} \mathcal{Q}_{1} \big[ -\tau^{2} (\lambda_{1}^{\sigma}(\sigma - \varrho_{\sigma})\gamma_{\mathbf{k}} - \gamma_{\mathbf{k},\sigma}') 
            + \lambda_{1}^{\sigma}\sigma\tau (2\varrho_{\tau}\gamma_{\mathbf{k}} + \lambda_{1}^{\tau}\gamma_{\mathbf{k},\tau}' + \lambda_{1}^{\tau}\gamma_{-\mathbf{k},\tau}'^{*}) \big]
        \end{multline}
        \begin{multline}
            \mathcal{F}_{\tau}(\mathbf{k}) = -\mathit{n} \mathcal{J}_{1}^{\tau} \left[ \lambda_{1}^{\tau}(\tau - \varrho_{\tau})\gamma_{\mathbf{k}} - \gamma_{\mathbf{k},\tau}' \right] \\
            -\mathit{n} \mathcal{Q}_{1} \big[ -\sigma^{2} (\lambda_{1}^{\tau}(\tau - \varrho_{\tau})\gamma_{\mathbf{k}} - \gamma_{\mathbf{k},\tau}') 
            + \lambda_{1}^{\tau}\sigma\tau (2\varrho_{\sigma}\gamma_{\mathbf{k}} + \lambda_{1}^{\sigma}\gamma_{\mathbf{k},\sigma}' + \lambda_{1}^{\sigma}\gamma_{-\mathbf{k},\sigma}'^{*}) \big]
        \end{multline}
\end{subequations}
Here, $\mathit{n}$ is the number of next-nearest (NN) neighbours, $\boldsymbol{\xi}$ is the NN neighbour coordination vector, $\boldsymbol{\zeta}$ is the next-to-next-nearest (NNN) neighbour coordination vector, $\Re(\Delta_{\sigma})$ and $\Re(\Delta_{\tau})$ denote the real part of $\Delta_{\sigma}$ and $\Delta_{\tau}$ respectively, and $\gamma_{\mathbf{k}}$ is the NN neighbour structure factor (also defined in Sec.~\ref{expct_val}):
\begin{equation}
    \gamma_{\mathbf{k}} = \frac{1}{\mathit{n}}\sum_{\boldsymbol{\xi}} e^{+\mathrm{i}\mathbf{k} \cdot \boldsymbol{\xi}}
\end{equation}
We further define the NN neighbour pairing structure factor for spins and isospins respectively:
\begin{subequations}
    \begin{gather}
        \gamma_{\mathbf{k},\sigma}' = \frac{1}{\mathit{n}}\sum_{\boldsymbol{\xi}} e^{+\mathrm{i}\mathbf{k} \cdot \boldsymbol{\xi}} \Delta_{\boldsymbol{\xi},\sigma}^{*} \\
        \gamma_{\mathbf{k},\tau}' = \frac{1}{\mathit{n}}\sum_{\boldsymbol{\xi}} e^{+\mathrm{i}\mathbf{k} \cdot \boldsymbol{\xi}} \Delta_{\boldsymbol{\xi},\tau}^{*}
    \end{gather}
\end{subequations}

The constant shift in the energy spectrum is given by $\langle\hat{\mathcal{H}_{0}}\rangle = E_{\sigma} + E_{\tau} + E_{\sigma\tau}$ which contains quantum corrections to the classical ground state energy of spins and isospins and the interaction energy between them.
\begin{subequations}
    \begin{multline}
            E_{\sigma} = \textstyle{\frac{\mathit{n} N_m \mathcal{J}_{1}^{\sigma}}{2}} \Big[\sigma^2 - \big(\varrho_{\sigma}^{2} + \textstyle{\frac{1}{\mathit{n}}\sum_{\boldsymbol{\xi}}} |\Delta_{\boldsymbol{\xi},\sigma}|^{2}\big) - 2\lambda_{1}^{\sigma}\varrho_{\sigma} \Re(\Delta_{\sigma})\Big] \\
            -\textstyle{\frac{N_m}{4}} \textstyle{\sum_{\boldsymbol{\zeta}}} \mathcal{J}_{2,\text{A}\boldsymbol{\zeta}}^{\sigma} \Big[\sigma^2 - (\varrho_{\sigma}^{2} + \mathit{t}_{\boldsymbol{\zeta},\sigma}^{2} - 2\lambda_{2}^{\sigma}\varrho_{\sigma} \mathit{t}_{\boldsymbol{\zeta},\sigma})\Big] 
            -\textstyle{\frac{N_m}{4}} \textstyle{\sum_{\boldsymbol{\zeta}}} \mathcal{J}_{2,\text{B}\boldsymbol{\zeta}}^{\sigma} \Big[\sigma^2 - (\varrho_{\sigma}^{2} + \mathit{t}_{\boldsymbol{\zeta},\sigma}^{2} - 2\lambda_{2}^{\sigma}\varrho_{\sigma} \mathit{t}_{\boldsymbol{\zeta},\sigma})\Big]
        \end{multline}
        \begin{multline}
            E_{\tau} = \textstyle{\frac{\mathit{n} N_m \mathcal{J}_{1}^{\tau}}{2}} \Big[\tau^2 - \big(\varrho_{\tau}^{2} + \textstyle{\frac{1}{\mathit{n}}\sum_{\boldsymbol{\xi}}} |\Delta_{\boldsymbol{\xi},\tau}|^{2}\big) - 2\lambda_{1}^{\tau}\varrho_{\tau} \Re(\Delta_{\tau})\Big] \\
            -\textstyle{\frac{N_m}{4} \sum_{\boldsymbol{\zeta}}} \mathcal{J}_{2,\text{A}\boldsymbol{\zeta}}^{\tau} \Big[\tau^2 - (\varrho_{\tau}^{2} + \mathit{t}_{\boldsymbol{\zeta},\tau}^{2} - 2\lambda_{2}^{\tau}\varrho_{\tau} \mathit{t}_{\boldsymbol{\zeta},\tau})\Big] 
            -\textstyle{\frac{N_m}{4} \sum_{\boldsymbol{\zeta}}} \mathcal{J}_{2,\text{B}\boldsymbol{\zeta}}^{\tau} \Big[\tau^2 - (\varrho_{\tau}^{2} + \mathit{t}_{\boldsymbol{\zeta},\tau}^{2} - 2\lambda_{2}^{\tau}\varrho_{\tau} \mathit{t}_{\boldsymbol{\zeta},\tau})\Big]
        \end{multline}
        \begin{multline}
            E_{\sigma\tau} = -\textstyle{\frac{\mathit{n} N_m \mathcal{Q}_{1}}{2}} \Big[ \sigma^2\tau^2 - \sigma^{2} \big(\varrho_{\tau}^{2} + \textstyle{\frac{1}{\mathit{n}}\sum_{\boldsymbol{\xi}}} |\Delta_{\boldsymbol{\xi},\tau}|^{2}\big) 
            - \tau^{2} \big(\varrho_{\sigma}^{2} + \textstyle{\frac{1}{\mathit{n}}\sum_{\boldsymbol{\xi}}} |\Delta_{\boldsymbol{\xi},\sigma}|^{2}\big) 
            - 2\lambda_{1}^{\tau} \sigma^{2}\varrho_{\tau} \Re(\Delta_{\tau}) - 2\lambda_{1}^{\sigma} \tau^{2}\varrho_{\sigma} \Re(\Delta_{\sigma}) \\
            - 4\sigma\tau \varrho_{\sigma}\varrho_{\tau} - 4\lambda_{1}^{\tau} \sigma\tau\varrho_{\sigma} \Re(\Delta_{\tau}) - 4\lambda_{1}^{\sigma} \sigma\tau\varrho_{\tau} \Re(\Delta_{\sigma}) 
            - 4\lambda_{1}^{\sigma}\lambda_{1}^{\tau} \sigma\tau \textstyle{\frac{1}{\mathit{n}}\sum_{\boldsymbol{\xi}}} \Re(\Delta_{\boldsymbol{\xi},\sigma}) \Re(\Delta_{\boldsymbol{\xi},\tau}) \Big] \\
            - \textstyle{\frac{N_m}{4} \sum_{\boldsymbol{\zeta}}} \mathcal{Q}_{2,\text{A}\boldsymbol{\zeta}} \Big[\sigma^2\tau^2 - \tau^2 \big( \varrho_{\sigma}^{2} + \mathit{t}_{\boldsymbol{\zeta},\sigma}^{2} - 2\lambda_{2}^{\sigma}\varrho_{\sigma} \mathit{t}_{\boldsymbol{\zeta},\sigma} \big) 
            - \sigma^2 \big( \varrho_{\tau}^{2} + \mathit{t}_{\boldsymbol{\zeta},\tau}^{2} - 2\lambda_{2}^{\tau}\varrho_{\tau} \mathit{t}_{\boldsymbol{\zeta},\tau} \big) 
            - 4\sigma\tau \big(\varrho_{\sigma} - \lambda_{2}^{\sigma}\mathit{t}_{\boldsymbol{\zeta},\sigma}\big) \big(\varrho_{\tau} - \lambda_{2}^{\tau} \mathit{t}_{\boldsymbol{\zeta},\tau}\big)\Big] \\
            - \textstyle{\frac{N_m}{4} \sum_{\boldsymbol{\zeta}}} \mathcal{Q}_{2,\text{B}\boldsymbol{\zeta}} \Big[\sigma^2\tau^2 - \tau^2 \big( \varrho_{\sigma}^{2} + \mathit{t}_{\boldsymbol{\zeta},\sigma}^{2} - 2\lambda_{2}^{\sigma}\varrho_{\sigma} \mathit{t}_{\boldsymbol{\zeta},\sigma} \big) 
            - \sigma^2 \big( \varrho_{\tau}^{2} + \mathit{t}_{\boldsymbol{\zeta},\tau}^{2} - 2\lambda_{2}^{\tau}\varrho_{\tau} \mathit{t}_{\boldsymbol{\zeta},\tau} \big) 
            - 4\sigma\tau \big(\varrho_{\sigma} - \lambda_{2}^{\sigma}\mathit{t}_{\boldsymbol{\zeta},\sigma}\big) \big(\varrho_{\tau} - \lambda_{2}^{\tau} \mathit{t}_{\boldsymbol{\zeta},\tau}\big)\Big]
        \end{multline}
\end{subequations}

\section{Chiral Magnon and Orbiton Eigenmodes}

The Hamiltonian in \textcolor{blue}{Eq.~\eqref{diagH} of main manuscript} is block diagonal in the Bogolyubov operators and may be rewritten in a matrix form:
\begin{equation}
    \hat{\mathcal{H}} = \mathcal{H}_0 \,\mathbb{I}_{4 \times 4} + \sum_{\mathbf{k}}
    \begin{pNiceArray}{cc:cc}
        \alpha_{\mathbf{k},\sigma}^{\dagger} & \beta_{-\mathbf{k},\sigma} & \alpha_{\mathbf{k},\tau}^{\dagger} & \beta_{-\mathbf{k},\tau}
    \end{pNiceArray}
    \begin{pNiceArray}{cc:cc}
        \omega_{\mathbf{k},\sigma}^{\alpha} & 0 & 0 & 0 \\
        0 & \omega_{-\mathbf{k},\sigma}^{\beta} & 0 & 0 \\
        \hdottedline
        0 & 0 & \omega_{\mathbf{k},\tau}^{\alpha} & 0 \\
        0 & 0 & 0 & \omega_{-\mathbf{k},\tau}^{\beta}
    \end{pNiceArray}
    \begin{pNiceArray}{c}
        \alpha_{\mathbf{k},\sigma} \\
        \beta_{-\mathbf{k},\sigma}^{\dagger} \\
        \hdottedline
        \alpha_{\mathbf{k},\tau} \\
        \beta_{-\mathbf{k},\tau}^{\dagger}
    \end{pNiceArray} 
\end{equation}
where, $\mathbb{I}_{4 \times 4}$ is a $4\times4$ identity matrix. The above matrix form clearly shows that $\alpha_{\mathbf{k},\sigma}$ and $\beta_{-\mathbf{k},\sigma}^{\dagger}$ are the magnon eigenmodes, whereas $\alpha_{\mathbf{k},\tau}$ and $\beta_{-\mathbf{k},\tau}^{\dagger}$ are the orbiton eigenmodes of $\hat{\mathcal{H}}$. The time-dependence of these magnon and orbiton eigenmodes can be viewed in the Heisenberg picture via the Heisenberg equations of motion. In \textcolor{blue}{Eq.~\eqref{HEOM} of the main manuscript}, the time dependence of the eigenmodes was not considered explicitly which we shall now consider:
\begin{subequations}
    \label{HEOM_full}
    \begin{gather}
        \dot{\alpha}_{\mathbf{k},\sigma} = -\mathrm{i}[\alpha_{\mathbf{k},\sigma}, \hat{\mathcal{H}}] = -\mathrm{i}\omega_{\mathbf{k},\sigma}^{\alpha} \alpha_{\mathbf{k},\sigma} \Rightarrow \alpha_{\mathbf{k},\sigma}(\mathrm{t}) = \alpha_{\mathbf{k},\sigma}(0) e^{-\mathrm{i} \omega_{\mathbf{k},\sigma}^{\alpha} \mathrm{t}} \\
        \dot{\beta}_{-\mathbf{k},\sigma}^{\dagger} = -\mathrm{i}[\beta_{-\mathbf{k},\sigma}^{\dagger}, \hat{\mathcal{H}}] = +\mathrm{i}\omega_{-\mathbf{k},\sigma}^{\beta} \beta_{-\mathbf{k},\sigma}^{\dagger} \Rightarrow \beta_{-\mathbf{k},\sigma}^{\dagger}(\mathrm{t}) = \beta_{-\mathbf{k},\sigma}^{\dagger}(0) e^{+\mathrm{i} \omega_{-\mathbf{k},\sigma}^{\beta} \mathrm{t}} \\
        \dot{\alpha}_{\mathbf{k},\tau} = -\mathrm{i}[\alpha_{\mathbf{k},\tau}, \hat{\mathcal{H}}] = -\mathrm{i}\omega_{\mathbf{k},\tau}^{\alpha} \alpha_{\mathbf{k},\tau} \Rightarrow \alpha_{\mathbf{k},\tau}(\mathrm{t}) = \alpha_{\mathbf{k},\tau}(0) e^{-\mathrm{i} \omega_{\mathbf{k},\tau}^{\alpha} \mathrm{t}} \\
        \dot{\beta}_{-\mathbf{k},\tau}^{\dagger} = -\mathrm{i}[\beta_{-\mathbf{k},\tau}^{\dagger}, \hat{\mathcal{H}}] = +\mathrm{i}\omega_{-\mathbf{k},\tau}^{\beta} \beta_{-\mathbf{k},\tau}^{\dagger} \Rightarrow \beta_{-\mathbf{k},\tau}^{\dagger}(\mathrm{t}) = \beta_{-\mathbf{k},\tau}^{\dagger}(0) e^{+\mathrm{i} \omega_{-\mathbf{k},\tau}^{\beta} \mathrm{t}}
    \end{gather}
\end{subequations}
Since the magnon eigenmodes $\alpha_{\mathbf{k},\sigma}$ and $\beta_{-\mathbf{k},\sigma}^{\dagger}$ have opposite phases, they can be viewed as spin precessions in the opposite sense. Similarly, the orbiton eigenmodes $\alpha_{\mathbf{k},\tau}$ and $\beta_{-\mathbf{k},\tau}^{\dagger}$ can be viewed as orbital isospin precessions in the opposite sense. Hence, magnons and orbitons possess a definite handedness or \emph{chirality} as stated in \textcolor{blue}{Sec.~\ref{mag_orb} of main manuscript}.

\section{Gauge Freedom of Bogolyubov Coefficients} \label{gauge}

Finding the magnon and orbiton spectrum as defined in \textcolor{blue}{Eq.~\eqref{dispersion} of main manuscript} requires the knowledge of Bogolyubov coefficients that enable the canonical transformations to diagonalize the Hamiltonian $\hat{\mathcal{H}}$ in \textcolor{blue}{Eq.~\eqref{nondiagH} of main manuscript}. The Bogolyubov operators in \textcolor{blue}{Eq.~\eqref{Bmatrix} of main manuscript} are linear combinations of the Holstein-Primakoff operators with complex coefficients (referred to as Bogolyubov coefficients) in general. Let us consider the following diagonal Hamiltonian:
\begin{equation}
    \hat{\mathcal{H}'} = \hat{\mathcal{H}}_{0} + \sum_{\mathbf{k}} \Big[ \omega_{\mathbf{k},\sigma}^{\alpha} {\alpha'}_{\mathbf{k},\sigma}^{\dagger}{\alpha'}_{\mathbf{k},\sigma} + \omega_{\mathbf{k},\sigma}^{\beta} {\beta'}_{\mathbf{k},\sigma}^{\dagger}{\beta'}_{\mathbf{k},\sigma} \Big] + \sum_{\mathbf{k}} \Big[ \omega_{\mathbf{k},\tau}^{\alpha} {\alpha'}_{\mathbf{k},\tau}^{\dagger}{\alpha'}_{\mathbf{k},\tau} + \omega_{\mathbf{k},\tau}^{\beta} {\beta'}_{\mathbf{k},\tau}^{\dagger}{\beta'}_{\mathbf{k},\tau} \Big]
    \label{diagH_prime}
\end{equation} 
where, the Bogolyubov operators $\alpha',{\alpha'}^{\dagger},\beta',{\beta'}^{\dagger}$ are defined in the following manner:
\begin{equation*}
    \begin{cases}
        \begin{gathered}
            \fbannihilatesa[\mathbf{k}] = u_{\mathbf{k},\sigma} \fannihilatesa[\mathbf{k}] - v_{\mathbf{k},\sigma}^{*} \fcreatesb[-\mathbf{k}] = |u_{\mathbf{k},\sigma}|e^{+\mathrm{i}\theta_{u}}\fannihilatesa[\mathbf{k}] - |v_{\mathbf{k},\sigma}|e^{-\mathrm{i}\theta_{v}}\fcreatesb[-\mathbf{k}] \\
            \Rightarrow \fbannihilatesa[\mathbf{k}] = e^{+\mathrm{i}\theta_{u}} \left(|u_{\mathbf{k},\sigma}|\fannihilatesa[\mathbf{k}] - |v_{\mathbf{k},\sigma}|e^{-\mathrm{i}(\theta_{u}+\theta_{v})}\fcreatesb[-\mathbf{k}]\right) = e^{+\mathrm{i}\theta_{u}} \fbannihilatesa[\mathbf{k}]' \\
            \therefore \boxed{\fbannihilatesa[\mathbf{k}]' = e^{-\mathrm{i}\theta_{u}} \fbannihilatesa[\mathbf{k}]}
        \end{gathered}
        \\ \\
        \begin{gathered}
            \fbcreatesa[\mathbf{k}] = u_{\mathbf{k},\sigma}^{*} \fcreatesa[\mathbf{k}] - v_{\mathbf{k},\sigma} \fannihilatesb[-\mathbf{k}] = |u_{\mathbf{k},\sigma}|e^{-\mathrm{i}\theta_{u}}\fcreatesa[\mathbf{k}] - |v_{\mathbf{k},\sigma}|e^{+\mathrm{i}\theta_{v}}\fannihilatesb[-\mathbf{k}] \\
            \Rightarrow \fbcreatesa[\mathbf{k}] =  \left(|u_{\mathbf{k},\sigma}|\fcreatesa[\mathbf{k}] - |v_{\mathbf{k},\sigma}|e^{+\mathrm{i}(\theta_{u}+\theta_{v})}\fannihilatesb[-\mathbf{k}]\right)e^{-\mathrm{i}\theta_{u}} =  {\alpha'}_{\mathbf{k},\sigma}^{\dagger}e^{-\mathrm{i}\theta_{u}} \\
            \therefore \boxed{{\alpha'}_{\mathbf{k},\sigma}^{\dagger} =  \fbcreatesa[\mathbf{k}]e^{+\mathrm{i}\theta_{u}}}
        \end{gathered}
    \end{cases}
\end{equation*}
\begin{equation*}
    \begin{cases}
        \begin{gathered}
            \fbannihilatesb[\mathbf{k}] = v_{\mathbf{k},\sigma} \fcreatesa[-\mathbf{k}] + u_{\mathbf{k},\sigma}^{*} \fannihilatesb[\mathbf{k}] = |v_{\mathbf{k},\sigma}|e^{+\mathrm{i}\theta_{v}}\fcreatesa[-\mathbf{k}] - |u_{\mathbf{k},\sigma}|e^{-\mathrm{i}\theta_{u}}\fannihilatesb[\mathbf{k}] \\
            \Rightarrow \fbannihilatesb[\mathbf{k}] = e^{-\mathrm{i}\theta_{u}} \left(|v_{\mathbf{k},\sigma}|e^{+\mathrm{i}(\theta_{u}+\theta_{v})}\fcreatesa[-\mathbf{k}] - |u_{\mathbf{k},\sigma}|\fannihilatesb[\mathbf{k}]\right) = e^{-\mathrm{i}\theta_{u}} \fbannihilatesb[\mathbf{k}]' \\
            \therefore \boxed{\fbannihilatesb[\mathbf{k}]' =  e^{+\mathrm{i}\theta_{u}} \fbannihilatesb[\mathbf{k}]}
        \end{gathered}
        \\ \\
        \begin{gathered}
            \fbcreatesb[\mathbf{k}] = v_{\mathbf{k},\sigma}^{*} \fannihilatesa[-\mathbf{k}] + u_{\mathbf{k},\sigma} \fcreatesb[\mathbf{k}] = |v_{\mathbf{k},\sigma}|e^{-\mathrm{i}\theta_{v}}\fannihilatesa[-\mathbf{k}] - |u_{\mathbf{k},\sigma}|e^{+\mathrm{i}\theta_{u}}\fcreatesb[\mathbf{k}] \\
            \Rightarrow \fbcreatesb[\mathbf{k}] =  \left(|v_{\mathbf{k},\sigma}|e^{-\mathrm{i}(\theta_{u}+\theta_{v})}\fannihilatesa[-\mathbf{k}] - |u_{\mathbf{k},\sigma}|\fcreatesb[\mathbf{k}]\right) e^{+\mathrm{i}\theta_{u}} = {\beta'}_{\mathbf{k},\sigma}^{\dagger} e^{+\mathrm{i}\theta_{u}} \\
            \therefore \boxed{{\beta'}_{\mathbf{k},\sigma}^{\dagger} = \fbcreatesb[\mathbf{k}] e^{-\mathrm{i}\theta_{u}}}
        \end{gathered}
    \end{cases}
\end{equation*}
\begin{equation*}
    \begin{cases}
        \begin{gathered}
            \fbannihilateta[\mathbf{k}] = u_{\mathbf{k},\tau} \fannihilateta[\mathbf{k}] - v_{\mathbf{k},\tau}^{*} \fcreatetb[-\mathbf{k}] = |u_{\mathbf{k},\tau}|e^{+\mathrm{i}\theta_{u}}\fannihilateta[\mathbf{k}] - |v_{\mathbf{k},\tau}|e^{-\mathrm{i}\theta_{v}}\fcreatetb[-\mathbf{k}] \\
            \Rightarrow \fbannihilateta[\mathbf{k}] = e^{+\mathrm{i}\theta_{u}} \left(|u_{\mathbf{k},\tau}|\fannihilateta[\mathbf{k}] - |v_{\mathbf{k},\tau}|e^{-\mathrm{i}(\theta_{u}+\theta_{v})}\fcreatetb[-\mathbf{k}]\right) = e^{+\mathrm{i}\theta_{u}} \fbannihilateta[\mathbf{k}]' \\
            \therefore \boxed{\fbannihilateta[\mathbf{k}]' = e^{-\mathrm{i}\theta_{u}} \fbannihilateta[\mathbf{k}]}
        \end{gathered}
        \\ \\
        \begin{gathered}
            \fbcreateta[\mathbf{k}] = u_{\mathbf{k},\tau}^{*} \fcreateta[\mathbf{k}] - v_{\mathbf{k},\tau} \fannihilatetb[-\mathbf{k}] = |u_{\mathbf{k},\tau}|e^{-\mathrm{i}\theta_{u}}\fcreateta[\mathbf{k}] - |v_{\mathbf{k},\tau}|e^{+\mathrm{i}\theta_{v}}\fannihilatetb[-\mathbf{k}] \\
            \Rightarrow \fbcreateta[\mathbf{k}] =  \left(|u_{\mathbf{k},\tau}|\fcreateta[\mathbf{k}] - |v_{\mathbf{k},\tau}|e^{+\mathrm{i}(\theta_{u}+\theta_{v})}\fannihilatetb[-\mathbf{k}]\right)e^{-\mathrm{i}\theta_{u}} =  {\alpha'}_{\mathbf{k},\tau}^{\dagger}e^{-\mathrm{i}\theta_{u}} \\
            \therefore \boxed{{\alpha'}_{\mathbf{k},\tau}^{\dagger} =  \fbcreateta[\mathbf{k}]e^{+\mathrm{i}\theta_{u}}}
        \end{gathered}
    \end{cases}
\end{equation*}
\begin{equation*}
    \begin{cases}
        \begin{gathered}
            \fbannihilatetb[\mathbf{k}] = v_{\mathbf{k},\tau} \fcreateta[-\mathbf{k}] + u_{\mathbf{k},\tau}^{*} \fannihilatetb[\mathbf{k}] = |v_{\mathbf{k},\tau}|e^{+\mathrm{i}\theta_{v}}\fcreateta[-\mathbf{k}] - |u_{\mathbf{k},\tau}|e^{-\mathrm{i}\theta_{u}}\fannihilatetb[\mathbf{k}] \\
            \Rightarrow \fbannihilatetb[\mathbf{k}] = e^{-\mathrm{i}\theta_{u}} \left(|v_{\mathbf{k},\tau}|e^{+\mathrm{i}(\theta_{u}+\theta_{v})}\fcreateta[-\mathbf{k}] - |u_{\mathbf{k},\tau}|\fannihilatetb[\mathbf{k}]\right) = e^{-\mathrm{i}\theta_{u}} \fbannihilatetb[\mathbf{k}]' \\
            \therefore \boxed{\fbannihilatetb[\mathbf{k}]' =  e^{+\mathrm{i}\theta_{u}} \fbannihilatetb[\mathbf{k}]}
        \end{gathered}
        \\ \\
        \begin{gathered}
            \fbcreatetb[\mathbf{k}] = v_{\mathbf{k},\tau}^{*} \fannihilateta[-\mathbf{k}] + u_{\mathbf{k},\tau} \fcreatetb[\mathbf{k}] = |v_{\mathbf{k},\tau}|e^{-\mathrm{i}\theta_{v}}\fannihilateta[-\mathbf{k}] - |u_{\mathbf{k},\tau}|e^{+\mathrm{i}\theta_{u}}\fcreatetb[\mathbf{k}] \\
            \Rightarrow \fbcreatetb[\mathbf{k}] =  \left(|v_{\mathbf{k},\tau}|e^{-\mathrm{i}(\theta_{u}+\theta_{v})}\fannihilateta[-\mathbf{k}] - |u_{\mathbf{k},\tau}|\fcreatetb[\mathbf{k}]\right) e^{+\mathrm{i}\theta_{u}} = {\beta'}_{\mathbf{k},\tau}^{\dagger} e^{+\mathrm{i}\theta_{u}} \\
            \therefore \boxed{{\beta'}_{\mathbf{k},\tau}^{\dagger} = \fbcreatetb[\mathbf{k}] e^{-\mathrm{i}\theta_{u}}}
        \end{gathered}
    \end{cases}
\end{equation*}
Then the quadratic operators in Eq.~\eqref{diagH_prime} become:
\begin{subequations}
    \begin{gather}
        \boxed{{\alpha'}_{\mathbf{k},\sigma}^{\dagger}{\alpha'}_{\mathbf{k},\sigma} = \fbcreatesa[\mathbf{k}]e^{+\mathrm{i}\theta_{u}} e^{-\mathrm{i}\theta_{u}} \fbannihilatesa[\mathbf{k}] = \fbnasigma{\mathbf{k}}{\mathbf{k}}} \\
        \boxed{{\beta'}_{\mathbf{k},\sigma}^{\dagger}{\beta'}_{\mathbf{k},\sigma} = \fbcreatesb[\mathbf{k}]e^{-\mathrm{i}\theta_{u}} e^{+\mathrm{i}\theta_{u}} \fbannihilatesb[\mathbf{k}] = \fbnbsigma{\mathbf{k}}{\mathbf{k}}} \\
        \boxed{{\alpha'}_{\mathbf{k},\tau}^{\dagger}{\alpha'}_{\mathbf{k},\tau} = \fbcreateta[\mathbf{k}]e^{+\mathrm{i}\theta_{u}} e^{-\mathrm{i}\theta_{u}} \fbannihilateta[\mathbf{k}] = \fbnatau{\mathbf{k}}{\mathbf{k}}} \\
        \boxed{{\beta'}_{\mathbf{k},\tau}^{\dagger}{\beta'}_{\mathbf{k},\tau} = \fbcreatetb[\mathbf{k}]e^{-\mathrm{i}\theta_{u}} e^{+\mathrm{i}\theta_{u}} \fbannihilatetb[\mathbf{k}] = \fbnbtau{\mathbf{k}}{\mathbf{k}}}
    \end{gather}
\end{subequations}
Thus it has been shown that the individual complex phases of the Bogolyubov coefficients results in an overall phase factor in the Bogolyubov operators, which gets nullified due to the operator structure of $\hat{\mathcal{H}'}$ and yields the same diagonal Hamiltonian $\hat{\mathcal{H}}$ as in \textcolor{blue}{Eq.~\eqref{diagH} of main manuscript}. Hence, even if the magnon or orbiton creation and annihilation operators acquire a global phase, the spectrum of $\hat{\mathcal{H}}$ remains unchanged, i.e., $\hat{\mathcal{H}}$ has an additional $U(1)$ gauge symmetry. In the main manuscript, we have chosen the gauge $u_{\mathbf{k},\sigma}, u_{\mathbf{k},\tau} \in \mathbb{R}$ and $v_{\mathbf{k},\sigma}, v_{\mathbf{k},\tau} \in \mathbb{C}$.

\section{Inversion-symmetric Magnon and Orbiton Dispersions}

Now, we shall show that the magnon and orbiton dispersions are symmetric under inversion, i.e., $\mathbf{k} \rightarrow -\mathbf{k}$. From the general expressions in Sec.~\ref{gen_coeff}, it is evident that $\mathcal{E}_{\sigma}^{\text{A}}(\mathbf{k}) = \mathcal{E}_{\sigma}^{\text{A}}(-\mathbf{k})$, $\mathcal{E}_{\sigma}^{\text{B}}(\mathbf{k}) = \mathcal{E}_{\sigma}^{\text{B}}(-\mathbf{k})$, $\mathcal{E}_{\tau}^{\text{A}}(\mathbf{k}) = \mathcal{E}_{\tau}^{\text{A}}(-\mathbf{k})$ and $\mathcal{E}_{\tau}^{\text{B}}(\mathbf{k}) = \mathcal{E}_{\tau}^{\text{B}}(-\mathbf{k})$ since they only contain cosine terms. But the situation is different for $\mathcal{F}_{\sigma}$ and $\mathcal{F}_{\tau}$. From \textcolor{blue}{Eq.~\eqref{HEOM} of main manuscript}, we have:
\begin{subequations}
    \begin{equation}
        \begin{cases}
            [\mathcal{E}_{\sigma}^{\text{A}}(\mathbf{k}) - \omega_{\mathbf{k},\sigma}^{\alpha}] u_{\mathbf{k},\sigma} + \mathcal{F}_{\sigma}(\mathbf{k}) v_{\mathbf{k},\sigma}^{*} = 0 \\
            \mathcal{F}_{\sigma}^{*}(\mathbf{k}) u_{\mathbf{k},\sigma} + [\mathcal{E}_{\sigma}^{\text{B}}(\mathbf{k}) + \omega_{\mathbf{k},\sigma}^{\alpha}] v_{\mathbf{k},\sigma}^{*} = 0
        \end{cases}
        \Rightarrow v_{\mathbf{k},\sigma}^{*} =  
        \begin{cases}
            -\frac{[\mathcal{E}_{\sigma}^{\text{A}}(\mathbf{k}) - \omega_{\mathbf{k},\sigma}^{\alpha}] u_{\mathbf{k},\sigma}}{\mathcal{F}_{\sigma}(\mathbf{k})} \\
            -\frac{\mathcal{F}_{\sigma}^{*}(\mathbf{k}) u_{\mathbf{k},\sigma}}{[\mathcal{E}_{\sigma}^{\text{B}}(\mathbf{k}) + \omega_{\mathbf{k},\sigma}^{\alpha}]}
        \end{cases}
        \label{vkstarsigma}
    \end{equation}
    \begin{equation}
        \begin{cases}
            [\mathcal{E}_{\sigma}^{\text{B}}(\mathbf{k}) - \omega_{\mathbf{k},\sigma}^{\beta}] u_{\mathbf{k},\sigma}^{*} + \mathcal{F}_{\sigma}(-\mathbf{k}) v_{\mathbf{k},\sigma} = 0 \\
            \mathcal{F}_{\sigma}^{*}(-\mathbf{k}) u_{\mathbf{k},\sigma}^{*} + [\mathcal{E}_{\sigma}^{\text{A}}(\mathbf{k}) +\omega_{\mathbf{k},\sigma}^{\beta}] v_{\mathbf{k},\sigma} = 0
        \end{cases}
        \Rightarrow v_{\mathbf{k},\sigma} = 
        \begin{cases}
            -\frac{[\mathcal{E}_{\sigma}^{\text{B}}(\mathbf{k}) - \omega_{\mathbf{k},\sigma}^{\beta}] u_{\mathbf{k},\sigma}^{*}}{\mathcal{F}_{\sigma}(-\mathbf{k})} \\
            -\frac{\mathcal{F}_{\sigma}^{*}(-\mathbf{k}) u_{\mathbf{k},\sigma}^{*}}{[\mathcal{E}_{\sigma}^{\text{A}}(\mathbf{k}) +\omega_{\mathbf{k},\sigma}^{\beta}]}
        \end{cases}
        \label{vksigma}
    \end{equation}
\end{subequations}
\begin{subequations}
    \begin{equation}
        \begin{cases}
            [\mathcal{E}_{\tau}^{\text{A}}(\mathbf{k}) - \omega_{\mathbf{k},\tau}^{\alpha}] u_{\mathbf{k},\tau} + \mathcal{F}_{\tau}(\mathbf{k}) v_{\mathbf{k},\tau}^{*} = 0 \\
            \mathcal{F}_{\tau}^{*}(\mathbf{k}) u_{\mathbf{k},\tau} + [\mathcal{E}_{\tau}^{\text{B}}(\mathbf{k}) + \omega_{\mathbf{k},\tau}^{\alpha}] v_{\mathbf{k},\tau}^{*} = 0
        \end{cases}
        \Rightarrow v_{\mathbf{k},\tau}^{*} =  
        \begin{cases}
            -\frac{[\mathcal{E}_{\tau}^{\text{A}}(\mathbf{k}) - \omega_{\mathbf{k},\tau}^{\alpha}] u_{\mathbf{k},\tau}}{\mathcal{F}_{\tau}(\mathbf{k})} \\
            -\frac{\mathcal{F}_{\tau}^{*}(\mathbf{k}) u_{\mathbf{k},\tau}}{[\mathcal{E}_{\tau}^{\text{B}}(\mathbf{k}) + \omega_{\mathbf{k},\tau}^{\alpha}]}
        \end{cases}
        \label{vkstartau}
    \end{equation}
    \begin{equation}
        \begin{cases}
            [\mathcal{E}_{\tau}^{\text{B}}(\mathbf{k}) - \omega_{\mathbf{k},\tau}^{\beta}] u_{\mathbf{k},\tau}^{*} + \mathcal{F}_{\tau}(-\mathbf{k}) v_{\mathbf{k},\tau} = 0 \\
            \mathcal{F}_{\tau}^{*}(-\mathbf{k}) u_{\mathbf{k},\tau}^{*} + [\mathcal{E}_{\tau}^{\text{A}}(\mathbf{k}) +\omega_{\mathbf{k},\tau}^{\beta}] v_{\mathbf{k},\tau} = 0
        \end{cases}
        \Rightarrow v_{\mathbf{k},\tau} = 
        \begin{cases}
            -\frac{[\mathcal{E}_{\tau}^{\text{B}}(\mathbf{k}) - \omega_{\mathbf{k},\tau}^{\beta}] u_{\mathbf{k},\tau}^{*}}{\mathcal{F}_{\tau}(-\mathbf{k})} \\
            -\frac{\mathcal{F}_{\tau}^{*}(-\mathbf{k}) u_{\mathbf{k},\tau}^{*}}{[\mathcal{E}_{\tau}^{\text{A}}(\mathbf{k}) +\omega_{\mathbf{k},\tau}^{\beta}]}
        \end{cases}
        \label{vktau}
    \end{equation}
\end{subequations}
Given the dispersion relations in \textcolor{blue}{Eq.~\eqref{dispersion} of main manuscript}, let us now evaluate the following expressions:
\begin{subequations}
    \begin{gather}
        \mathcal{E}_{\sigma}^{\text{B}}(\mathbf{k}) + \omega_{\mathbf{k},\sigma}^{\alpha} = \mathcal{E}_{\sigma}^{+}(\mathbf{k}) + \sqrt{(\mathcal{E}_{\sigma}^{+}(\mathbf{k}))^{2} - \left|\mathcal{F}_{\sigma}(\mathbf{k})\right|^{2}} \\
        \mathcal{E}_{\sigma}^{\text{A}}(\mathbf{k}) + \omega_{\mathbf{k},\sigma}^{\beta} = \mathcal{E}_{\sigma}^{+}(\mathbf{k}) + \sqrt{(\mathcal{E}_{\sigma}^{+}(\mathbf{k}))^{2} - \left|\mathcal{F}_{\sigma}(-\mathbf{k})\right|^{2}} \\
        \mathcal{E}_{\sigma}^{\text{A}}(\mathbf{k}) - \omega_{\mathbf{k},\sigma}^{\alpha} = \mathcal{E}_{\sigma}^{+}(\mathbf{k}) - \sqrt{(\mathcal{E}_{\sigma}^{+}(\mathbf{k}))^{2} - \left|\mathcal{F}_{\sigma}(\mathbf{k})\right|^{2}} \\
        \mathcal{E}_{\sigma}^{\text{B}}(\mathbf{k}) - \omega_{\mathbf{k},\sigma}^{\beta} = \mathcal{E}_{\sigma}^{+}(\mathbf{k}) - \sqrt{(\mathcal{E}_{\sigma}^{+}(\mathbf{k}))^{2} - \left|\mathcal{F}_{\sigma}(-\mathbf{k})\right|^{2}}
    \end{gather}
    \label{dispersionsigma1}
\end{subequations}
\begin{subequations}
    \begin{gather}
        \mathcal{E}_{\tau}^{\text{B}}(\mathbf{k}) + \omega_{\mathbf{k},\tau}^{\alpha} = \mathcal{E}_{\tau}^{+}(\mathbf{k}) + \sqrt{(\mathcal{E}_{\tau}^{+}(\mathbf{k}))^{2} - \left|\mathcal{F}_{\tau}(\mathbf{k})\right|^{2}} \\
        \mathcal{E}_{\tau}^{\text{A}}(\mathbf{k}) + \omega_{\mathbf{k},\tau}^{\beta} = \mathcal{E}_{\tau}^{+}(\mathbf{k}) + \sqrt{(\mathcal{E}_{\tau}^{+}(\mathbf{k}))^{2} - \left|\mathcal{F}_{\tau}(-\mathbf{k})\right|^{2}} \\
        \mathcal{E}_{\tau}^{\text{A}}(\mathbf{k}) - \omega_{\mathbf{k},\tau}^{\alpha} = \mathcal{E}_{\tau}^{+}(\mathbf{k}) - \sqrt{(\mathcal{E}_{\tau}^{+}(\mathbf{k}))^{2} - \left|\mathcal{F}_{\tau}(\mathbf{k})\right|^{2}} \\
        \mathcal{E}_{\tau}^{\text{B}}(\mathbf{k}) - \omega_{\mathbf{k},\tau}^{\beta} = \mathcal{E}_{\tau}^{+}(\mathbf{k}) - \sqrt{(\mathcal{E}_{\tau}^{+}(\mathbf{k}))^{2} - \left|\mathcal{F}_{\tau}(-\mathbf{k})\right|^{2}}
    \end{gather}
    \label{dispersiontau1}
\end{subequations}
Using Eqs.\eqref{vkstarsigma} and \eqref{vksigma}:
\begin{subequations}
    \begin{equation}
        \begin{gathered}
            v_{\mathbf{k},\sigma}^{*} = -\frac{\mathcal{F}_{\sigma}^{*}(\mathbf{k}) u_{\mathbf{k},\sigma}}{[\mathcal{E}_{\sigma}^{\text{B}}(\mathbf{k}) + \omega_{\mathbf{k},\sigma}^{\alpha}]} \Rightarrow (v_{\mathbf{k},\sigma}^{*})^{*} = -\frac{[\mathcal{F}_{\sigma}^{*}(\mathbf{k}) u_{\mathbf{k},\sigma}]^{*}}{[\mathcal{E}_{\sigma}^{\text{B}}(\mathbf{k}) + \omega_{\mathbf{k},\sigma}^{\alpha}]^{*}} \Rightarrow v_{\mathbf{k},\sigma} = -\frac{\mathcal{F}_{\sigma}(\mathbf{k}) u_{\mathbf{k},\sigma}^{*}}{[\mathcal{E}_{\sigma}^{\text{B}} + \omega_{\mathbf{k},\sigma}^{\alpha}]} \\
            \Rightarrow -\frac{\mathcal{F}_{\sigma}^{*}(-\mathbf{k}) u_{\mathbf{k},\sigma}^{*}}{[\mathcal{E}_{\sigma}^{\text{A}}(\mathbf{k}) +\omega_{\mathbf{k},\sigma}^{\beta}]} = -\frac{\mathcal{F}_{\sigma}(\mathbf{k}) u_{\mathbf{k},\sigma}^{*}}{[\mathcal{E}_{\sigma}^{\text{B}}(\mathbf{k}) + \omega_{\mathbf{k},\sigma}^{\alpha}]} \Rightarrow \frac{\mathcal{F}_{\sigma}^{*}(-\mathbf{k})}{[\mathcal{E}_{\sigma}^{\text{A}}(\mathbf{k}) +\omega_{\mathbf{k},\sigma}^{\beta}]} = \frac{\mathcal{F}_{\sigma}(\mathbf{k})}{[\mathcal{E}_{\sigma}^{\text{B}}(\mathbf{k}) + \omega_{\mathbf{k},\sigma}^{\alpha}]} \\
            \Rightarrow \frac{\mathcal{F}_{\sigma}(\mathbf{k})}{\mathcal{F}_{\sigma}^{*}(-\mathbf{k})} = \frac{\mathcal{E}_{\sigma}^{\text{B}}(\mathbf{k}) + \omega_{\mathbf{k},\sigma}^{\alpha}}{\mathcal{E}_{\sigma}^{\text{A}}(\mathbf{k}) + \omega_{\mathbf{k},\sigma}^{\beta}} \Rightarrow \left(\frac{\mathcal{F}_{\sigma}(\mathbf{k})}{\mathcal{F}_{\sigma}^{*}(-\mathbf{k})}\right)^{*} = \left(\frac{\mathcal{E}_{\sigma}^{\text{B}}(\mathbf{k}) + \omega_{\mathbf{k},\sigma}^{\alpha}}{\mathcal{E}_{\sigma}^{\text{A}}(\mathbf{k}) + \omega_{\mathbf{k},\sigma}^{\beta}}\right)^{*} \Rightarrow \frac{\mathcal{F}_{\sigma}^{*}(\mathbf{k})}{\mathcal{F}_{\sigma}(-\mathbf{k})} = \frac{\mathcal{E}_{\sigma}^{\text{B}}(\mathbf{k}) + \omega_{\mathbf{k},\sigma}^{\alpha}}{\mathcal{E}_{\sigma}^{\text{A}}(\mathbf{k}) + \omega_{\mathbf{k},\sigma}^{\beta}} \\
            \therefore \boxed{
            \frac{\mathcal{F}_{\sigma}(\mathbf{k})}{\mathcal{F}_{\sigma}^{*}(-\mathbf{k})} = \frac{\mathcal{E}_{\sigma}^{\text{B}}(\mathbf{k}) + \omega_{\mathbf{k},\sigma}^{\alpha}}{\mathcal{E}_{\sigma}^{\text{A}}(\mathbf{k}) + \omega_{\mathbf{k},\sigma}^{\beta}} = \frac{\mathcal{F}_{\sigma}^{*}(\mathbf{k})}{\mathcal{F}_{\sigma}(-\mathbf{k})}
            }
        \end{gathered}
        \label{pairingsfsigma1}
    \end{equation}
    \begin{equation}
        \begin{gathered}
            v_{\mathbf{k},\sigma}^{*} = -\frac{[\mathcal{E}_{\sigma}^{\text{A}}(\mathbf{k}) - \omega_{\mathbf{k},\sigma}^{\alpha}] u_{\mathbf{k},\sigma}}{\mathcal{F}_{\sigma}(\mathbf{k})} \Rightarrow (v_{\mathbf{k},\sigma}^{*})^{*} = -\frac{[\mathcal{E}_{\sigma}^{\text{A}}(\mathbf{k}) - \omega_{\mathbf{k},\sigma}^{\alpha}]^{*} u_{\mathbf{k},\sigma}^{*}}{\mathcal{F}_{\sigma}^{*}(\mathbf{k})} \Rightarrow v_{\mathbf{k},\sigma} = -\frac{[\mathcal{E}_{\sigma}^{\text{A}}(\mathbf{k}) - \omega_{\mathbf{k},\sigma}^{\alpha}] u_{\mathbf{k},\sigma}^{*}}{\mathcal{F}_{\sigma}^{*}(\mathbf{k})} \\
            \Rightarrow -\frac{[\mathcal{E}_{\sigma}^{\text{B}}(\mathbf{k}) - \omega_{\mathbf{k},\sigma}^{\beta}] u_{\mathbf{k},\sigma}^{*}}{\mathcal{F}_{\sigma}(-\mathbf{k})} = -\frac{[\mathcal{E}_{\sigma}^{\text{A}}(\mathbf{k}) - \omega_{\mathbf{k},\sigma}^{\alpha}] u_{\mathbf{k},\sigma}^{*}}{\mathcal{F}_{\sigma}^{*}(\mathbf{k})} \Rightarrow \frac{[\mathcal{E}_{\sigma}^{\text{B}}(\mathbf{k}) - \omega_{\mathbf{k},\sigma}^{\beta}]}{\mathcal{F}_{\sigma}(-\mathbf{k})} = \frac{[\mathcal{E}_{\sigma}^{\text{A}}(\mathbf{k}) - \omega_{\mathbf{k},\sigma}^{\alpha}]}{\mathcal{F}_{\sigma}^{*}(\mathbf{k})} \\
            \Rightarrow \frac{\mathcal{F}_{\sigma}^{*}(\mathbf{k})}{\mathcal{F}_{\sigma}(-\mathbf{k})} = \frac{\mathcal{E}_{\sigma}^{\text{A}}(\mathbf{k}) - \omega_{\mathbf{k},\sigma}^{\alpha}}{\mathcal{E}_{\sigma}^{\text{B}}(\mathbf{k}) - \omega_{\mathbf{k},\sigma}^{\beta}} \Rightarrow \left(\frac{\mathcal{F}_{\sigma}^{*}(\mathbf{k})}{\mathcal{F}_{\sigma}(-\mathbf{k})}\right)^{*} = \left(\frac{\mathcal{E}_{\sigma}^{\text{A}}(\mathbf{k}) - \omega_{\mathbf{k},\sigma}^{\alpha}}{\mathcal{E}_{\sigma}^{\text{B}}(\mathbf{k}) - \omega_{\mathbf{k},\sigma}^{\beta}}\right)^{*} \Rightarrow \frac{\mathcal{F}_{\sigma}(\mathbf{k})}{\mathcal{F}_{\sigma}^{*}(-\mathbf{k})} = \frac{\mathcal{E}_{\sigma}^{\text{A}}(\mathbf{k}) - \omega_{\mathbf{k},\sigma}^{\alpha}}{\mathcal{E}_{\sigma}^{\text{B}}(\mathbf{k}) - \omega_{\mathbf{k},\sigma}^{\beta}} \\
            \therefore \boxed{
            \frac{\mathcal{F}_{\sigma}^{*}(\mathbf{k})}{\mathcal{F}_{\sigma}(-\mathbf{k})} = \frac{\mathcal{E}_{\sigma}^{\text{A}}(\mathbf{k}) - \omega_{\mathbf{k},\sigma}^{\alpha}}{\mathcal{E}_{\sigma}^{\text{B}}(\mathbf{k}) - \omega_{\mathbf{k},\sigma}^{\beta}} = \frac{\mathcal{F}_{\sigma}(\mathbf{k})}{\mathcal{F}_{\sigma}^{*}(-\mathbf{k})}
            }
        \end{gathered}
        \label{pairingsfsigma2}
    \end{equation}
\end{subequations}
Similarly, using Eqs.\eqref{vkstartau} and \eqref{vktau}:
\begin{subequations}
    \begin{equation}
        \begin{gathered}
            v_{\mathbf{k},\tau}^{*} = -\frac{\mathcal{F}_{\tau}^{*}(\mathbf{k}) u_{\mathbf{k},\tau}}{[\mathcal{E}_{\tau}^{\text{B}}(\mathbf{k}) + \omega_{\mathbf{k},\tau}^{\alpha}]} \Rightarrow (v_{\mathbf{k},\tau}^{*})^{*} = -\frac{[\mathcal{F}_{\tau}^{*}(\mathbf{k}) u_{\mathbf{k},\tau}]^{*}}{[\mathcal{E}_{\tau}^{\text{B}}(\mathbf{k}) + \omega_{\mathbf{k},\tau}^{\alpha}]^{*}} \Rightarrow v_{\mathbf{k},\tau} = -\frac{\mathcal{F}_{\tau}(\mathbf{k}) u_{\mathbf{k},\tau}^{*}}{[\mathcal{E}_{\tau}^{\text{B}}(\mathbf{k}) + \omega_{\mathbf{k},\tau}^{\alpha}]} \\
            \Rightarrow -\frac{\mathcal{F}_{\tau}^{*}(-\mathbf{k}) u_{\mathbf{k},\tau}^{*}}{[\mathcal{E}_{\tau}^{\text{A}}(\mathbf{k}) +\omega_{\mathbf{k},\tau}^{\beta}]} = -\frac{\mathcal{F}_{\tau}(\mathbf{k}) u_{\mathbf{k},\tau}^{*}}{[\mathcal{E}_{\tau}^{\text{B}}(\mathbf{k}) + \omega_{\mathbf{k},\tau}^{\alpha}]} \Rightarrow \frac{\mathcal{F}_{\tau}^{*}(-\mathbf{k})}{[\mathcal{E}_{\tau}^{\text{A}}(\mathbf{k}) +\omega_{\mathbf{k},\tau}^{\beta}]} = \frac{\mathcal{F}_{\tau}(\mathbf{k})}{[\mathcal{E}_{\tau}^{\text{B}}(\mathbf{k}) + \omega_{\mathbf{k},\tau}^{\alpha}]} \\
            \Rightarrow \frac{\mathcal{F}_{\tau}(\mathbf{k})}{\mathcal{F}_{\tau}^{*}(-\mathbf{k})} = \frac{\mathcal{E}_{\tau}^{\text{B}}(\mathbf{k}) + \omega_{\mathbf{k},\tau}^{\alpha}}{\mathcal{E}_{\tau}^{\text{A}}(\mathbf{k}) + \omega_{\mathbf{k},\tau}^{\beta}} \Rightarrow \left(\frac{\mathcal{F}_{\tau}(\mathbf{k})}{\mathcal{F}_{\tau}^{*}(-\mathbf{k})}\right)^{*} = \left(\frac{\mathcal{E}_{\tau}^{\text{B}}(\mathbf{k}) + \omega_{\mathbf{k},\tau}^{\alpha}}{\mathcal{E}_{\tau}^{\text{A}}(\mathbf{k}) + \omega_{\mathbf{k},\tau}^{\beta}}\right)^{*} \Rightarrow \frac{\mathcal{F}_{\tau}^{*}(\mathbf{k})}{\mathcal{F}_{\tau}(-\mathbf{k})} = \frac{\mathcal{E}_{\tau}^{\text{B}}(\mathbf{k}) + \omega_{\mathbf{k},\tau}^{\alpha}}{\mathcal{E}_{\tau}^{\text{A}}(\mathbf{k}) + \omega_{\mathbf{k},\tau}^{\beta}} \\
            \therefore \boxed{
            \frac{\mathcal{F}_{\tau}(\mathbf{k})}{\mathcal{F}_{\tau}^{*}(-\mathbf{k})} = \frac{\mathcal{E}_{\tau}^{\text{B}}(\mathbf{k}) + \omega_{\mathbf{k},\tau}^{\alpha}}{\mathcal{E}_{\tau}^{\text{A}}(\mathbf{k}) + \omega_{\mathbf{k},\tau}^{\beta}} = \frac{\mathcal{F}_{\tau}^{*}(\mathbf{k})}{\mathcal{F}_{\tau}(-\mathbf{k})}
            }
        \end{gathered}
        \label{pairingsftau1}
    \end{equation}
    \begin{equation}
        \begin{gathered}
            v_{\mathbf{k},\tau}^{*} = -\frac{[\mathcal{E}_{\tau}^{\text{A}} - \omega_{\mathbf{k},\tau}^{\alpha}] u_{\mathbf{k},\tau}}{\mathcal{F}_{\tau}(\mathbf{k})} \Rightarrow (v_{\mathbf{k},\tau}^{*})^{*} = -\frac{[\mathcal{E}_{\tau}^{\text{A}}(\mathbf{k}) - \omega_{\mathbf{k},\tau}^{\alpha}]^{*} u_{\mathbf{k},\tau}^{*}}{\mathcal{F}_{\tau}^{*}(\mathbf{k})} \Rightarrow v_{\mathbf{k},\tau} = -\frac{[\mathcal{E}_{\tau}^{\text{A}}(\mathbf{k}) - \omega_{\mathbf{k},\tau}^{\alpha}] u_{\mathbf{k},\tau}^{*}}{\mathcal{F}_{\tau}^{*}(\mathbf{k})} \\
            \Rightarrow -\frac{[\mathcal{E}_{\tau}^{\text{B}}(\mathbf{k}) - \omega_{\mathbf{k},\tau}^{\beta}] u_{\mathbf{k},\tau}^{*}}{\mathcal{F}_{\tau}(-\mathbf{k})} = -\frac{[\mathcal{E}_{\tau}^{\text{A}}(\mathbf{k}) - \omega_{\mathbf{k},\tau}^{\alpha}] u_{\mathbf{k},\tau}^{*}}{\mathcal{F}_{\tau}^{*}(\mathbf{k})} \Rightarrow \frac{[\mathcal{E}_{\tau}^{\text{B}}(\mathbf{k}) - \omega_{\mathbf{k},\tau}^{\beta}]}{\mathcal{F}_{\tau}(-\mathbf{k})} = \frac{[\mathcal{E}_{\tau}^{\text{A}}(\mathbf{k}) - \omega_{\mathbf{k},\tau}^{\alpha}]}{\mathcal{F}_{\tau}^{*}(\mathbf{k})} \\
            \Rightarrow \frac{\mathcal{F}_{\tau}^{*}(\mathbf{k})}{\mathcal{F}_{\tau}(-\mathbf{k})} = \frac{\mathcal{E}_{\tau}^{\text{A}}(\mathbf{k}) - \omega_{\mathbf{k},\tau}^{\alpha}}{\mathcal{E}_{\tau}^{\text{B}}(\mathbf{k}) - \omega_{\mathbf{k},\tau}^{\beta}} \Rightarrow \left(\frac{\mathcal{F}_{\tau}^{*}(\mathbf{k})}{\mathcal{F}_{\tau}(-\mathbf{k})}\right)^{*} = \left(\frac{\mathcal{E}_{\tau}^{\text{A}}(\mathbf{k}) - \omega_{\mathbf{k},\tau}^{\alpha}}{\mathcal{E}_{\tau}^{\text{B}}(\mathbf{k}) - \omega_{\mathbf{k},\tau}^{\beta}}\right)^{*} \Rightarrow \frac{\mathcal{F}_{\tau}(\mathbf{k})}{\mathcal{F}_{\tau}^{*}(-\mathbf{k})} = \frac{\mathcal{E}_{\tau}^{\text{A}}(\mathbf{k}) - \omega_{\mathbf{k},\tau}^{\alpha}}{\mathcal{E}_{\tau}^{\text{B}}(\mathbf{k}) - \omega_{\mathbf{k},\tau}^{\beta}} \\
            \therefore \boxed{
            \frac{\mathcal{F}_{\tau}^{*}(\mathbf{k})}{\mathcal{F}_{\tau}(-\mathbf{k})} = \frac{\mathcal{E}_{\tau}^{\text{A}}(\mathbf{k}) - \omega_{\mathbf{k},\tau}^{\alpha}}{\mathcal{E}_{\tau}^{\text{B}}(\mathbf{k}) - \omega_{\mathbf{k},\tau}^{\beta}} = \frac{\mathcal{F}_{\tau}(\mathbf{k})}{\mathcal{F}_{\tau}^{*}(-\mathbf{k})}
            }
        \end{gathered}
        \label{pairingsftau2}
    \end{equation}
\end{subequations}
Equating Eqs.\eqref{pairingsfsigma1} and \eqref{pairingsfsigma2} by substituting Eq.\eqref{dispersionsigma1} and using Componendo-Dividendo rules, we have:
\begin{gather*}
    \frac{\mathcal{E}_{\sigma}^{\text{B}}(\mathbf{k}) + \omega_{\mathbf{k},\sigma}^{\alpha}}{\mathcal{E}_{\sigma}^{\text{A}}(\mathbf{k}) + \omega_{\mathbf{k},\sigma}^{\beta}} = \frac{\mathcal{E}_{\sigma}^{\text{A}}(\mathbf{k}) - \omega_{\mathbf{k},\sigma}^{\alpha}}{\mathcal{E}_{\sigma}^{\text{B}}(\mathbf{k}) - \omega_{\mathbf{k},\sigma}^{\beta}} \Rightarrow \frac{\mathcal{E}_{\sigma}^{\text{B}}(\mathbf{k}) + \omega_{\mathbf{k},\sigma}^{\alpha}}{\mathcal{E}_{\sigma}^{\text{A}}(\mathbf{k}) - \omega_{\mathbf{k},\sigma}^{\alpha}} = \frac{\mathcal{E}_{\sigma}^{\text{A}}(\mathbf{k}) + \omega_{\mathbf{k},\sigma}^{\beta}}{\mathcal{E}_{\sigma}^{\text{B}}(\mathbf{k}) - \omega_{\mathbf{k},\sigma}^{\beta}} \\
    \Rightarrow \frac{\mathcal{E}_{\sigma}^{+}(\mathbf{k}) + \sqrt{(\mathcal{E}_{\sigma}^{+}(\mathbf{k}))^{2} - \left|\mathcal{F}_{\sigma}(\mathbf{k})\right|^{2}}}{\mathcal{E}_{\sigma}^{+}(\mathbf{k}) - \sqrt{(\mathcal{E}_{\sigma}^{+}(\mathbf{k}))^{2} - \left|\mathcal{F}_{\sigma}(\mathbf{k})\right|^{2}}} = \frac{\mathcal{E}_{\sigma}^{+}(\mathbf{k}) + \sqrt{(\mathcal{E}_{\sigma}^{+}(\mathbf{k}))^{2} + \left|\mathcal{F}_{\sigma}(-\mathbf{k})\right|^{2}}}{\mathcal{E}_{\sigma}^{+}(\mathbf{k}) - \sqrt{(\mathcal{E}_{\sigma}^{+}(\mathbf{k}))^{2} - \left|\mathcal{F}_{\sigma}(-\mathbf{k})\right|^{2}}} \\
    \begin{multlined}
        \Rightarrow \frac{\left[\mathcal{E}_{\sigma}^{+}(\mathbf{k}) + \sqrt{(\mathcal{E}_{\sigma}^{+}(\mathbf{k}))^{2} - \left|\mathcal{F}_{\sigma}(\mathbf{k})\right|^{2}}\right] + \left[\mathcal{E}_{\sigma}^{+}(\mathbf{k}) - \sqrt{(\mathcal{E}_{\sigma}^{+}(\mathbf{k}))^{2} - \left|\mathcal{F}_{\sigma}(\mathbf{k})\right|^{2}}\right]}{\left[\mathcal{E}_{\sigma}^{+}(\mathbf{k}) + \sqrt{(\mathcal{E}_{\sigma}^{+}(\mathbf{k}))^{2} - \left|\mathcal{F}_{\sigma}(\mathbf{k})\right|^{2}}\right] - \left[\mathcal{E}_{\sigma}^{+}(\mathbf{k}) - \sqrt{(\mathcal{E}_{\sigma}^{+}(\mathbf{k}))^{2} - \left|\mathcal{F}_{\sigma}(\mathbf{k})\right|^{2}}\right]} \\
        = \frac{\left[\mathcal{E}_{\sigma}^{+}(\mathbf{k}) + \sqrt{(\mathcal{E}_{\sigma}^{+}(\mathbf{k}))^{2} - \left|\mathcal{F}_{\sigma}(-\mathbf{k})\right|^{2}}\right] + \left[\mathcal{E}_{\sigma}^{+}(\mathbf{k}) - \sqrt{(\mathcal{E}_{\sigma}^{+}(\mathbf{k}))^{2} - \left|\mathcal{F}_{\sigma}(-\mathbf{k})\right|^{2}}\right]}{\left[\mathcal{E}_{\sigma}^{+}(\mathbf{k}) + \sqrt{(\mathcal{E}_{\sigma}^{+}(\mathbf{k}))^{2} - \left|\mathcal{F}_{\sigma}(-\mathbf{k})\right|^{2}}\right] - \left[\mathcal{E}_{\sigma}^{+}(\mathbf{k}) - \sqrt{(\mathcal{E}_{\sigma}^{+}(\mathbf{k}))^{2} - \left|\mathcal{F}_{\sigma}(-\mathbf{k})\right|^{2}}\right]}
    \end{multlined} \\
    \Rightarrow \frac{2\mathcal{E}_{\sigma}^{+}(\mathbf{k})}{2\sqrt{(\mathcal{E}_{\sigma}^{+}(\mathbf{k}))^{2} - \left|\mathcal{F}_{\sigma}(\mathbf{k})\right|^{2}}} = \frac{2\mathcal{E}_{\sigma}^{+}(\mathbf{k})}{2\sqrt{(\mathcal{E}_{\sigma}^{+}(\mathbf{k}))^{2} - \left|\mathcal{F}_{\sigma}(-\mathbf{k})\right|^{2}}} \Rightarrow \sqrt{(\mathcal{E}_{\sigma}^{+}(\mathbf{k}))^{2} - \left|\mathcal{F}_{\sigma}(\mathbf{k})\right|^{2}} = \sqrt{(\mathcal{E}_{\sigma}^{+}(\mathbf{k}))^{2} - \left|\mathcal{F}_{\sigma}(-\mathbf{k})\right|^{2}} \\
    \Rightarrow (\mathcal{E}_{\sigma}^{+}(\mathbf{k}))^{2} - \left|\mathcal{F}_{\sigma}(\mathbf{k})\right|^{2} = (\mathcal{E}_{\sigma}^{+}(\mathbf{k}))^{2} - \left|\mathcal{F}_{\sigma}(-\mathbf{k})\right|^{2}
\end{gather*}
\begin{equation}
    \therefore \boxed{
    \left|\mathcal{F}_{\sigma}(\mathbf{k})\right|^{2} = \left|\mathcal{F}_{\sigma}(-\mathbf{k})\right|^{2}
    }
    \label{pairingsfsigma3}
\end{equation}
Similarly, equating Eqs.\eqref{pairingsftau1} and \eqref{pairingsftau2} by substituting Eq.\eqref{dispersiontau1} and using Componendo-Dividendo rules, we have:
\begin{gather*}
    \frac{\mathcal{E}_{\tau}^{\text{B}}(\mathbf{k}) + \omega_{\mathbf{k},\tau}^{\alpha}}{\mathcal{E}_{\tau}^{\text{A}}(\mathbf{k}) + \omega_{\mathbf{k},\tau}^{\beta}} = \frac{\mathcal{E}_{\tau}^{\text{A}}(\mathbf{k}) - \omega_{\mathbf{k},\tau}^{\alpha}}{\mathcal{E}_{\tau}^{\text{B}}(\mathbf{k}) - \omega_{\mathbf{k},\tau}^{\beta}} \Rightarrow \frac{\mathcal{E}_{\tau}^{\text{B}}(\mathbf{k}) + \omega_{\mathbf{k},\tau}^{\alpha}}{\mathcal{E}_{\tau}^{\text{A}}(\mathbf{k}) - \omega_{\mathbf{k},\tau}^{\alpha}} = \frac{\mathcal{E}_{\tau}^{\text{A}}(\mathbf{k}) + \omega_{\mathbf{k},\tau}^{\beta}}{\mathcal{E}_{\tau}^{\text{B}}(\mathbf{k}) - \omega_{\mathbf{k},\tau}^{\beta}} \\
    \Rightarrow \frac{\mathcal{E}_{\tau}^{+}(\mathbf{k}) + \sqrt{(\mathcal{E}_{\tau}^{+}(\mathbf{k}))^{2} - \left|\mathcal{F}_{\tau}(\mathbf{k})\right|^{2}}}{\mathcal{E}_{\tau}^{+}(\mathbf{k}) - \sqrt{(\mathcal{E}_{\tau}^{+}(\mathbf{k}))^{2} - \left|\mathcal{F}_{\tau}(\mathbf{k})\right|^{2}}} = \frac{\mathcal{E}_{\tau}^{+}(\mathbf{k}) + \sqrt{(\mathcal{E}_{\tau}^{+}(\mathbf{k}))^{2} + \left|\mathcal{F}_{\tau}(-\mathbf{k})\right|^{2}}}{\mathcal{E}_{\tau}^{+}(\mathbf{k}) - \sqrt{(\mathcal{E}_{\tau}^{+}(\mathbf{k}))^{2} - \left|\mathcal{F}_{\tau}(-\mathbf{k})\right|^{2}}} \\
    \begin{multlined}
        \Rightarrow \frac{\left[\mathcal{E}_{\tau}^{+}(\mathbf{k}) + \sqrt{(\mathcal{E}_{\tau}^{+}(\mathbf{k}))^{2} - \left|\mathcal{F}_{\tau}(\mathbf{k})\right|^{2}}\right] + \left[\mathcal{E}_{\tau}^{+}(\mathbf{k}) - \sqrt{(\mathcal{E}_{\tau}^{+}(\mathbf{k}))^{2} - \left|\mathcal{F}_{\tau}(\mathbf{k})\right|^{2}}\right]}{\left[\mathcal{E}_{\tau}^{+}(\mathbf{k}) + \sqrt{(\mathcal{E}_{\tau}^{+}(\mathbf{k}))^{2} - \left|\mathcal{F}_{\tau}(\mathbf{k})\right|^{2}}\right] - \left[\mathcal{E}_{\tau}^{+}(\mathbf{k}) - \sqrt{(\mathcal{E}_{\tau}^{+}(\mathbf{k}))^{2} - \left|\mathcal{F}_{\tau}(\mathbf{k})\right|^{2}}\right]} \\
        = \frac{\left[\mathcal{E}_{\tau}^{+}(\mathbf{k}) + \sqrt{(\mathcal{E}_{\tau}^{+}(\mathbf{k}))^{2} - \left|\mathcal{F}_{\tau}(-\mathbf{k})\right|^{2}}\right] + \left[\mathcal{E}_{\tau}^{+}(\mathbf{k}) - \sqrt{(\mathcal{E}_{\tau}^{+}(\mathbf{k}))^{2} - \left|\mathcal{F}_{\tau}(-\mathbf{k})\right|^{2}}\right]}{\left[\mathcal{E}_{\tau}^{+}(\mathbf{k}) + \sqrt{(\mathcal{E}_{\tau}^{+}(\mathbf{k}))^{2} - \left|\mathcal{F}_{\tau}(-\mathbf{k})\right|^{2}}\right] - \left[\mathcal{E}_{\tau}^{+}(\mathbf{k}) - \sqrt{(\mathcal{E}_{\tau}^{+}(\mathbf{k}))^{2} - \left|\mathcal{F}_{\tau}(-\mathbf{k})\right|^{2}}\right]}
    \end{multlined} \\
    \Rightarrow \frac{2\mathcal{E}_{\tau}^{+}(\mathbf{k})}{2\sqrt{(\mathcal{E}_{\tau}^{+}(\mathbf{k}))^{2} - \left|\mathcal{F}_{\tau}(\mathbf{k})\right|^{2}}} = \frac{2\mathcal{E}_{\tau}^{+}(\mathbf{k})}{2\sqrt{(\mathcal{E}_{\tau}^{+}(\mathbf{k}))^{2} - \left|\mathcal{F}_{\tau}(-\mathbf{k})\right|^{2}}} \Rightarrow \sqrt{(\mathcal{E}_{\tau}^{+}(\mathbf{k}))^{2} - \left|\mathcal{F}_{\tau}(\mathbf{k})\right|^{2}} = \sqrt{(\mathcal{E}_{\tau}^{+}(\mathbf{k}))^{2} - \left|\mathcal{F}_{\tau}(-\mathbf{k})\right|^{2}} \\
    \Rightarrow (\mathcal{E}_{\tau}^{+}(\mathbf{k}))^{2} - \left|\mathcal{F}_{\tau}(\mathbf{k})\right|^{2} = (\mathcal{E}_{\tau}^{+}(\mathbf{k}))^{2} - \left|\mathcal{F}_{\tau}(-\mathbf{k})\right|^{2}
\end{gather*}
\begin{equation}
    \therefore \boxed{
    \left|\mathcal{F}_{\tau}(\mathbf{k})\right|^{2} = \left|\mathcal{F}_{\tau}(-\mathbf{k})\right|^{2}
    }
    \label{pairingsftau3}
\end{equation}
Thus, one finally has:
\begin{gather}
    \boxed{\omega_{\mathbf{k},\sigma}^{\alpha} = \omega_{-\mathbf{k},\sigma}^{\alpha}} \;;\; \boxed{\omega_{\mathbf{k},\sigma}^{\beta} = \omega_{-\mathbf{k},\sigma}^{\beta}} \\
    \boxed{\omega_{\mathbf{k},\tau}^{\alpha} = \omega_{-\mathbf{k},\tau}^{\alpha}} \;;\; \boxed{\omega_{\mathbf{k},\tau}^{\beta} = \omega_{-\mathbf{k},\tau}^{\beta}}
\end{gather}
Eqs.~\eqref{dispersionsigma1}, \eqref{pairingsfsigma1}, \eqref{pairingsfsigma2} and \eqref{pairingsfsigma3} imply the following relation between the two magnon eigenmodes:
\begin{equation}
    \begin{drcases}
        \mathcal{E}_{\sigma}^{\text{A}} + \omega_{\mathbf{k},\sigma}^{\beta} = \mathcal{E}_{\sigma}^{\text{B}} + \omega_{\mathbf{k},\sigma}^{\alpha} \\
        \mathcal{E}_{\sigma}^{\text{B}} - \omega_{\mathbf{k},\sigma}^{\beta} = \mathcal{E}_{\sigma}^{\text{A}} - \omega_{\mathbf{k},\sigma}^{\alpha}
    \end{drcases}
    \implies \boxed{
    \mathcal{F}_{\sigma}^{*}(\mathbf{k}) = \mathcal{F}_{\sigma}(-\mathbf{k})
    }
\end{equation}
Eqs.~\eqref{dispersiontau1}, \eqref{pairingsftau1}, \eqref{pairingsftau2} and \eqref{pairingsftau3} imply the following relation between the two orbiton eigenmodes:
\begin{equation}
    \begin{drcases}
        \mathcal{E}_{\tau}^{\text{A}} + \omega_{\mathbf{k},\tau}^{\beta} = \mathcal{E}_{\tau}^{\text{B}} + \omega_{\mathbf{k},\tau}^{\alpha} \\
        \mathcal{E}_{\tau}^{\text{B}} - \omega_{\mathbf{k},\tau}^{\beta} = \mathcal{E}_{\tau}^{\text{A}} - \omega_{\mathbf{k},\tau}^{\alpha}
    \end{drcases}
    \implies \boxed{
    \mathcal{F}_{\tau}^{*}(\mathbf{k}) = \mathcal{F}_{\tau}(-\mathbf{k})
    }
\end{equation}
Using \textcolor{blue}{Eq.~\eqref{Bcoeff} of main manuscript}, one obtains the following relations between Bogolyubov coefficients in our chosen gauge (see Sec.~\ref{gauge}):
\begin{subequations}
    \begin{gather}
        u_{\mathbf{k},\sigma} = u_{-\mathbf{k},\sigma} \;;\; v_{\mathbf{k},\sigma}^{*} = v_{-\mathbf{k},\sigma} \\
        u_{\mathbf{k},\tau} = u_{-\mathbf{k},\tau} \;;\; v_{\mathbf{k},\tau}^{*} = v_{-\mathbf{k},\tau}
    \end{gather}
\end{subequations}

\section{Quadratic Form HP Operators and Bogolyubov Coefficients}

The expectation values of quadratic form HP operators defined in \textcolor{blue}{Eq.~\eqref{HPavg} of main manuscript} and Sec.~\ref{expct_val} are defined with respect to the ground state of $\hat{\mathcal{H}}$, i.e., the Bogolyubov vacuum. Let us now express these expectation values with Bogolyubov coefficients:
\begin{subequations}
    \begin{equation}
        \label{rho_sA_k}
        \begin{gathered}
            \varrho_{\mathbf{k},\sigma}^{\text{A}} = \langle\fnasigma{\mathbf{k}}{\mathbf{k}}\rangle \Rightarrow \varrho_{\mathbf{k},\sigma}^{\text{A}} = \langle(u_{\mathbf{k},\sigma} \fbcreatesa[\mathbf{k}] + v_{\mathbf{k},\sigma} \fbannihilatesb[-\mathbf{k}])(u_{\mathbf{k},\sigma} \fbannihilatesa[\mathbf{k}] + v_{-\mathbf{k},\sigma} \fbcreatesb[-\mathbf{k}])\rangle \\
            \Rightarrow \varrho_{\mathbf{k},\sigma}^{\text{A}} = \langle u_{\mathbf{k},\sigma}^{2} \fbcreatesa[\mathbf{k}]\fbannihilatesa[\mathbf{k}] + v_{\mathbf{k},\sigma}u_{\mathbf{k},\sigma} \fbannihilatesb[-\mathbf{k}]\fbannihilatesa[\mathbf{k}] + u_{\mathbf{k},\sigma}v_{-\mathbf{k},\sigma} \fbcreatesa[\mathbf{k}]\fbcreatesb[-\mathbf{k}] + |v_{\mathbf{k},\sigma}|^{2} \fbannihilatesb[-\mathbf{k}]\fbcreatesb[-\mathbf{k}]\rangle \\
            \Rightarrow \varrho_{\mathbf{k},\sigma}^{\text{A}} = u_{\mathbf{k},\sigma}^{2} \langle\fbcreatesa[\mathbf{k}]\fbannihilatesa[\mathbf{k}]\rangle + v_{\mathbf{k},\sigma}u_{\mathbf{k},\sigma} \langle\fbannihilatesb[-\mathbf{k}]\fbannihilatesa[\mathbf{k}]\rangle + u_{\mathbf{k},\sigma}v_{-\mathbf{k},\sigma} \langle\fbcreatesa[\mathbf{k}]\fbcreatesb[-\mathbf{k}]\rangle + |v_{\mathbf{k},\sigma}|^{2} \langle\fbannihilatesb[-\mathbf{k}]\fbcreatesb[-\mathbf{k}]\rangle \\
            \therefore \boxed{
            \varrho_{\mathbf{k},\sigma}^{\text{A}} = |v_{\mathbf{k},\sigma}|^{2}
            }
        \end{gathered}
    \end{equation}
    \begin{equation}
        \label{rho_sB_k}
        \begin{gathered}
            \varrho_{\mathbf{k},\sigma}^{\text{B}} = \langle\fnbsigma{\mathbf{k}}{\mathbf{k}}\rangle \Rightarrow \varrho_{\mathbf{k},\sigma}^{\text{B}} = \langle(u_{\mathbf{k},\sigma} \fbcreatesb[\mathbf{k}] + v_{-\mathbf{k},\sigma} \fbannihilatesa[-\mathbf{k}])(u_{\mathbf{k},\sigma} \fbannihilatesb[\mathbf{k}] + v_{\mathbf{k},\sigma} \fbcreatesa[-\mathbf{k}])\rangle \\
            \Rightarrow \varrho_{\mathbf{k},\sigma}^{\text{B}} = \langle u_{\mathbf{k},\sigma}^{2} \fbcreatesb[\mathbf{k}]\fbannihilatesb[\mathbf{k}] + v_{-\mathbf{k},\sigma}u_{\mathbf{k},\sigma} \fbannihilatesa[-\mathbf{k}]\fbannihilatesb[\mathbf{k}] + u_{\mathbf{k},\sigma}v_{\mathbf{k},\sigma} \fbcreatesb[\mathbf{k}]\fbcreatesa[-\mathbf{k}] + |v_{\mathbf{k},\sigma}|^{2} \fbannihilatesa[-\mathbf{k}]\fbcreatesa[-\mathbf{k}]\rangle \\
            \Rightarrow \varrho_{\mathbf{k},\sigma}^{\text{B}} = u_{\mathbf{k},\sigma}^{2} \langle\fbcreatesb[\mathbf{k}]\fbannihilatesb[\mathbf{k}]\rangle + v_{-\mathbf{k},\sigma}u_{\mathbf{k},\sigma} \langle\fbannihilatesa[-\mathbf{k}]\fbannihilatesb[\mathbf{k}]\rangle + u_{\mathbf{k},\sigma}v_{\mathbf{k},\sigma} \langle\fbcreatesb[\mathbf{k}]\fbcreatesa[-\mathbf{k}]\rangle + |v_{\mathbf{k},\sigma}|^{2} \langle\fbannihilatesa[-\mathbf{k}]\fbcreatesa[-\mathbf{k}]\rangle \\
            \therefore \boxed{
            \varrho_{\mathbf{k},\sigma}^{\text{B}} = |v_{\mathbf{k},\sigma}|^{2}
            }
        \end{gathered}
    \end{equation}
    \begin{equation}
        \begin{gathered}
            \Delta_{\mathbf{k},\sigma} = \langle\fannihilatesa[\mathbf{k}]\fannihilatesb[-\mathbf{k}]\rangle \Rightarrow \Delta_{\mathbf{k},\sigma} = \langle(u_{\mathbf{k},\sigma} \fbannihilatesa[\mathbf{k}] + v_{-\mathbf{k},\sigma} \fbcreatesb[-\mathbf{k}])(u_{\mathbf{k},\sigma} \fbannihilatesb[-\mathbf{k}] + v_{-\mathbf{k},\sigma} \fbcreatesa[\mathbf{k}])\rangle \\
            \Rightarrow \Delta_{\mathbf{k},\sigma} = \langle u_{\mathbf{k},\sigma}^{2} \fbannihilatesa[\mathbf{k}]\fbannihilatesb[-\mathbf{k}] + v_{-\mathbf{k},\sigma}u_{\mathbf{k},\sigma} \fbcreatesb[-\mathbf{k}]\fbannihilatesb[-\mathbf{k}] + u_{\mathbf{k},\sigma}v_{-\mathbf{k},\sigma} \fbannihilatesa[\mathbf{k}]\fbcreatesa[\mathbf{k}] + v_{-\mathbf{k},\sigma}^{2} \fbcreatesb[-\mathbf{k}]\fbcreatesa[\mathbf{k}]\rangle \\
            \Rightarrow \Delta_{\mathbf{k},\sigma} = u_{\mathbf{k},\sigma}^{2} \langle\fbannihilatesa[\mathbf{k}]\fbannihilatesb[-\mathbf{k}]\rangle + v_{-\mathbf{k},\sigma}u_{\mathbf{k},\sigma} \langle\fbcreatesb[-\mathbf{k}]\fbannihilatesb[-\mathbf{k}]\rangle + u_{\mathbf{k},\sigma}v_{-\mathbf{k},\sigma} \langle\fbannihilatesa[\mathbf{k}]\fbcreatesa[\mathbf{k}]\rangle + v_{-\mathbf{k},\sigma}^{2} \langle\fbcreatesb[-\mathbf{k}]\fbcreatesa[\mathbf{k}]\rangle \\
            \therefore \boxed{
            \Delta_{\mathbf{k},\sigma} = u_{\mathbf{k},\sigma}v_{-\mathbf{k},\sigma}
            }
        \end{gathered}
    \end{equation}
    \begin{equation}
        \begin{gathered}
            \Delta_{\mathbf{k},\sigma}^{*} = \langle\fcreatesa[\mathbf{k}]\fcreatesb[-\mathbf{k}]\rangle \Rightarrow \Delta_{\mathbf{k},\sigma}^{*} = \langle(u_{\mathbf{k},\sigma} \fbcreatesa[\mathbf{k}] + v_{\mathbf{k},\sigma} \fbannihilatesb[-\mathbf{k}])(u_{\mathbf{k},\sigma} \fbcreatesb[-\mathbf{k}] + v_{\mathbf{k},\sigma} \fbannihilatesa[\mathbf{k}])\rangle \\
            \Rightarrow \Delta_{\mathbf{k},\sigma}^{*} = \langle u_{\mathbf{k},\sigma}^{2} \fbcreatesa[\mathbf{k}]\fbcreatesb[-\mathbf{k}] + v_{\mathbf{k},\sigma}u_{\mathbf{k},\sigma} \fbannihilatesb[-\mathbf{k}]\fbcreatesb[-\mathbf{k}] + u_{\mathbf{k},\sigma}v_{\mathbf{k},\sigma} \fbcreatesa[\mathbf{k}]\fbannihilatesa[\mathbf{k}] + v_{\mathbf{k},\sigma}^{2} \fbannihilatesb[-\mathbf{k}]\fbannihilatesa[\mathbf{k}]\rangle \\
            \Rightarrow \Delta_{\mathbf{k},\sigma}^{*} = u_{\mathbf{k},\sigma}^{2} \langle\fbcreatesa[\mathbf{k}]\fbcreatesb[-\mathbf{k}]\rangle + v_{\mathbf{k},\sigma}u_{\mathbf{k},\sigma} \langle\fbannihilatesb[-\mathbf{k}]\fbcreatesb[-\mathbf{k}]\rangle + u_{\mathbf{k},\sigma}v_{\mathbf{k},\sigma} \langle\fbcreatesa[\mathbf{k}]\fbannihilatesa[\mathbf{k}]\rangle + v_{\mathbf{k},\sigma}^{2} \langle\fbannihilatesb[-\mathbf{k}]\fbannihilatesa[\mathbf{k}]\rangle \\
            \therefore \boxed{
            \Delta_{\mathbf{k},\sigma}^{*} = u_{\mathbf{k},\sigma}v_{\mathbf{k},\sigma} = \Delta_{-\mathbf{k},\sigma}
            }
        \end{gathered}
    \end{equation}
    \begin{equation}
        \label{rho_tA_k}
        \begin{gathered}
            \varrho_{\mathbf{k},\tau}^{\text{A}} = \langle\fnatau{\mathbf{k}}{\mathbf{k}}\rangle \Rightarrow \varrho_{\mathbf{k},\tau}^{\text{A}} = \langle(u_{\mathbf{k},\tau} \fbcreateta[\mathbf{k}] + v_{\mathbf{k},\tau} \fbannihilatetb[-\mathbf{k}])(u_{\mathbf{k},\tau} \fbannihilateta[\mathbf{k}] + v_{-\mathbf{k},\tau} \fbcreatetb[-\mathbf{k}])\rangle = \varrho_{\mathbf{k},\tau} \\
            \Rightarrow \varrho_{\mathbf{k},\tau}^{\text{A}} = \langle u_{\mathbf{k},\tau}^{2} \fbcreateta[\mathbf{k}]\fbannihilateta[\mathbf{k}] + v_{\mathbf{k},\tau}u_{\mathbf{k},\tau} \fbannihilatetb[-\mathbf{k}]\fbannihilateta[\mathbf{k}] + u_{\mathbf{k},\tau}v_{-\mathbf{k},\tau} \fbcreateta[\mathbf{k}]\fbcreatetb[-\mathbf{k}] + |v_{\mathbf{k},\tau}|^{2} \fbannihilatetb[-\mathbf{k}]\fbcreatetb[-\mathbf{k}]\rangle \\
            \Rightarrow \varrho_{\mathbf{k},\tau}^{\text{A}} = u_{\mathbf{k},\tau}^{2} \langle\fbcreateta[\mathbf{k}]\fbannihilateta[\mathbf{k}]\rangle + v_{\mathbf{k},\tau}u_{\mathbf{k},\tau} \langle\fbannihilatetb[-\mathbf{k}]\fbannihilateta[\mathbf{k}]\rangle + u_{\mathbf{k},\tau}v_{-\mathbf{k},\tau} \langle\fbcreateta[\mathbf{k}]\fbcreatetb[-\mathbf{k}]\rangle + |v_{\mathbf{k},\tau}|^{2} \langle\fbannihilatetb[-\mathbf{k}]\fbcreatetb[-\mathbf{k}]\rangle \\
            \therefore \boxed{
            \varrho_{\mathbf{k},\tau}^{\text{A}} = |v_{\mathbf{k},\tau}|^{2}
            }
        \end{gathered}
    \end{equation}
    \begin{equation}
        \label{rho_tB_k}
        \begin{gathered}
            \varrho_{\mathbf{k},\tau}^{\text{B}} = \langle\fnbtau{\mathbf{k}}{\mathbf{k}}\rangle \Rightarrow \varrho_{\mathbf{k},\tau}^{\text{B}} = \langle(u_{\mathbf{k},\tau} \fbcreatetb[\mathbf{k}] + v_{-\mathbf{k},\tau} \fbannihilateta[-\mathbf{k}])(u_{\mathbf{k},\tau} \fbannihilatetb[\mathbf{k}] + v_{\mathbf{k},\tau} \fbcreateta[-\mathbf{k}])\rangle \\
            \Rightarrow \varrho_{\mathbf{k},\tau}^{\text{B}} = \langle u_{\mathbf{k},\tau}^{2} \fbcreatetb[\mathbf{k}]\fbannihilatetb[\mathbf{k}] + v_{-\mathbf{k},\tau}u_{\mathbf{k},\tau} \fbannihilateta[-\mathbf{k}]\fbannihilatetb[\mathbf{k}] + u_{\mathbf{k},\tau}v_{\mathbf{k},\tau} \fbcreatetb[\mathbf{k}]\fbcreateta[-\mathbf{k}] + |v_{\mathbf{k},\tau}|^{2} \fbannihilateta[-\mathbf{k}]\fbcreateta[-\mathbf{k}]\rangle \\
            \Rightarrow \varrho_{\mathbf{k},\tau}^{\text{B}} = u_{\mathbf{k},\tau}^{2} \langle\fbcreatetb[\mathbf{k}]\fbannihilatetb[\mathbf{k}]\rangle + v_{-\mathbf{k},\tau}u_{\mathbf{k},\tau} \langle\fbannihilateta[-\mathbf{k}]\fbannihilatetb[\mathbf{k}]\rangle + u_{\mathbf{k},\tau}v_{\mathbf{k},\tau} \langle\fbcreatetb[\mathbf{k}]\fbcreateta[-\mathbf{k}]\rangle + |v_{\mathbf{k},\tau}|^{2} \langle\fbannihilateta[-\mathbf{k}]\fbcreateta[-\mathbf{k}]\rangle \\
            \therefore \boxed{
            \varrho_{\mathbf{k},\tau}^{\text{B}} = |v_{\mathbf{k},\tau}|^{2}
            }
        \end{gathered}
    \end{equation}
    \begin{equation}
        \begin{gathered}
            \Delta_{\mathbf{k},\tau} = \langle\fannihilateta[\mathbf{k}]\fannihilatetb[-\mathbf{k}]\rangle \Rightarrow \Delta_{\mathbf{k},\tau} = \langle(u_{\mathbf{k},\tau} \fbannihilateta[\mathbf{k}] + v_{-\mathbf{k},\tau} \fbcreatetb[-\mathbf{k}])(u_{\mathbf{k},\tau} \fbannihilatetb[-\mathbf{k}] + v_{-\mathbf{k},\tau} \fbcreateta[\mathbf{k}])\rangle \\
            \Rightarrow \Delta_{\mathbf{k},\tau} = \langle u_{\mathbf{k},\tau}^{2} \fbannihilateta[\mathbf{k}]\fbannihilatetb[-\mathbf{k}] + v_{-\mathbf{k},\tau}u_{\mathbf{k},\tau} \fbcreatetb[-\mathbf{k}]\fbannihilatetb[-\mathbf{k}] + u_{\mathbf{k},\tau}v_{-\mathbf{k},\tau} \fbannihilateta[\mathbf{k}]\fbcreateta[\mathbf{k}] + v_{-\mathbf{k},\tau}^{2} \fbcreatetb[-\mathbf{k}]\fbcreateta[\mathbf{k}]\rangle \\
            \Rightarrow \Delta_{\mathbf{k},\tau} = u_{\mathbf{k},\tau}^{2} \langle\fbannihilateta[\mathbf{k}]\fbannihilatetb[-\mathbf{k}]\rangle + v_{-\mathbf{k},\tau}u_{\mathbf{k},\tau} \langle\fbcreatetb[-\mathbf{k}]\fbannihilatetb[-\mathbf{k}]\rangle + u_{\mathbf{k},\tau}v_{-\mathbf{k},\tau} \langle\fbannihilateta[\mathbf{k}]\fbcreateta[\mathbf{k}]\rangle + v_{-\mathbf{k},\tau}^{2} \langle\fbcreatetb[-\mathbf{k}]\fbcreateta[\mathbf{k}]\rangle \\
            \therefore \boxed{
            \Delta_{\mathbf{k},\tau} = u_{\mathbf{k},\tau}v_{-\mathbf{k},\tau}
            }
        \end{gathered}
    \end{equation}
    \begin{equation}
        \begin{gathered}
            \Delta_{\mathbf{k},\tau}^{*} = \langle\fcreateta[\mathbf{k}]\fcreatetb[-\mathbf{k}]\rangle \Rightarrow \Delta_{\mathbf{k},\tau}^{*} = \langle(u_{\mathbf{k},\tau} \fbcreateta[\mathbf{k}] + v_{\mathbf{k},\tau} \fbannihilatetb[-\mathbf{k}])(u_{\mathbf{k},\tau} \fbcreatetb[-\mathbf{k}] + v_{\mathbf{k},\tau} \fbannihilateta[\mathbf{k}])\rangle \\
            \Rightarrow \Delta_{\mathbf{k},\tau}^{*} = \langle u_{\mathbf{k},\tau}^{2} \fbcreateta[\mathbf{k}]\fbcreatetb[-\mathbf{k}] + v_{\mathbf{k},\tau}u_{\mathbf{k},\tau} \fbannihilatetb[-\mathbf{k}]\fbcreatetb[-\mathbf{k}] + u_{\mathbf{k},\tau}v_{\mathbf{k},\tau} \fbcreateta[\mathbf{k}]\fbannihilateta[\mathbf{k}] + v_{\mathbf{k},\tau}^{2} \fbannihilatetb[-\mathbf{k}]\fbannihilateta[\mathbf{k}]\rangle \\
            \Rightarrow \Delta_{\mathbf{k},\tau}^{*} = u_{\mathbf{k},\tau}^{2} \langle\fbcreateta[\mathbf{k}]\fbcreatetb[-\mathbf{k}]\rangle + v_{\mathbf{k},\tau}u_{\mathbf{k},\tau} \langle\fbannihilatetb[-\mathbf{k}]\fbcreatetb[-\mathbf{k}]\rangle + u_{\mathbf{k},\tau}v_{\mathbf{k},\tau} \langle\fbcreateta[\mathbf{k}]\fbannihilateta[\mathbf{k}]\rangle + v_{\mathbf{k},\tau}^{2} \langle\fbannihilatetb[-\mathbf{k}]\fbannihilateta[\mathbf{k}]\rangle \\
            \therefore \boxed{
            \Delta_{\mathbf{k},\tau}^{*} = u_{\mathbf{k},\tau}v_{\mathbf{k},\tau} = \Delta_{-\mathbf{k},\tau}
            }
        \end{gathered}
    \end{equation}
\end{subequations}

\section{Effect of Symmetry on the General Coefficients of HP Operators}

The 2D decorated square lattice considered in the main manuscript belongs to $p4mm$ wallpaper group. The momentum space symmetries, as outlined in \textcolor{blue}{Eq.~\eqref{ksym} of main manuscript}, has the following effects:
\begin{subequations}
    \begin{gather}
        \gamma_{\mathbf{k}}^{*} = \frac{1}{\mathit{n}} \sum_{\boldsymbol{\xi}} e^{-\mathrm{i} \mathbf{k} \cdot \boldsymbol{\xi}} = \gamma_{-\mathbf{k}} = \gamma_{\mathbf{k}} \Rightarrow \boxed{\gamma_{\mathbf{k}} \in \mathbb{R}} \\
        \gamma_{\mathbf{k},\sigma}' = \frac{1}{\mathit{n}} \sum_{\boldsymbol{\xi}} e^{+\mathrm{i} \mathbf{k} \cdot \boldsymbol{\xi}} \Delta_{\boldsymbol{\xi},\sigma}^{*} = \Delta_{\sigma} \frac{1}{\mathit{n}} \sum_{\boldsymbol{\xi}} e^{+\mathrm{i} \mathbf{k} \cdot \boldsymbol{\xi}} \Rightarrow \boxed{\gamma_{\mathbf{k},\sigma}' = \Delta_{\sigma}\gamma_{\mathbf{k}}} \;;\; \gamma_{\mathbf{k},\sigma}' \in \mathbb{R} \Rightarrow v_{\mathbf{k},\sigma} \in \mathbb{R} \\
        \gamma_{\mathbf{k},\tau}' = \frac{1}{\mathit{n}} \sum_{\boldsymbol{\xi}} e^{+\mathrm{i} \mathbf{k} \cdot \boldsymbol{\xi}} \Delta_{\boldsymbol{\xi},\tau}^{*} = \Delta_{\tau} \frac{1}{\mathit{n}} \sum_{\boldsymbol{\xi}} e^{+\mathrm{i} \mathbf{k} \cdot \boldsymbol{\xi}} \Rightarrow \boxed{\gamma_{\mathbf{k},\tau}' = \Delta_{\tau}\gamma_{\mathbf{k}}} \;;\; \gamma_{\mathbf{k},\tau}' \in \mathbb{R} \Rightarrow v_{\mathbf{k},\tau} \in \mathbb{R} \\
        \boxed{\Delta_{\mathbf{k}',\sigma} = \Delta_{\mathbf{k},\sigma}} \;(\mathbf{k}' = R\mathbf{k}) \;;\; \Delta_{\mathbf{k},\sigma}^{*} = \Delta_{-\mathbf{k},\sigma} = \Delta_{\mathbf{k},\sigma} \Rightarrow \boxed{\Delta_{\mathbf{k},\sigma} \in \mathbb{R}} \\
        \boxed{\Delta_{\mathbf{k}',\tau} = \Delta_{\mathbf{k},\tau}} \;(\mathbf{k}' = R\mathbf{k}) \;;\; \Delta_{\mathbf{k},\tau}^{*} = \Delta_{-\mathbf{k},\tau} = \Delta_{\mathbf{k},\tau} \Rightarrow \boxed{\Delta_{\mathbf{k},\tau} \in \mathbb{R}} \\
        \begin{gathered}
            \Delta_{\boldsymbol{\xi},\sigma} =
            \begin{cases}
                \frac{2}{N_m} \sum_{\mathbf{k}} e^{+\mathrm{i} (k_{x}a + k_{y}a)} \Delta_{(k_{x},k_{y}),\sigma} \;&;\; \boldsymbol{\xi_{1}} = (a,a) \\
                \frac{2}{N_m} \sum_{\mathbf{k}} e^{+\mathrm{i} (k_{x}a - k_{y}a)} \Delta_{(k_{x},k_{y}),\sigma} \;&;\; \boldsymbol{\xi_{2}} = (a,-a) \\
                \frac{2}{N_m} \sum_{\mathbf{k}} e^{-\mathrm{i} (k_{x}a - k_{y}a)} \Delta_{(k_{x},k_{y}),\sigma} \;&;\; \boldsymbol{\xi_{3}} = (-a,a) \\
                \frac{2}{N_m} \sum_{\mathbf{k}} e^{-\mathrm{i} (k_{x}a + k_{y}a)} \Delta_{(k_{x},k_{y}),\sigma} \;&;\; \boldsymbol{\xi_{4}} = (-a,-a) \\
            \end{cases}
            =
            \begin{cases}
                \frac{2}{N_m} \sum_{\mathbf{k}} e^{+\mathrm{i} (k_{x}a + k_{y}a)} \Delta_{(k_{x},k_{y}),\sigma} \\
                \frac{2}{N_m} \sum_{\mathbf{k}} e^{+\mathrm{i} (k_{x}a + k_{y}a)} \Delta_{(k_{x},-k_{y}),\sigma} \\
                \frac{2}{N_m} \sum_{\mathbf{k}} e^{+\mathrm{i} (k_{x}a + k_{y}a)} \Delta_{(-k_{x},k_{y}),\sigma} \\
                \frac{2}{N_m} \sum_{\mathbf{k}} e^{+\mathrm{i} (k_{x}a + k_{y}a)} \Delta_{(-k_{x},-k_{y}),\sigma} \\
            \end{cases} \\
            \Rightarrow \boxed{\Delta_{\boldsymbol{\xi},\sigma} = \Delta_{\sigma}} \;\forall\; \boldsymbol{\xi}
        \end{gathered}
        \\
        \begin{gathered}
            \Delta_{\boldsymbol{\xi},\tau} =
            \begin{cases}
                \frac{2}{N_m} \sum_{\mathbf{k}} e^{+\mathrm{i} (k_{x}a + k_{y}a)} \Delta_{(k_{x},k_{y}),\tau} \;&;\; \boldsymbol{\xi_{1}} = (a,a) \\
                \frac{2}{N_m} \sum_{\mathbf{k}} e^{+\mathrm{i} (k_{x}a - k_{y}a)} \Delta_{(k_{x},k_{y}),\tau} \;&;\; \boldsymbol{\xi_{2}} = (a,-a) \\
                \frac{2}{N_m} \sum_{\mathbf{k}} e^{-\mathrm{i} (k_{x}a - k_{y}a)} \Delta_{(k_{x},k_{y}),\tau} \;&;\; \boldsymbol{\xi_{3}} = (-a,a) \\
                \frac{2}{N_m} \sum_{\mathbf{k}} e^{-\mathrm{i} (k_{x}a + k_{y}a)} \Delta_{(k_{x},k_{y}),\tau} \;&;\; \boldsymbol{\xi_{4}} = (-a,-a) \\
            \end{cases}
            =
            \begin{cases}
                \frac{2}{N_m} \sum_{\mathbf{k}} e^{+\mathrm{i} (k_{x}a + k_{y}a)} \Delta_{(k_{x},k_{y}),\tau} \\
                \frac{2}{N_m} \sum_{\mathbf{k}} e^{+\mathrm{i} (k_{x}a + k_{y}a)} \Delta_{(k_{x},-k_{y}),\tau} \\
                \frac{2}{N_m} \sum_{\mathbf{k}} e^{+\mathrm{i} (k_{x}a + k_{y}a)} \Delta_{(-k_{x},k_{y}),\tau} \\
                \frac{2}{N_m} \sum_{\mathbf{k}} e^{+\mathrm{i} (k_{x}a + k_{y}a)} \Delta_{(-k_{x},-k_{y}),\tau} \\
            \end{cases} \\
            \Rightarrow \boxed{\Delta_{\boldsymbol{\xi},\tau} = \Delta_{\tau}} \;\forall\; \boldsymbol{\xi}
        \end{gathered}
    \end{gather} \label{sym}
\end{subequations}

\section{Finite Temperature Magnon and Orbiton Number Density}

At finite temperature $T$, magnons and orbitons follow the Bose-Einstein distribution \cite{Bose1924_sm, Fazekas1999_sm}:
\begin{subequations}
    \begin{gather}
        \langle\fbcreatesa[\mathbf{k}]\fbannihilatesa[\mathbf{k}]\rangle = \frac{1}{e^{\sfrac{\omega_{\mathbf{k},\sigma}^{\alpha}}{k_{B}T}} - 1} = f(\omega_{\mathbf{k},\sigma}^{\alpha},T) \\
        \langle\fbcreatesb[\mathbf{k}]\fbannihilatesb[\mathbf{k}]\rangle = \frac{1}{e^{\sfrac{\omega_{\mathbf{k},\sigma}^{\beta}}{k_{B}T}} - 1} = f(\omega_{\mathbf{k},\sigma}^{\beta},T) \\
        \langle\fbcreateta[\mathbf{k}]\fbannihilateta[\mathbf{k}]\rangle = \frac{1}{e^{\sfrac{\omega_{\mathbf{k},\tau}^{\alpha}}{k_{B}T}} - 1} = f(\omega_{\mathbf{k},\tau}^{\alpha},T) \\
        \langle\fbcreatetb[\mathbf{k}]\fbannihilatetb[\mathbf{k}]\rangle = \frac{1}{e^{\sfrac{\omega_{\mathbf{k},\tau}^{\beta}}{k_{B}T}} - 1} = f(\omega_{\mathbf{k},\tau}^{\beta},T)
    \end{gather}
\end{subequations}
The number density of magnon and orbiton modes at $T = 0$ derived in Eqs.~\eqref{rho_sA_k}, \eqref{rho_sB_k}, \eqref{rho_tA_k} and \eqref{rho_tB_k} get modified at finite $T$ in the following way:
\begin{subequations}
    \begin{equation}
        \label{rho_sA_k_T}
        \begin{gathered}
            \varrho_{\mathbf{k},\sigma}^{\text{A}} = \langle\fnasigma{\mathbf{k}}{\mathbf{k}}\rangle \\
            \Rightarrow \varrho_{\mathbf{k},\sigma}^{\text{A}} = u_{\mathbf{k},\sigma}^{2} \langle\fbcreatesa[\mathbf{k}]\fbannihilatesa[\mathbf{k}]\rangle + |v_{\mathbf{k},\sigma}|^{2} \langle\fbannihilatesb[-\mathbf{k}]\fbcreatesb[-\mathbf{k}]\rangle \\
            \Rightarrow \varrho_{\mathbf{k},\sigma}^{\text{A}}(T) = u_{\mathbf{k},\sigma}^{2}  f(\omega_{\mathbf{k},\sigma}^{\alpha},T) + |v_{\mathbf{k},\sigma}|^{2} [1 + f(\omega_{-\mathbf{k},\sigma}^{\beta},T)] \\
            \Rightarrow \varrho_{\mathbf{k},\sigma}^{\text{A}}(T) = [1 + |v_{\mathbf{k},\sigma}|^{2}]  f(\omega_{\mathbf{k},\sigma}^{\alpha},T) + |v_{\mathbf{k},\sigma}|^{2} [1 + f(\omega_{\mathbf{k},\sigma}^{\beta},T)] \\
            \therefore \boxed{
            \varrho_{\mathbf{k},\sigma}^{\text{A}}(T) = f(\omega_{\mathbf{k},\sigma}^{\alpha},T) + |v_{\mathbf{k},\sigma}|^{2} [1 + f(\omega_{\mathbf{k},\sigma}^{\alpha},T) + f(\omega_{\mathbf{k},\sigma}^{\beta},T)]
            }
        \end{gathered}
    \end{equation}
    \begin{equation}
        \label{rho_sB_k_T}
        \begin{gathered}
            \varrho_{\mathbf{k},\sigma}^{\text{B}} = \langle\fnbsigma{\mathbf{k}}{\mathbf{k}}\rangle \\
            \Rightarrow \varrho_{\mathbf{k},\sigma}^{\text{B}} = u_{\mathbf{k},\sigma}^{2} \langle\fbcreatesb[\mathbf{k}]\fbannihilatesb[\mathbf{k}]\rangle + |v_{\mathbf{k},\sigma}|^{2} \langle\fbannihilatesa[-\mathbf{k}]\fbcreatesa[-\mathbf{k}]\rangle \\
            \Rightarrow \varrho_{\mathbf{k},\sigma}^{\text{B}}(T) = u_{\mathbf{k},\sigma}^{2} f(\omega_{\mathbf{k},\sigma}^{\beta},T) + |v_{\mathbf{k},\sigma}|^{2} [1 + f(\omega_{-\mathbf{k},\sigma}^{\alpha},T)] \\
            \Rightarrow \varrho_{\mathbf{k},\sigma}^{\text{B}}(T) = [1 + |v_{\mathbf{k},\sigma}|^{2}] f(\omega_{\mathbf{k},\sigma}^{\beta},T) + |v_{\mathbf{k},\sigma}|^{2} [1 + f(\omega_{\mathbf{k},\sigma}^{\alpha},T)] \\
            \therefore \boxed{
            \varrho_{\mathbf{k},\sigma}^{\text{B}}(T) = f(\omega_{\mathbf{k},\sigma}^{\beta},T) + |v_{\mathbf{k},\sigma}|^{2} [1 + f(\omega_{\mathbf{k},\sigma}^{\alpha},T) + f(\omega_{\mathbf{k},\sigma}^{\beta},T)]
            }
        \end{gathered}
    \end{equation}
    \begin{equation}
        \label{rho_tA_k_T}
        \begin{gathered}
            \varrho_{\mathbf{k},\tau}^{\text{A}} = \langle\fnatau{\mathbf{k}}{\mathbf{k}}\rangle \\
            \Rightarrow \varrho_{\mathbf{k},\tau}^{\text{A}} = u_{\mathbf{k},\tau}^{2} \langle\fbcreateta[\mathbf{k}]\fbannihilateta[\mathbf{k}]\rangle + |v_{\mathbf{k},\tau}|^{2} \langle\fbannihilatetb[-\mathbf{k}]\fbcreatetb[-\mathbf{k}]\rangle \\
            \Rightarrow \varrho_{\mathbf{k},\tau}^{\text{A}}(T) = u_{\mathbf{k},\tau}^{2} f(\omega_{\mathbf{k},\tau}^{\alpha},T) + |v_{\mathbf{k},\tau}|^{2} [1 + f(\omega_{-\mathbf{k},\tau}^{\beta},T)] \\
            \Rightarrow \varrho_{\mathbf{k},\tau}^{\text{A}}(T) = [1 + |v_{\mathbf{k},\tau}|^{2}] f(\omega_{\mathbf{k},\tau}^{\alpha},T) + |v_{\mathbf{k},\tau}|^{2} [1 + f(\omega_{\mathbf{k},\tau}^{\beta},T)] \\
            \therefore \boxed{
            \varrho_{\mathbf{k},\tau}^{\text{A}}(T) = f(\omega_{\mathbf{k},\tau}^{\alpha},T) + |v_{\mathbf{k},\tau}|^{2} [1 + f(\omega_{\mathbf{k},\tau}^{\alpha},T) + f(\omega_{\mathbf{k},\tau}^{\beta},T)]
            }
        \end{gathered}
    \end{equation}
    \begin{equation}
        \label{rho_tB_k_T}
        \begin{gathered}
            \varrho_{\mathbf{k},\tau}^{\text{B}} = \langle\fnbtau{\mathbf{k}}{\mathbf{k}}\rangle \\
            \Rightarrow \varrho_{\mathbf{k},\tau}^{\text{B}} = u_{\mathbf{k},\tau}^{2} \langle\fbcreatetb[\mathbf{k}]\fbannihilatetb[\mathbf{k}]\rangle + |v_{\mathbf{k},\tau}|^{2} \langle\fbannihilateta[-\mathbf{k}]\fbcreateta[-\mathbf{k}]\rangle \\
            \Rightarrow \varrho_{\mathbf{k},\tau}^{\text{B}}(T) = u_{\mathbf{k},\tau}^{2} f(\omega_{\mathbf{k},\tau}^{\beta},T) + |v_{\mathbf{k},\tau}|^{2} [1 + f(\omega_{-\mathbf{k},\tau}^{\alpha},T)] \\
            \Rightarrow \varrho_{\mathbf{k},\tau}^{\text{B}}(T) = [1 + |v_{\mathbf{k},\tau}|^{2}] f(\omega_{\mathbf{k},\tau}^{\beta},T) + |v_{\mathbf{k},\tau}|^{2} [1 + f(\omega_{\mathbf{k},\tau}^{\alpha},T)] \\
            \therefore \boxed{
            \varrho_{\mathbf{k},\tau}^{\text{B}}(T) = f(\omega_{\mathbf{k},\tau}^{\beta},T) + |v_{\mathbf{k},\tau}|^{2} [1 + f(\omega_{\mathbf{k},\tau}^{\alpha},T) + f(\omega_{\mathbf{k},\tau}^{\beta},T)]
            }
        \end{gathered}
    \end{equation}
\end{subequations}
From the above equations, it should be clear that $\varrho_{\mathbf{k},\sigma}^{\text{A}}(T) \neq \varrho_{\mathbf{k},\sigma}^{\text{B}}(T)$ and $\varrho_{\mathbf{k},\tau}^{\text{A}}(T) \neq \varrho_{\mathbf{k},\tau}^{\text{B}}(T)$ for general $\mathbf{k}$. This means at finite $T$, the population of magnon/orbiton modes at A and B sublattices are not same in general.

In the case of 2D decorated square lattice considered in the main manuscript, \textcolor{blue}{Eqs.~\eqref{ksym} and \eqref{HPcoeff} of main manuscript} yield $\mathcal{E}_{\sigma}^{\text{A}}(k_x,k_y) = \mathcal{E}_{\sigma}^{\text{B}}(k_y,k_x)$ and $\mathcal{E}_{\tau}^{\text{A}}(k_x,k_y) = \mathcal{E}_{\tau}^{\text{B}}(k_y,k_x)$ due to rotation and mirror symmetries. This further results in $\omega_{(k_x,k_y),\sigma}^{\alpha} = \omega_{(k_y,k_x),\sigma}^{\beta}$ and $\omega_{(k_x,k_y),\tau}^{\alpha} = \omega_{(k_y,k_x),\tau}^{\beta}$ respectively. Then:
\begin{subequations}
    \begin{gather}
        \begin{gathered}
            \sum_{\mathbf{k}} \omega_{\mathbf{k},\sigma}^{\alpha} = \sum_{k_x} \sum_{k_y} \omega_{(k_x,k_y),\sigma}^{\alpha} = \underbrace{\sum_{k_y} \sum_{k_x} \omega_{(k_y,k_x),\sigma}^{\alpha}}_{\substack{k_x \leftrightarrow k_y \\ \text{(dummy indices)}}} = \sum_{k_x} \sum_{k_y} \omega_{(k_x,k_y),\sigma}^{\beta} = \sum_{\mathbf{k}} \omega_{\mathbf{k},\sigma}^{\beta} \\
            \Rightarrow \boxed{\sum_{\mathbf{k}} \varrho_{\mathbf{k},\sigma}^{\text{A}}(T) = \sum_{\mathbf{k}} \varrho_{\mathbf{k},\sigma}^{\text{B}}(T)}
        \end{gathered} \\
        \begin{gathered}
            \sum_{\mathbf{k}} \omega_{\mathbf{k},\tau}^{\alpha} = \sum_{k_x} \sum_{k_y} \omega_{(k_x,k_y),\tau}^{\alpha} = \underbrace{\sum_{k_y} \sum_{k_x} \omega_{(k_y,k_x),\tau}^{\alpha}}_{\substack{k_x \leftrightarrow k_y \\ \text{(dummy indices)}}} = \sum_{k_x} \sum_{k_y} \omega_{(k_x,k_y),\tau}^{\beta} = \sum_{\mathbf{k}} \omega_{\mathbf{k},\tau}^{\beta} \\
            \Rightarrow \boxed{\sum_{\mathbf{k}} \varrho_{\mathbf{k},\tau}^{\text{A}}(T) = \sum_{\mathbf{k}} \varrho_{\mathbf{k},\tau}^{\text{B}}(T)}
        \end{gathered}
    \end{gather}
\end{subequations}
Thus the global magnon and orbiton population, $\varrho_{\sigma}(T)$ and $\varrho_{\tau}(T)$, can still be evaluated by summing the local magnon and orbiton population respectively, over the Brillouin zone, at either A or B sublattice. But the magnon and orbiton hopping amplitudes begin to differ at finite $T$ as mentioned in \textcolor{blue}{Sec.~\ref{TDSWOW} of main manuscript}.

\section{Classical Monte-Carlo Method}

Classical Monte-Carlo simulations were performed on the 2D decorated square lattice using the Metropolis-Hastings algorithm \cite{Metropolis1953_sm, Hastings1970_sm}. Lattice sizes were varied and a $16\times16$ square lattice was chosen for minimizing finite-size effects. Temperature of the system was varied from 0.05 to 5.05 in steps of 0.03 (dimensionless units) and $10^4$ Monte-Carlo steps were performed at each temperature step for thermalizing the system. The equilibrium data of observables at each temperature was averaged over $10^5$ Monte-Carlo steps. The spins and orbital isospins were normalized to unity using the quantum spin and isospin length $\sqrt{\sigma(\sigma + 1)}$ and $\sqrt{\tau(\tau + 1)}$ respectively. To simulate the thermal fluctuations and proper uniform sphere sampling, the spins and orbital isospins were chosen to lie randomly on a cone around the initial spins and orbital isospins following Ref.~\cite{Serena1993_sm}.

\putbib[sm]    

\end{bibunit}

\end{document}